\documentclass[aps,prb,superscriptaddress,floatfix,showkeys]{revtex4}

\usepackage{amsmath} 
\usepackage{graphicx} 
\usepackage{dcolumn} 
\usepackage{subfigure} 
\usepackage{mathrsfs} 
\usepackage{multirow}
\usepackage{mciteplus} 
\usepackage{color}
\usepackage{blkarray}

\AtBeginDocument{\nocite{achemso-control}}

\newcommand{\erf}[1]{Eq.~(\ref{#1})} 
\newcommand{\bra}[1]{\left\langle{#1}\right\vert} 
\newcommand{\ket}[1]{\left\vert{#1}\right\rangle}

\begin{document}

\title{\textit{Ab-Initio} Calculation of Molecular Aggregation Effects: a Coumarin-343 Case Study}
\author{Donghyun Lee}
\affiliation{Department of Chemistry, University of California, Berkeley, California, USA}
\author{Loren Greenman}
\affiliation{Department of Chemistry, University of California, Berkeley, California, USA} 
\author{Mohan Sarovar}
\affiliation{Scalable and Secure Systems Research, Sandia National Laboratories, MS 9158, 7011 East Avenue, Livermore, California 94550 USA} 
\author{K. Birgitta Whaley}
\email[{\bf Corresponding Author}\\E-mail: \ \ whaley@berkeley.edu\\]{}
\affiliation{Department of Chemistry, University of California, Berkeley, California, USA}
\date{\today}

\begin{abstract}
\noindent
{\bf ABSTRACT:}
We present time-dependent density functional theory (TDDFT) calculations for single and dimerized Coumarin-343 molecules in order to investigate the quantum mechanical effects of chromophore aggregation in extended systems designed to function as a new generation of sensors and light-harvesting devices.
Using the single-chromophore results, we describe the construction of effective Hamiltonians to predict the excitonic properties of aggregate systems.
We compare the electronic coupling properties predicted by such effective Hamiltonians to those obtained from TDDFT calculations of dimers, and to the coupling predicted by the transition density cube (TDC) method.
We determine the accuracy of the dipole-dipole approximation and TDC with respect to the separation distance and orientation of the dimers.
In particular, we investigate the effects of including Coulomb coupling terms ignored in the typical tight-binding effective Hamiltonian.
We also examine effects of orbital relaxation which cannot be captured by either of these models.
\end{abstract}


\maketitle

\section{Introduction}
The geometry of chromophore aggregates influences how they couple to one another, which in turn determines the electronic properties of the extended system.
Deliberate control over the positions and orientations of chromophores can thereby be used to achieve specific properties such as efficient energy transfer~\cite{Scholes2005}.
Light harvesting in photosynthesis is an example from nature of how chromophores embedded in proteins have been optimized by evolution to capture light over a specific spectrum and to efficiently transfer the excitation energy to the photosynthetic reaction center~\cite{Cheng_Fleming_2009, Scholes:2011qf, Bla-2002, Ame.Val.etal-2000}.
In designing synthetic light harvesting antennae for organic sensors or photovoltaics, it is critical to understand how the structural and orientational properties of the chromophore arrays affects their coupling, and hence their excitonic and optical properties~\cite{Scholes2003, Scholes1996a, Cao:2009kx, Didraga:2002zr, Didraga2004, Sarovar:2013tc}.
Importantly, recent work has suggested that quantum mechanical effects play a key role in the high efficiency of biological excitation energy transfer (EET)~\cite{Sch-2010, Collini2010, Engel2007, Sarovar:2010hc}.
Understanding the details of the quantum mechanical coupling between chromophores is an important step towards probing and exploiting any beneficial effects of quantum mechanical coherence in energy transfer.

Due to its spectroscopic attributes and small size, coumarin-343 (see Fig.~\ref{fig:chromophores}) is a molecule of particular interest for use in virus-templated synthetic light harvesting complexes, using \textit{e.g.} the tobacco mosaic virus (TMV) protein scaffold~\cite{Ma2008}.
The TMV protein monomers undergo self-assembly to form complex structures such as helices and stacked disks~\cite{Klu-1999}.
The assembled TMV can be made into a light harvesting array by covalently attaching chromophores such as coumarin-343 onto the protein~\cite{End.Fuj.etal-2007,Ma2008,Nam:2010fw, Miller2007}.
The stacked disks and helical arrays of chromophores act as aggregate systems with optical properties (spectral width, peak absorption frequency) that differ from those of the free chromophore.
It is possible to exercise a fine degree of control over these optical properties by changing the concentration of chromophores and the positions at which they are attached~\cite{Witus:2011wz}.

Conventionally the interaction between excited states of nearby chromophores is modeled using a tight-binding Hamiltonian
of the form~\cite{Ame.Val.etal-2000, Scholes2006a}, 
\begin{equation}
	H = \sum_{i=1}^N \epsilon_i \sigma_i^\dagger \sigma_i + \sum_{i\ne j}^N J_{i,j}\sigma_i^\dagger \sigma_j ,\label{eqn:frenkel_hamiltonian} 
\end{equation}
where $\epsilon_i$ is the transition energy of chromophore $i$, $J_{i,j}$ is the coupling between chromophores $i$ and $j$, $\sigma_i^\dagger \equiv \ket{\psi_1^i}\bra{\psi_0^i}$ is a Pauli creation operator for an excitation on chromophore $i$, and $\sigma_i \equiv \ket{\psi_0^i}\bra{\psi_1^i}$ is a Pauli annihilation operator for an excitation on chromophore $i$. 
In this approach each molecule is treated as a two-level system and the number of excitations is conserved.
This truncated description of the intermolecular Hamiltonian is often referred to as the Frenkel Hamiltonian (or Frenkel exciton model) in the Heitler-London approximation \cite{Knoester2002a}.
Such effective Hamiltonian descriptions constitute the only feasible approach to study large molecular aggregates, since \textit{ab-initio} methods cannot be scaled to such large systems.
However, some of the parameters entering the effective Hamiltonian descriptions can be calculated using \textit{ab-initio} methods.
For instance, the coupling between chromophores ($J_{i,j}$), which is well-approximated by a Coulombic interaction between the transition densities of chromophores $i$ and $j$, can be calculated exactly using the transition density cube (TDC) method~\cite{Krueger:1998vn}.
The most common way to approximate this coupling is by the ideal dipole approximation (IDA)~\cite{Kasha1965} which treats the $1/r$ interaction between the two densities as an interaction between the dipole moments of the transition densities; several papers have compared the accuracy of the IDA to the more complete TDC description in various molecules ~\cite{Howard:2004kx, Aurora-Mun-oz-Losa:sb, Scholes2001a}.
The effective Hamiltonian in \erf{eqn:frenkel_hamiltonian} relies on the approximation that the aggregate wavefunction can be reasonably constructed from the product of monomer wavefunctions (a Heitler-London-type picture).
However such effective Hamiltonian descriptions may also neglect other potentially important aspects of the intermolecular interactions.
In particular, \erf{eqn:frenkel_hamiltonian} does not include electron exchange between the chromophores, nor does it allow for relaxation of the monomer wavefunctions.
To go beyond this approximate description requires making detailed electronic structure calculations on the aggregate, and this is what we undertake in this study.

In the following sections we present time-dependent density functional theory (TDDFT) results for coumarin-343 and compare them to results from the IDA and TDC.
We also compare to an effective Hamiltonian that we obtain from expressing the molecular Hamiltonian in a basis constructed from products of monomer wavefunctions.
TDDFT provides a middle ground between accuracy and computational expense, and we have performed some benchmarking calculations to determine the quality of TDDFT for coumarin.

The results presented here are similar in spirit to a number of recent \emph{ab-initio} studies of excited states of molecular aggregates, and we briefly review some of these studies here.
Firstly, Ref. \cite{Madjet2006}, which perhaps has the most overlap with the results in this work, develops a method referred to as TrEsp for using \emph{ab-initio} calculations of charge and transition densities for monomers to determine energies of excited states of coupled chromophores.
Next, some recent papers, \emph{e.g.}, \cite{Hsu:2001el, Curutchet:2007ji}, have examined the various components of electronic coupling in condensed media using \emph{ab-initio} methods, separately characterized through-bond and through-space contributions, and analyzed the effects of solvent properties on these.
Finally, several works have examined the aggregation mechanisms and subsequent excited states of molecular aggregates of dimers and extended systems using \emph{ab-initio} methods \cite{Sinnokrot:2006ka, Fink:2008he,Fink:2008jv,Zhao:2009kd,Liu:2011fk}.
These studies concentrate on aggregates common in molecular crystals (\emph{e.g.}, perylene bisimide (PBI) aggregates) and as a result focus on very densely packed systems; typical inter-molecular separation distances studied in these works are in the range $2$-$8$\AA.
In such self-assembling aggregates the mechanisms that dictate aggregation geometries and intermolecular potentials -- \emph{e.g.}, dispersion forces -- are critical to understanding excited state energies.
In contrast, in the papers cited above and in this work the focus is on molecular aggregates found in biomimetic or biological systems.
In such systems the inter-molecular separation is typically larger, and critically, the forces that dictate aggregation are mostly due to external influences such as protein scaffolds.
Therefore in such systems the intermolecular potentials play a smaller role although as we show below they cannot be ignored completely.

An outline of the remainder of the paper follows. In section~\ref{sec:onedye}, we present these benchmarking calculations and give the TDDFT transition energies and transition dipoles for a single coumarin-343 chromophore. 
These are the simplest parameters which can be used in \erf{eqn:frenkel_hamiltonian}.
Then, in section~\ref{sec:twodye}, we explore using TDDFT the energetics resulting from coupling between two dyes at a number of separation distances and orientations.
We compare the TDDFT exciton splitting energies to splittings calculated by approximate methods, and also examine the effects of aggregation on exciton energies and wave functions, in particular, the role of orbital relaxation and deviations from the solutions to the Heitler-London description of monomer coupling.

\section{Methods}
The geometries of all the monomers considered in this report have been minimized using the B3LYP/6-31G* level of theory~\cite{B3LYP,631g_h,631g_cnof,631g_polarization}.
Single-point ground and excited-state energy calculations were then performed using TDDFT~\cite{TDDFT,TDDFT_Rev} with the M06-HF~\cite{M06HF}, M11~\cite{Peverati2011}, M11-L~\cite{Peverati2012}, and PBE~\cite{PBE,PBE_Err} functionals and the 6-31G*~\cite{631g_h,631g_cnof,631g_polarization}, 6-31+G*~\cite{631g_h,631g_cnof,631g_polarization,321G_diffuse}, and 6-311G**~\cite{6311g,6311g_polarization} basis sets.
EOM-CCSD~\cite{EOMCCSD1,EOMCCSD2,EOMCCSD3} calculations were performed to benchmark the density functionals, and multi-configurational self-consistent-field (MCSCF)~\cite{MCSCF_Direct} calculations were used to examine the presence of multireference character.
The polarized continuum model (PCM)~\cite{TDDFT_CPCM} was used to model the effects of water solvation.
Range-corrected TDDFT~\cite{uBLYP} was also used to determine the energetic order of the charge-transfer states, which TDDFT predicts to be too low in energy.
All calculations were performed using the GAMESS electronic structure package~\cite{GAMESS,GAMESS2}, except for the transition density cube files (the one particle transition density projected onto a 3 dimensional cartesian grid) which were obtained from Q-Chem~\cite{QCHEM}.

The calculations in this paper use a modified coumarin-343 molecule which has an amide group that is necessary for attaching the dye on to the TMV substrate.
For simplicity, in this work we shall refer to this modified molecule as coumarin-343-MA (coumarin-343-methylamide).

\section{Results and Discussion}
\subsection{Single-chromophore benchmarking studies and parameter determintation\label{sec:onedye}}
\begin{figure}
	\begin{center}
		\subfigure[~Coumarin-343-MA]{
		\includegraphics[width=2.3in]{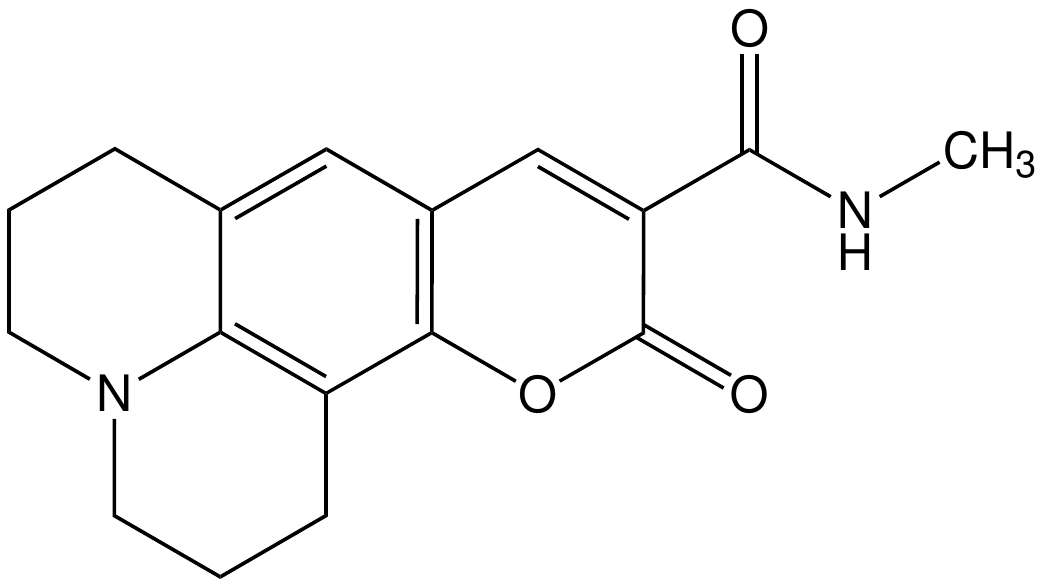}} \qquad \subfigure[~Coumarin]{ 
		\includegraphics[width=1.3in]{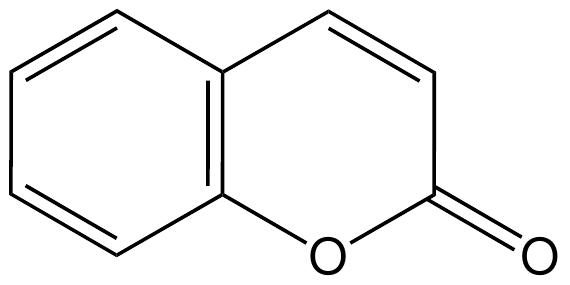}} 
	\end{center}
	\caption{\small
	(a) Structure of coumarin-343-MA, modified for attachment to TMV substrate.
	(b) The smaller coumarin molecule, on which we performed larger and more accurate calculations to benchmark the density functionals we used.
	\label{fig:chromophores}}
\end{figure}

In order to choose an appropriate density functional and basis set for coumarin-343-MA, we first ran benchmarking calculations on coumarin (see Fig.~\ref{fig:chromophores}).
We use the transition energy and the transition dipole moment to gauge accuracy, since these are the key quantities relevant in many effective Hamiltonian approaches.
Four TDDFT functionals were compared with the more accurate EOM-CCSD method, and the results are summarized in Table~\ref{tbl:coumarin_energy}. 

\begingroup
 \squeezetable 
\begin{table}[h]
	\caption{\small
	Density functional benchmarking for coumarin in the gas phase.
	The TDDFT excitation energies and oscillator strengths ($f$) for five functionals and three basis sets were compared with the EOM-CCSD results.
	The B3LYP functional matched both the excitation energy and the oscillator strength of EOM-CCSD within our tolerance.
    The results show only a very weak dependency to the choice of basis set.
	\label{tbl:coumarin_energy}} 
 \begin{ruledtabular}
	\begin{tabular}
		{ccddddd} \multicolumn{3}{c}{}&\multicolumn{2}{c}{$1^{st}$ Excited State}&\multicolumn{2}{c}{$2^{nd}$ Excited State}\\
		\cline{4-5}\cline{6-7} Basis Set & Functional & \multicolumn{1}{c}{Ground State Energy [Ha]} & \multicolumn{1}{c}{Exc. Energy [eV]} & \multicolumn{1}{c}{$f$} & \multicolumn{1}{c}{Exc. Energy [eV]} & \multicolumn{1}{c}{$f$}\\
		\hline 6-31G* & M06-HF & -496.8747 & 4.415 & 0.000 & 4.780 & 0.215\\
		& M11 & -496.7467 & 4.606 & 0.180 & 4.809 & 0.000\\
		& M11-L & -496.8800 & 4.152 & 0.052 & 4.407 & 0.000\\
		& PBE & -496.4380 & 3.625 & 0.000 & 3.822 & 0.060\\
		& B3LYP & -496.7418 & 4.182 & 0.111 & 4.396 & 0.000\\
		& EOM-CCSD & -495.5607 & 4.563 & 0.089 & 5.126 & 0.000\\
		\hline 6-31+G* & M06-HF & -496.8921 & 4.457 & 0.000 & 4.699 & 0.241\\
		& M11 & -496.7619 & 4.548 & 0.197 & 4.848 & 0.000\\
		& M11-L & -496.8908 & 4.078 & 0.055 & 4.414 & 0.000\\
		& PBE & -496.4571 & 3.705 & 0.000 & 3.782 & 0.062\\
		& B3LYP & -496.7612 & 4.129 & 0.119 & 4.453 & 0.000\\
		& EOM-CCSD & -495.5870 & 4.497 & 0.098 & 5.121 & 0.000\\
		\hline 6-311G** & M06-HF & -497.0182 & 4.481 & 0.000 & 4.707 & 0.239\\
		& M11 & -496.8850 & 4.552 & 0.181 & 4.775 & 0.000\\
		& M11-L & -497.0357 & 4.090 & 0.054 & 4.320 & 0.000\\
		& PBE & -496.5658 & 3.618 & 0.000 & 3.781 & 0.060\\
		& B3LYP & -496.8725 & 4.132 & 0.111 & 4.383 & 0.000\\
		& EOM-CCSD & -495.7996 & 4.502 & 0.095 & 5.068 & 0.000\\
	\end{tabular}
 \end{ruledtabular}
\end{table}
\endgroup

For each basis set, the EOM-CCSD method predicts a first excitation energy of 4.5-4.6 eV and an oscillator strength of 0.09-0.10.
The M11 functional is able to reproduce the excitation energy well, but it significantly overestimates the oscillator strength and transition dipoles.
Since the accuracy of the transition dipole directly affects the accuracy of the chromophore couplings, the B3LYP functional offered a better combination of accuracy in the energetics and the transition dipoles.
There is not a large basis set effect for the basis sets considered, and so we performed further calculations at the B3LYP/6-31G* level of theory.

We also check for multireference character, which occurs when multiple Slater determinants are needed to accurately express the zeroth-order wavefunction.
This is common for large or conjugated molecules.
MCSCF calculations were run on coumarin-343-MA using an active space of 7 electrons in 7 $\pi$-orbitals.
The natural orbital occupation numbers, which are the eigenvalues of the one-electron reduced density operator~\cite{Lowdin1955}, 
\begin{equation}\label{occ_num}
	\gamma(r_1, r_1^\prime) = N \int\dots\int\psi^*(r_1,r_2\dots r_N)\psi(r_1^\prime,r_2\dots r_N)\mathrm{d}r_2\dots\mathrm{d}r_N,
\end{equation}
are useful for determining multireference character; occupation numbers that deviate significantly from 0.0 or 2.0 indicate that a multireference wavefunction is needed.
We find that the occupation numbers of the occupied and unoccupied orbitals were all greater than 1.9 or less than 0.12.
This indicates that there is little multireference character for this dye and so the benchmark calculations and TDDFT calculations are both adequate.

\begin{table}[htp!]
\caption{\small
TDDFT results for Coumarin-343.
The ground-state and excitation energies, transition dipoles, and oscillator strengths are given for each functional using the 6-31G* basis set.
The PBE functional is not included as it performed poorly in the benchmark calculations on the coumarin molecule.
The final column gives the B3LYP result using a continuum model to describe solvation in water.
\label{tbl:coumarin343}}
\begin{ruledtabular}
\begin{tabular}
	{ccddddd} \multicolumn{2}{c}{} & \multicolumn{1}{c}{M06-HF} & \multicolumn{1}{c}{M11} & \multicolumn{1}{c}{M11-L}& \multicolumn{1}{c}{B3LYP }&\multicolumn{1}{c}{B3LYP (Water)}\\
	\hline Ground State Energy [Ha]&&-993.5938&-993.3408&-993.5986&-993.2967&-993.3128\\
	1$^{st}$ Excitation Energy [eV]&&3.927&3.839&3.445&3.465&3.199\\
	\hline &x& 0.062 & 0.070 &  -0.014 &  0.000 & -0.076\\
	Transition Dipole [Debye]&y& -0.022 &  0.114 & -0.018 &  0.000 &  0.034\\
	&z&  7.228 &  7.029 &  6.101 &  6.706 &  7.611\\
	\hline Oscillator Strength&&0.778&0.729&0.486&0.591&0.703

\end{tabular}
\end{ruledtabular}
\end{table}

\begin{figure}[htp!]
\begin{center}
\subfigure[~transition dipole moment]{\label{fig:coumarin343tdm}
\includegraphics[width=2in]{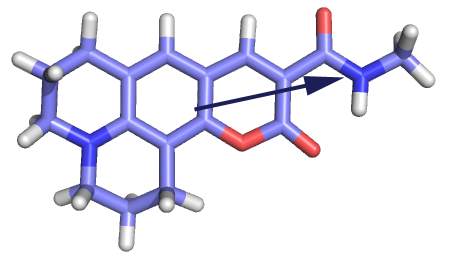} } 
\subfigure[~HOMO]{ 
\includegraphics[width=2.1in]{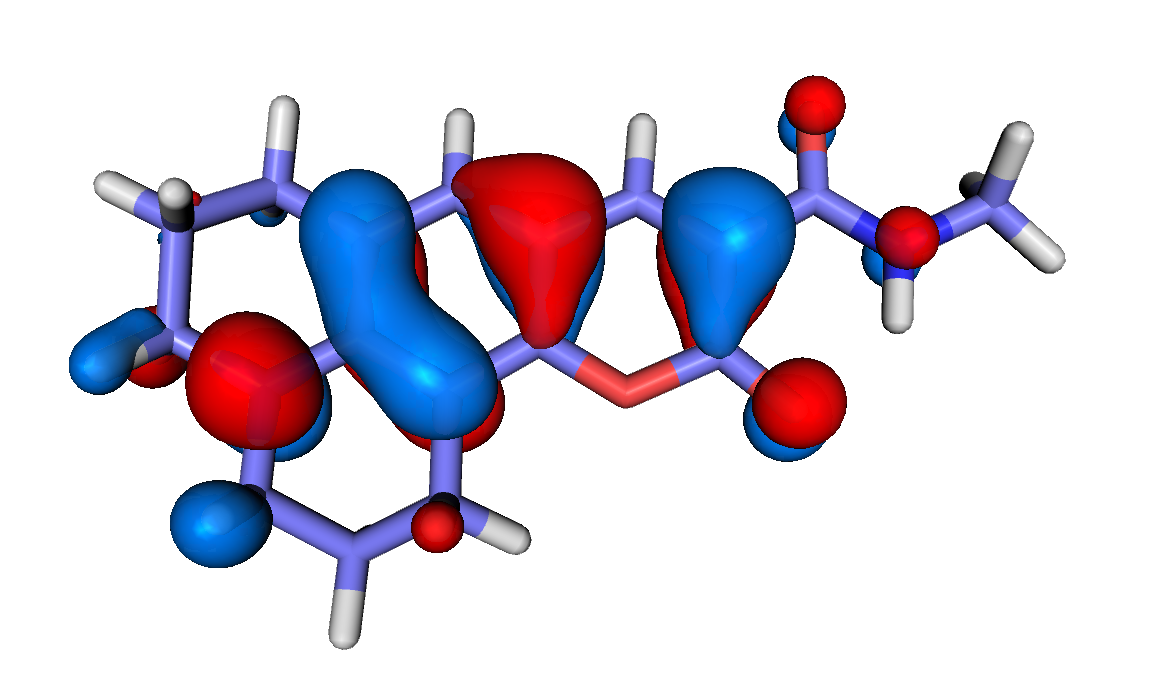} }
\subfigure[~LUMO]{ 
\includegraphics[width=2.1in]{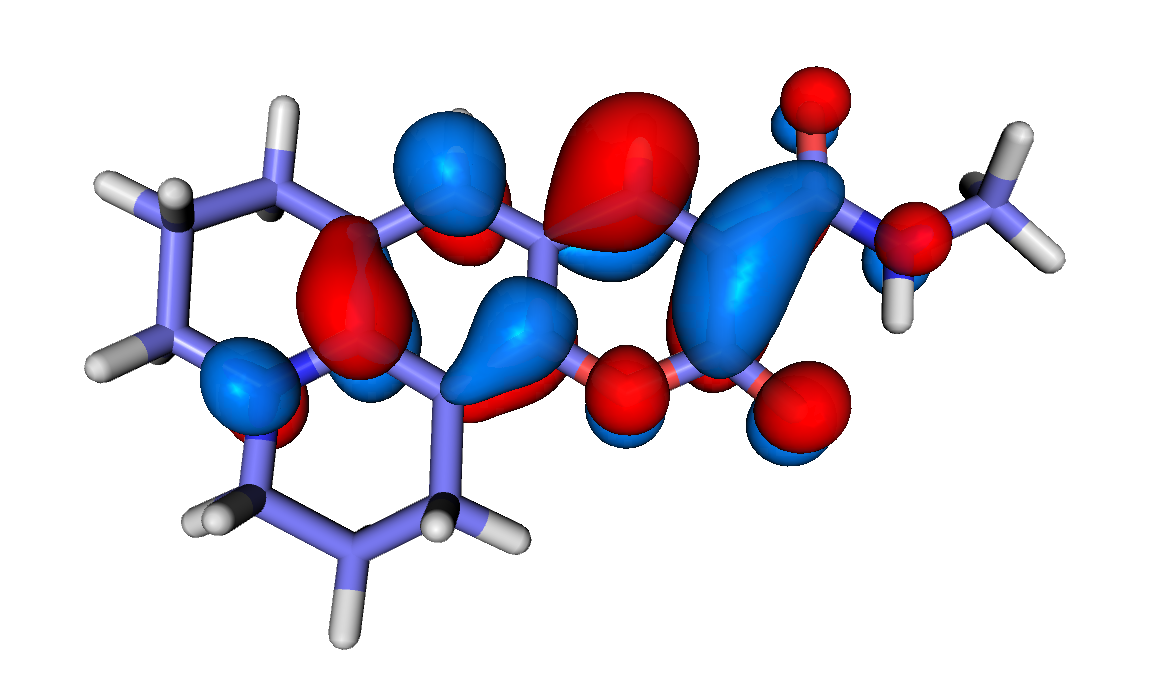}} 
\end{center}
\caption{\small
	(a) The transition dipole of coumarin-343-MA calculated at the TD-B3LYP/6-31G* level of theory.
	(b) The highest occupied molecular orbital (HOMO) of coumarin-343-MA. (c) The lowest unoccupied molecular orbital (LUMO) of coumarin-343-MA.
	\label{fig:coumarin343}}
\end{figure}

The B3LYP/6-31G* excitation energies and transition dipoles for coumarin-343-MA are given in Table~\ref{tbl:coumarin343}, together with results for the other functionals.
At this level of theory, the first excitation energy is 3.465 eV and the oscillator strength is 0.591.
The transition dipole between the ground and first excited states of coumarin-343-MA is pictured in Fig.~\ref{fig:coumarin343tdm}.
We see that the transition dipole lies flat along the plane of the molecule, on the axis formed between the center of the molecule and the nitrogen atom of the amide group.

\subsection{Two-dye coupling calculations\label{sec:twodye}}
Now we turn to assessing the magnitude of the coupling between excited states of pairs of molecules, where these are denoted as A and B.
Using the benchmarking studies described in the previous subsection, we calculated the energetics of two coumarin-343-MA molecules at various separation distances and different relative orientations, using the B3LYP/6-31G* functional and basis set.
With one molecule aligned with its molecular axis along the $\phi_A = 0$ azimuth in the $xy$-plane, the relative orientations are defined by the two polar angles $\theta_A, \theta_B$ and the second azimuthal angle $\phi_B$, where the polar angle is the angle between the long axis of the molecule and a laboratory fixed $z$-axis (see Appendix A).

In the following we will use $E_i ~(\tilde{E}_i)$ to denote the monomer (dimer) energies calculated using TDDFT.

For a homo-dimer, the tight-binding effective Hamiltonian of \erf{eqn:frenkel_hamiltonian} reduces to
\begin{align}
	\hat{H} &= \begin{blockarray}{ccc}\label{eqn:frenkel_hmat}
	& \ket{10} & \ket{01}\\
	\begin{block}{c (cc)}
	 	\bra{10} \ \	&	\epsilon_0 & J\\
	 	\bra{01} \ \	&	J & \epsilon_0\\
	\end{block}
	\end{blockarray}
\end{align}
in a basis of excitations on the left ($\ket{10}$) and right ($\ket{01}$) chromophores, where $\epsilon_0 = E_1-E_0$ is the first excitation energy of the monomer and $J$ is the coulombic coupling between monomers.
As mentioned in the Introduction, in the absence of exchange this coupling can be captured as the Coulomb interaction between transition densities (e.g. in the TDC method), and this can be further approximated as a dipole-dipole interaction (as in the IDA).
Diagonalizing the matrix in \erf{eqn:frenkel_hmat} results in delocalized exciton states with symmetrically split energies, also known as Davydov splitting~\cite{Davydov1964},
\begin{align}\label{eqn:frenkel_eig}
	\ket{\psi_\pm} &= \frac{\ket{10} \pm \ket{01}}{\sqrt{2}}	&	\epsilon_\pm &= \epsilon_0 \pm J .
\end{align}
W emphasize that this theory predicts a symmetric splitting of the exciton energies according to \erf{eqn:frenkel_eig}.
In this section we will assess the degree to which TDDFT calculations agree with the tight-binding effective Hamiltonian description of the excited states of the coumarin-343-MA homo-dimer.
In particular, we will examine three specific aspects, for various chromophore separation distances and chromophore orientations: (i) we will compare the TDDFT description of the Coulombic coupling magnitude $J$ to the magnitudes provided by IDA and TDC, (ii) we will assess the validity of the form of the symmetric eigenstates given in \erf{eqn:frenkel_eig}, and (iii) we will assess the validity of the symmetric splitting of eigenenergies given in \erf{eqn:frenkel_eig}.

\subsubsection{Coulomb coupling energy}
\begin{figure}[htp!]
\begin{minipage}{5.in}
	\includegraphics{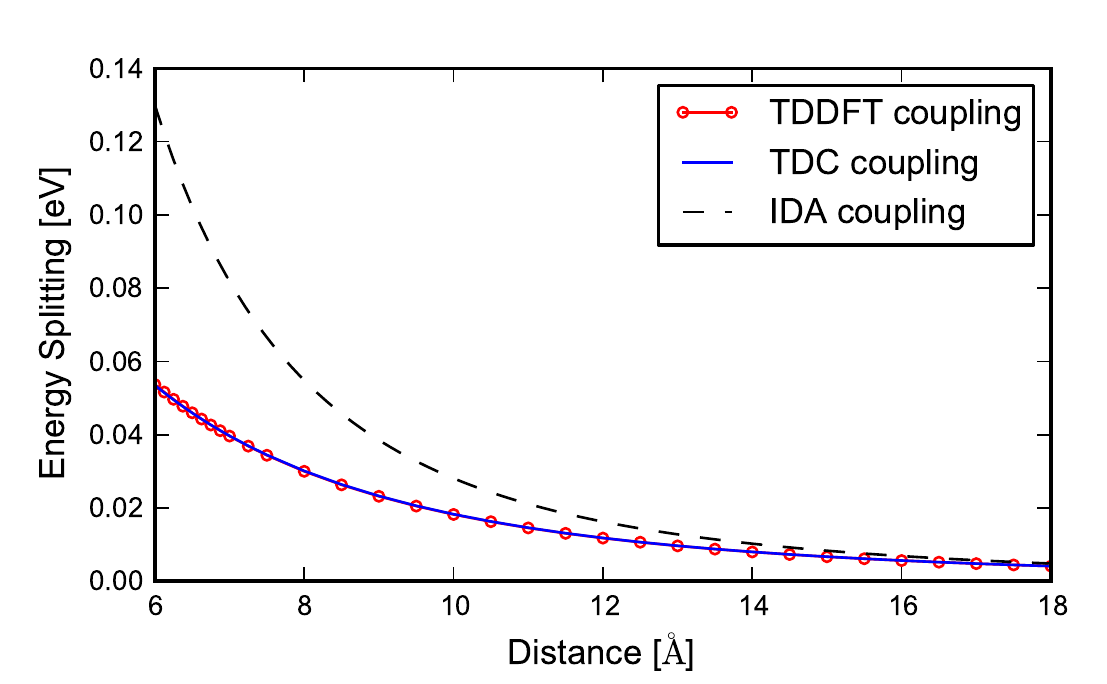}\label{fig:dimer0_rel_coupl} 
\end{minipage}
\begin{minipage}{1.4in}
    \includegraphics[width=1.4in]{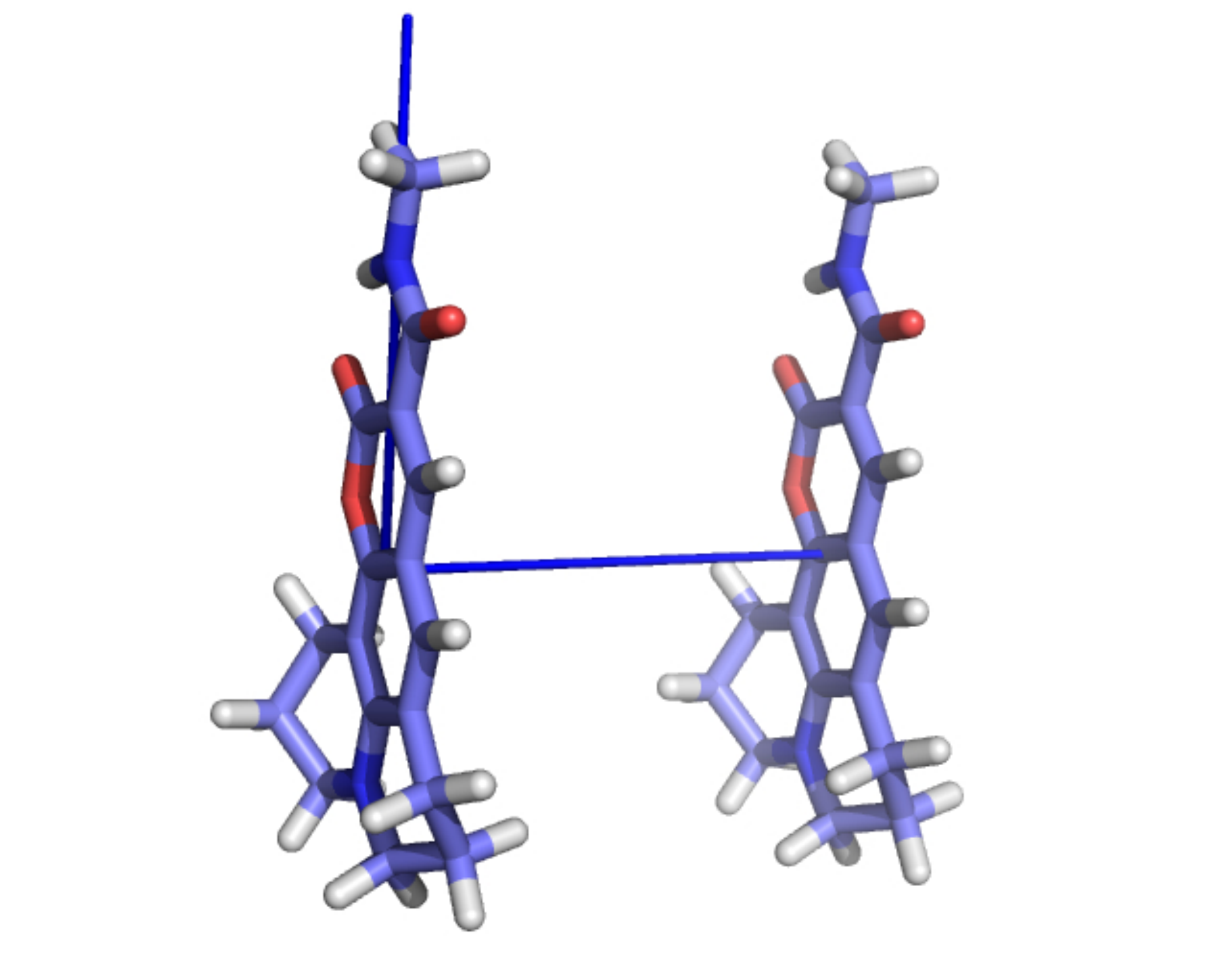}
	Parallel-$0^\circ$
\end{minipage}
\begin{minipage}{5.in}
	\includegraphics{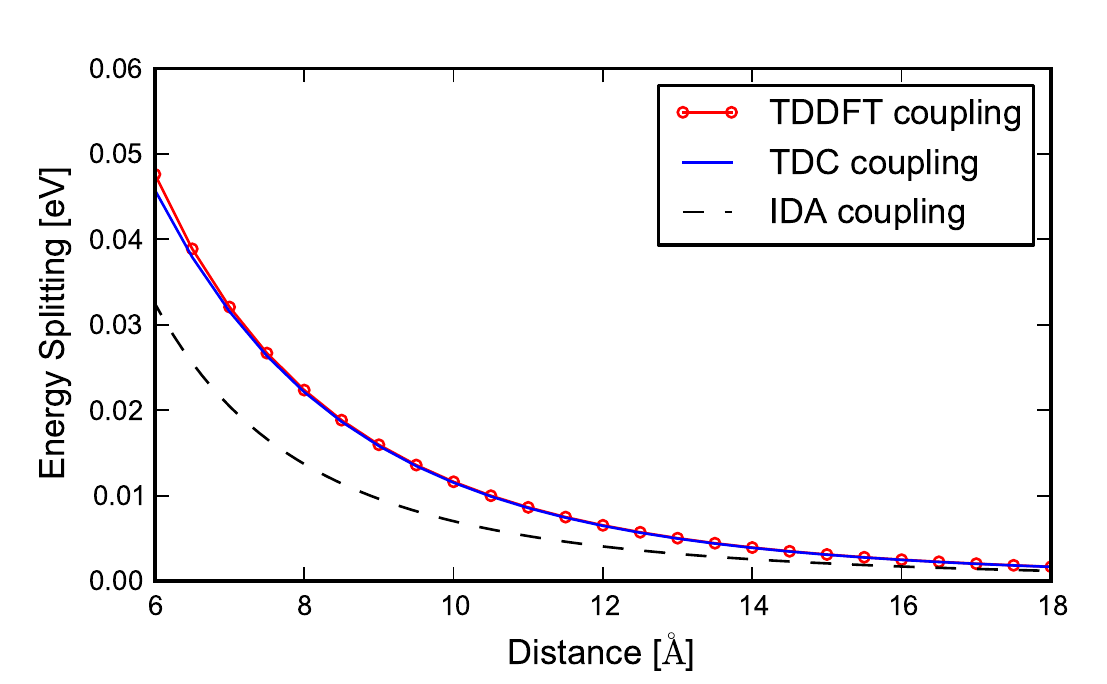}\label{fig:dimer30_rel_coupl} 
\end{minipage}
\begin{minipage}{1.4in}
    \includegraphics[width=1.4in]{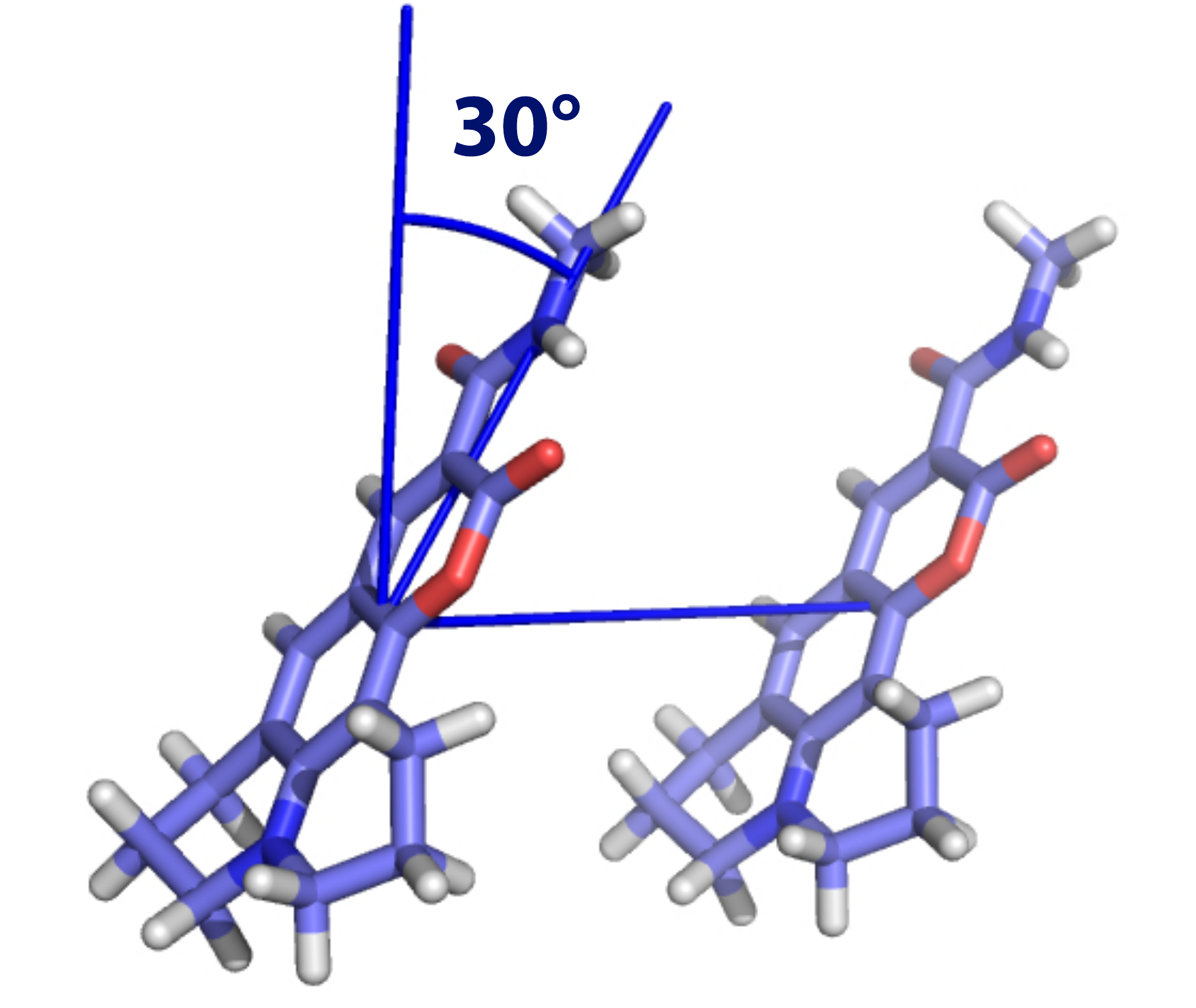}
	Parallel-$30^\circ$
\end{minipage}
\begin{minipage}{5.in}
	\includegraphics{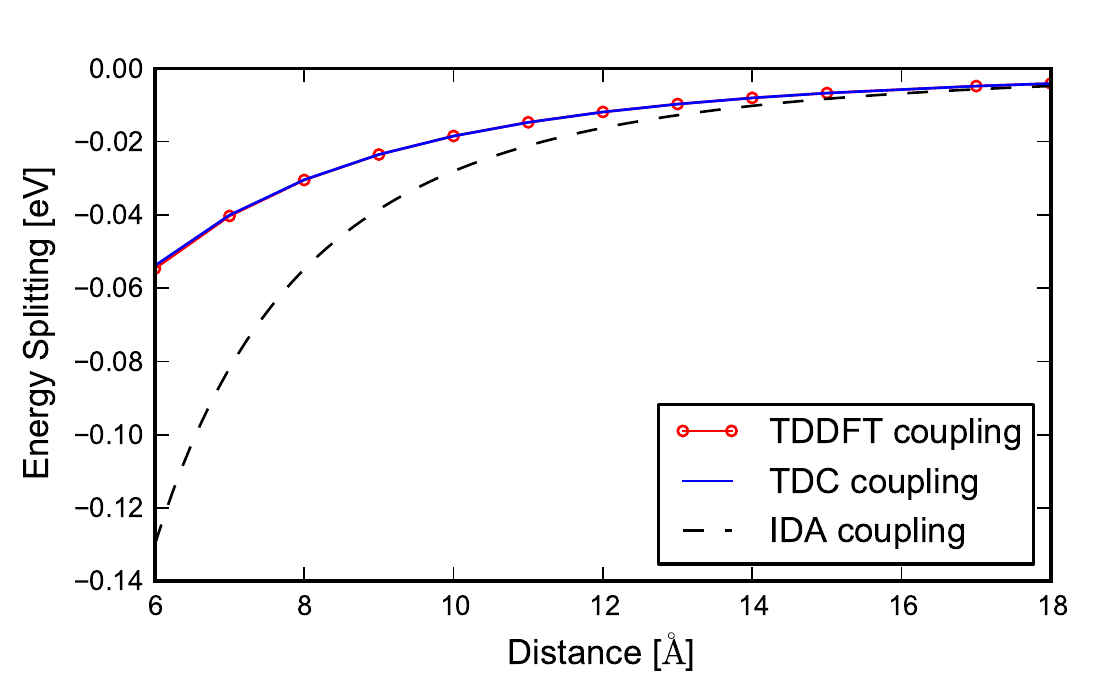}\label{fig:dimer0_180_rel_coupl} 
\end{minipage}
\begin{minipage}{1.4in}
    \includegraphics[width=1.4in]{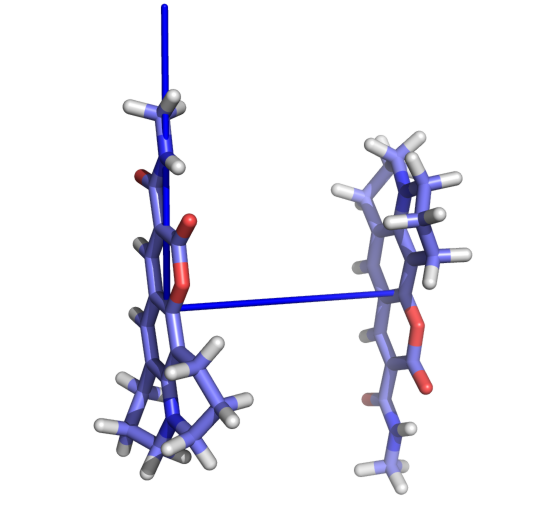}
	Anti-Parallel-$0^\circ$
\end{minipage}

\caption{\small
	The energy splittings between the dimer excited states are shown for TDDFT, TDC and IDA calculations.
	The TDDFT points show  $\frac{\tilde{E_2}-\tilde{E_1}}{2}$, while the TDC and IDA lines show $\frac{\epsilon_+ - \epsilon_-}{2}$ with the $J$ coupling calculated using the respective approximation.
	The TDC and TDDFT predictions mostly overlap.
\label{fig:dimer_coupling}}
\end{figure}

Fig.~\ref{fig:dimer_coupling} shows the splitting of the excited state energies ($\frac{\tilde{E}_2-\tilde{E}_1}{2}$) for TDDFT as a function of the inter-chromophore separation distance for three different relative orientations.
For comparison, we have also plotted the energetic splitting predicted by \erf{eqn:frenkel_eig} when the Coulomb coupling $J$ is calculated using the IDA and TDC methods.
Comparison of this predicted energetic splitting is the most consistent methods for comparing the three methods.

From Fig.~\ref{fig:dimer_coupling} we see that the numerically integrated TDC method agrees very well with the TDDFT calculations, while the IDA over-predicts for the $0^\circ$ relative orientation (parallel and anti-parallel) and under-predicts for the $30^\circ$ relative orientation.
As the intermolecular separation increases, the IDA values begin to qualitatively match the TDDFT/TDC values after 12\AA~separation, however the convergence of the percent error is still quite slow.
The percent error of the IDA splitting decreases to 10\% only after 30\AA~separation (averaged over the three orientations).
This shows that the IDA can be a poor description of Coulombic coupling for inter-chromophoric distances that are less than 30\AA.
More sophisticated methods such as TDC should be used in such cases.
This conclusion is in agreement with Refs. ~\cite{Howard:2004kx, Aurora-Mun-oz-Losa:sb, Scholes2001a}.

Calculations where the relative orientation between dimers is explored while keeping the distance separation fixed were also done.
The results confirm that the TDC can reliably predict the energetic splittings. TDC systematically outperforms IDA, and is also able to predict the correct splitting in geometries where the molecules are nearly in contact with one another.
Details of this analysis may be found in the appendix.

\subsubsection{Exciton energies and site-energy shifts}
\begin{figure}[htp!]
	\begin{minipage}{5.in}
		\includegraphics{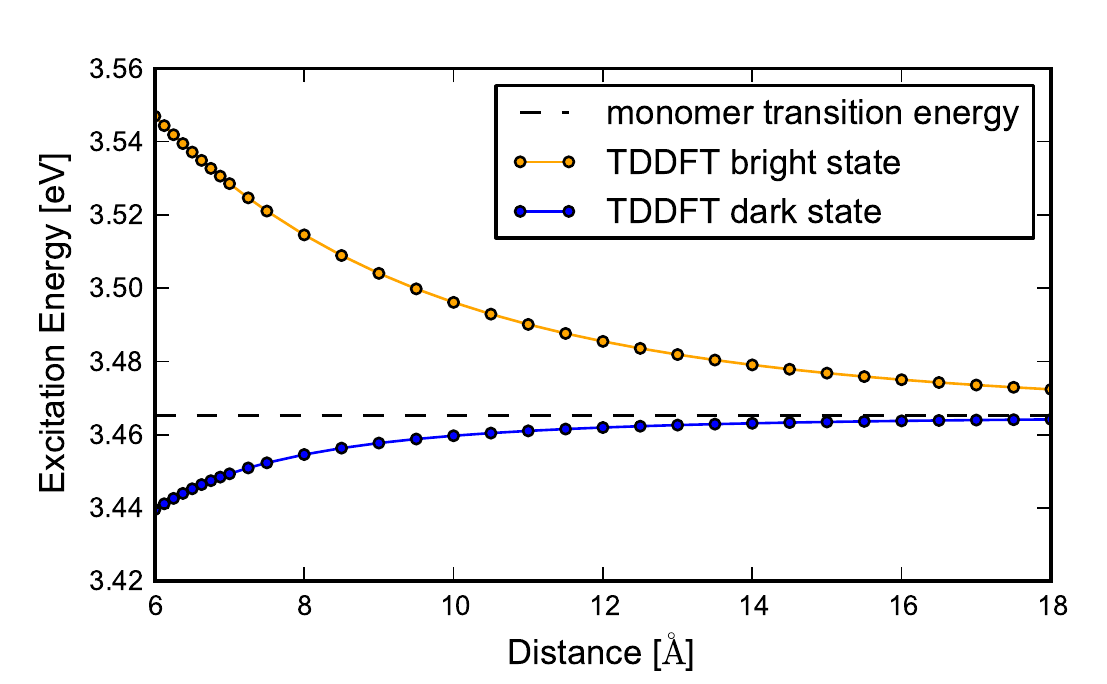}\label{fig:dimer0} 
	\end{minipage}
	\begin{minipage}{1.4in}
	    \includegraphics[width=1.4in]{figures/two_dye_par_tilted0_sep9.pdf}
		Parallel-$0^\circ$
	\end{minipage}
	\begin{minipage}{5.in}
		\includegraphics{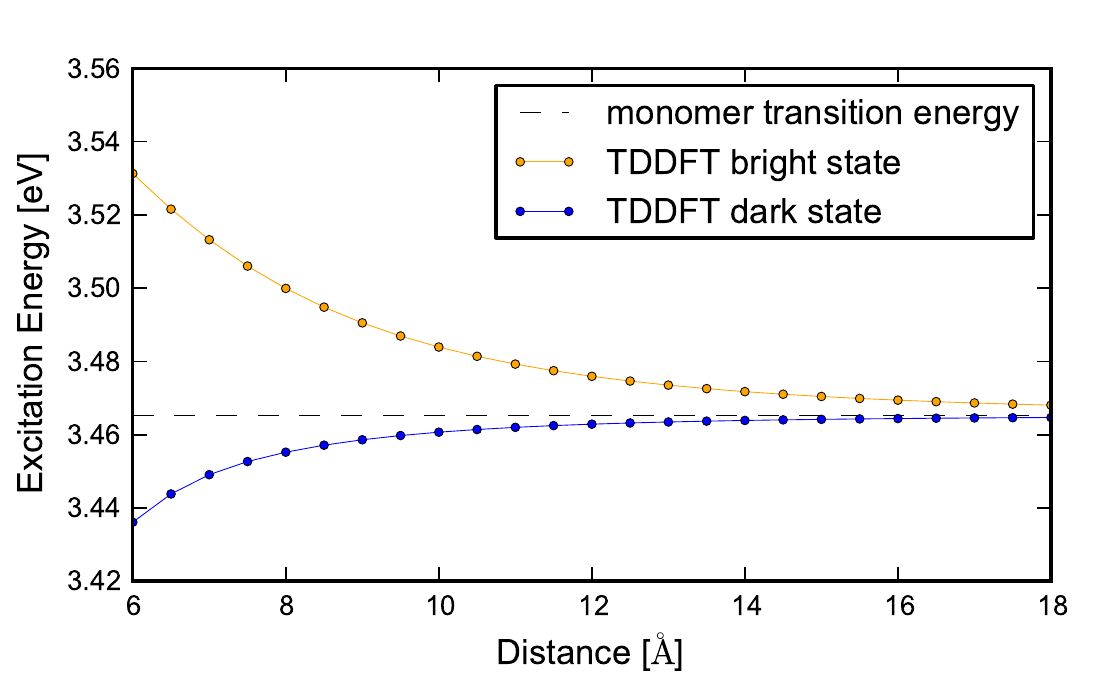}\label{fig:dimer30} 
	\end{minipage}
	\begin{minipage}{1.4in}
		\includegraphics[width=1.4in]{figures/two_dye_par_tilted30_sep9.pdf}
		Parallel-$30^\circ$
	\end{minipage}
	\begin{minipage}{5.in}
		\includegraphics{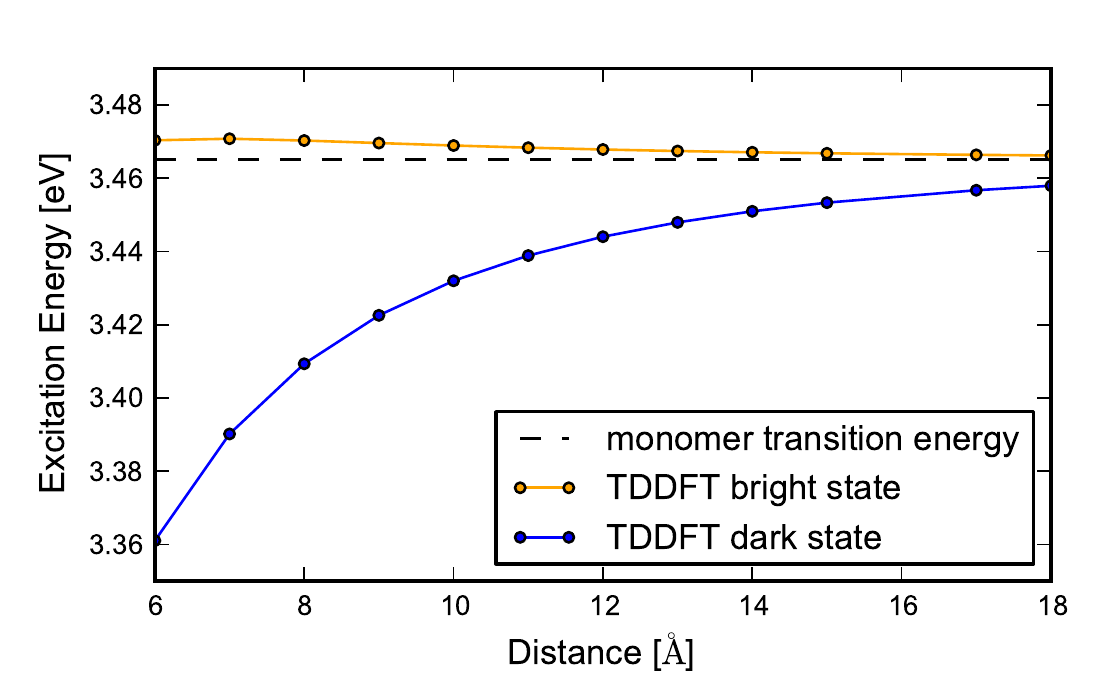}\label{fig:dimer0_180} 
	\end{minipage}
	\begin{minipage}{1.4in}
		\includegraphics[width=1.4in]{figures/anti_parallel.png}
		Anti-Parallel-$0^\circ$
	\end{minipage}
	\caption{\small
	TDDFT excitation energies ($\tilde{E_1}-\tilde{E_0}$ and $\tilde{E_2}-\tilde{E_0}$) for the coupled excited states of the coumarin-343-MA dimer.
	Orientation of the dimer is shown to the right.
	Erroneous charge transfer states are not shown, since they do not mix into the optical states (see appendix for details).
	\label{fig:energy_lvls}}
\end{figure}

Fig.~\ref{fig:energy_lvls} shows TDDFT excitation energies as a function of inter-chromophore distance at three different orientations.
The most striking feature of these plots is that the excitation energies split asymmetrically from the energy of the monomer excited state (indicated by the dashed line).
This is in disagreement with the tight-binding effective Hamiltonian which predicts a symmetric splitting of the energies around the monomer energy (see \erf{eqn:frenkel_eig}).

To explain this discrepancy we must re-examine the full molecular Hamiltonian of the coupled chromophore system.
In a basis of single-chromphore ground states and single excitations,
\begin{align*}
	\ket{00} &= \ket{\psi_0^A} \otimes \ket{\psi_0^B}	&	\ket{10} &= \ket{\psi_1^A} \otimes \ket{\psi_0^B}\\
	\ket{11} &= \ket{\psi_1^A} \otimes \ket{\psi_1^B}	&	\ket{01} &= \ket{\psi_0^A} \otimes \ket{\psi_1^B}\\
\end{align*}
where $\ket{\psi_i^A} \otimes \ket{\psi_j^B}$ indicates a direct product wavefunction between molecule A in state $i$ and molecule B in state $j$,
the Born-Oppenheimer molecular Hamiltonian is written as~\cite{Agranovich2008}
\begin{align}
	\hat{H} &= \begin{blockarray}{ccccc}\label{eqn:hamiltonian_matrix}
	& \ket{00}	& \ket{10}& \ket{01} & \ket{11}\\
	\begin{block}{c (cccc)}
	 	\bra{00} \ \	&	2E_0 + J_\text{gs}^\text{gs}&	J_\text{trans}^\text{gs}			&	J_\text{gs}^\text{trans}			&	J_\text{trans}^\text{trans}\\
	 	\bra{10} \ \	&	J_\text{trans}^\text{gs}	&	E_0 + E_1 + c J_\text{es}^\text{gs}	&	J_\text{trans}^\text{trans}			&	J_\text{es}^\text{trans}\\
	 	\bra{01} \ \	&	J_\text{gs}^\text{trans}	&	J_\text{trans}^\text{trans}			&	E_0 + E_1 + c J_\text{gs}^\text{es}	&	J_\text{trans}^\text{es}\\
	 	\bra{11} \ \	&	J_\text{trans}^\text{trans}	&	J_\text{es}^\text{trans}			&	J_\text{trans}^\text{es}			&	2 E_1 + J_\text{es}^\text{es}\\
	\end{block}
	\end{blockarray}
\end{align}
where $E_i$ are the relevant monomer energies.
Ignoring the effects of quantum mechanical exchange, which we have determined from the TDDFT calculations to be small at these distances, the coupling terms $J_i^j$ indicate coulomb integrals between either a charge or transition density on molecule A and either a charge or transition density on molecule B, i.e. 
\begin{align}
	J_\text{i}^\text{j} &= \int\rho_{i}^A(r_1) \frac{1}{|r_1 - r_2|}\rho_{j}^B(r_2)\mathrm{d}r_1\mathrm{d}r_2\label{eqn:trans_coulomb_integral}.
\end{align}
with $i,j$ equal to either $gs$ (ground state charge density), $es$ (excited state charge density), or $trans$ (transition density).
The charge densities and transition densities are defined for molecule A as
\begin{align}
	\rho_{\text{gs}}^A(r_1) &= N\int\dots\int\psi_0^{A*}(r_1,r_2\dots r_N)\psi_0^{A}(r_1,r_2\dots r_N) \mathrm{d}r_2\dots\mathrm{d}r_N \times \sum_{n \in A} Z_n\delta(R_n - r_1)\\
	\rho_{\text{es}}^A(r_1) &= N\int\dots\int\psi_1^{B*}(r_1,r_2\dots r_N)\psi_1^{B}(r_1,r_2\dots r_N) \mathrm{d}r_2\dots\mathrm{d}r_N \times \sum_{n \in B} Z_n\delta(R_n - r_1)\\
	\rho_{\text{trans}}^A(r_1) &= \int\dots\int\psi_0^{A*}(r_1,r_2\dots r_N)\psi_1^{A}(r_1,r_2\dots r_N)\mathrm{d}r_2\dots\mathrm{d}r_N,
\end{align}
and similarly for molecule B, where $Z_n$ and $R_n$ correspond to the charge and positions of the nuclei in their respective molecules, and $N$ is the total number of electrons in a molecule.
The coefficient $c$ is a parameter used to scale the magnitude of the ground-state/excited-state Coulomb integral (see discussion below).

In order to reduce this Hamiltonian into the tight-binding effective Hamiltonian in \erf{eqn:frenkel_hmat} a number of approximations must be made.
Firstly, the Coulomb integrals are assumed to be much smaller than the energetic differences and therefore the matrix in \erf{eqn:hamiltonian_matrix} is approximated as block diagonal, with each block labeled by the number of excited states.
Explicitly, $J_i^j \ll E_1-E_0$ but $J_\text{es}^\text{gs} \approx J_\text{gs}^\text{es}$, and therefore we can ignore all off-diagonal terms except for those coupling $\ket{10}$ and $\ket{01}$.
This approximation is sometimes referred to as the Heitler-London approximation in the literature ~\cite{Agr.Bas-2000}.
Typically, the intermolecular Coulomb interaction terms are ignored i.e., one assumes that $J_\text{gs}^\text{gs} \approx J_\text{es}^\text{gs} \approx J_\text{gs}^\text{es} \approx 0$, and then the only Coulomb interaction terms that remain are the $J_\text{trans}^\text{trans}$ terms.
Under this approximation the Hamiltonian in the single excitation subspace (after a shift of the diagonal energies by $2 E_0$) is the one given in \erf{eqn:frenkel_hmat},
\begin{align}
	\hat{H} - 2E_0 &= \begin{blockarray}{ccc}\label{eqn:frenkel_hmat2}
	& \ket{10} & \ket{01}\\
	\begin{block}{c (cc)}
	 	\bra{10} \ \	&	\epsilon_0	&	J_\text{trans}^\text{trans}\\
	 	\bra{01} \ \	&	J_\text{trans}^\text{trans}	&	\epsilon_0\\
	\end{block}
	\end{blockarray}
\end{align}
with $\epsilon_0 = E_1-E_0$.

\begin{figure}[htp]
\begin{minipage}{5in}
	\includegraphics{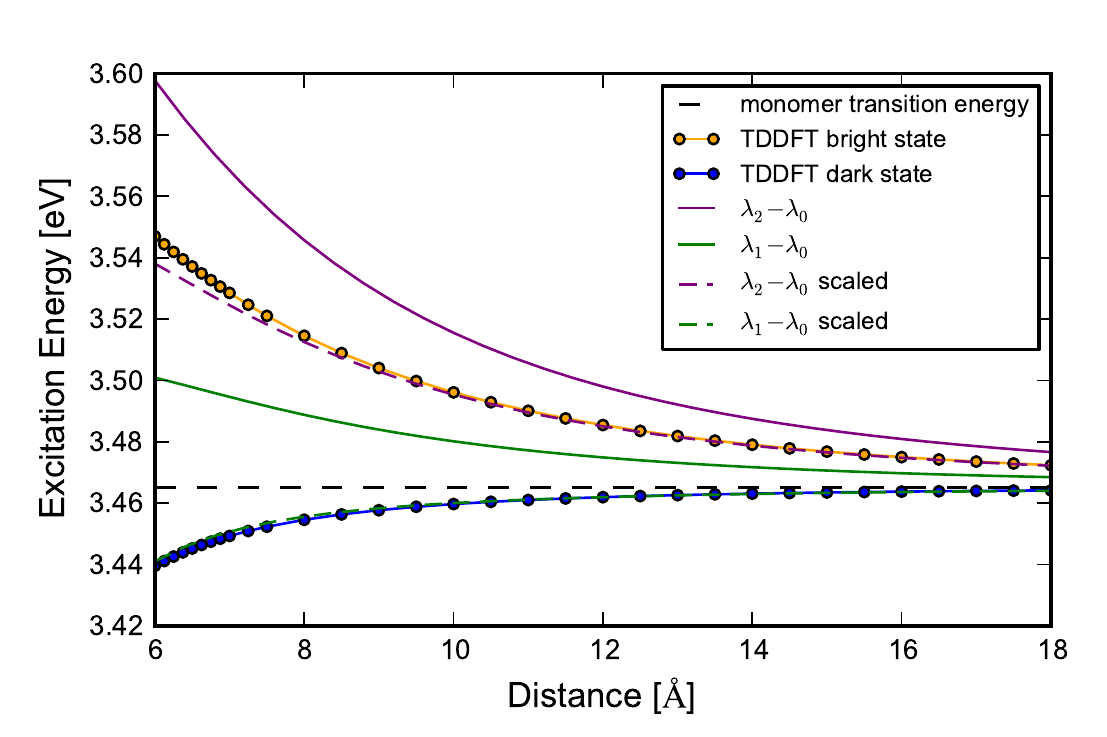}
\end{minipage}
\begin{minipage}{1.4in}
    \includegraphics[width=1.4in]{figures/two_dye_par_tilted0_sep9.pdf}
	Parallel-$0^\circ$
\end{minipage}
\begin{minipage}{5.in}
	\includegraphics{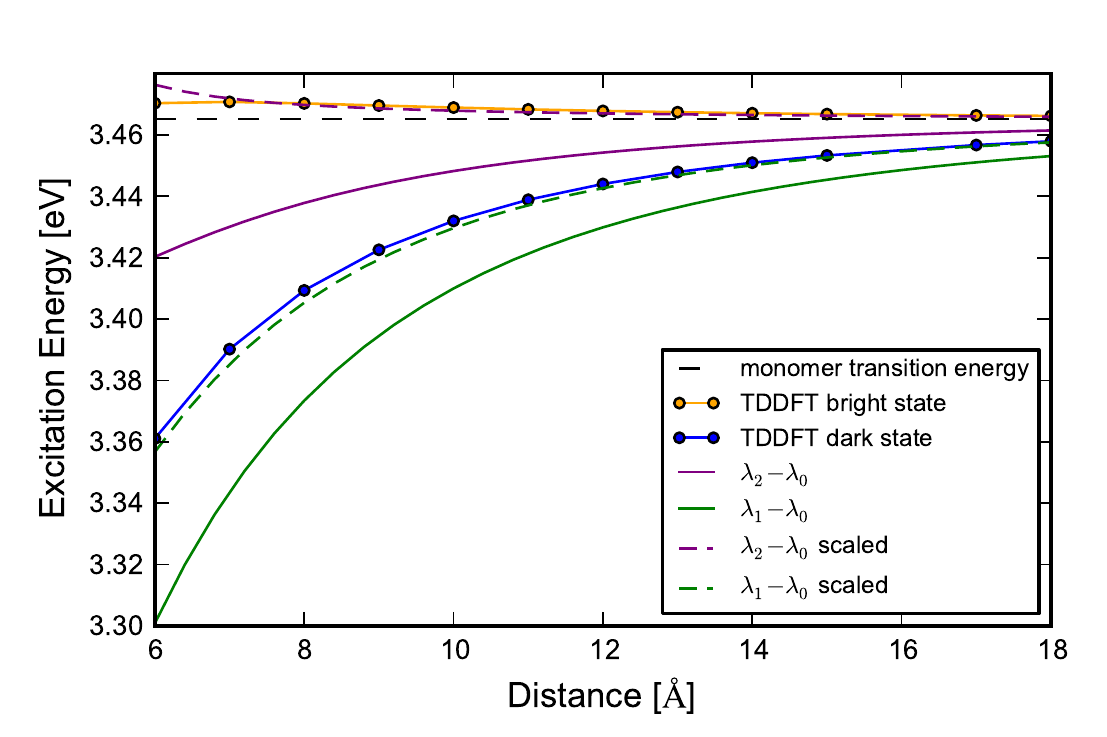}
\end{minipage}
\begin{minipage}{1.4in}
    \includegraphics[width=1.5in]{figures/anti_parallel.png}
	Anti-Parallel-$0^\circ$
\end{minipage}

\caption{\small
	The energy levels calculated using TDDFT are together with the excited states of the 4x4 Hamiltonian in \erf{eqn:hamiltonian_matrix}.
	For the latter, the excited states are plotted as $\lambda_1-\lambda_0$ and $\lambda_2 - \lambda_0$ where $\lambda_i$ is the $i^\text{th}$ eigenvalue.
	\label{eigenvalue_energies}}
\end{figure}

We assess the validity of these approximations by evaluating the expanded $4\times 4$ effective Hamiltonian \erf{eqn:hamiltonian_matrix}.
In order to calculate the matrix elements of this larger Hamiltonian we use Mulliken partial atomic charges~\cite{Mulliken1955} for the ground state and excited state densities instead of density cubes.
This is because density cube calculations are not constrained to reproduce the correct multipole expansion of the electron densities.
These errors can be corrected for the transition density, as shown by Kreuger et al.~\cite{Krueger:1998vn}, however the errors are more pronounced in the ground state and excited state densities, making them sensitive to the choice of grid resolution.
In Fig.~\ref{eigenvalue_energies}, we show the results of using this expanded effective Hamiltonian.
It is evident from Fig.~\ref{eigenvalue_energies} that we are able to reproduce the asymmetric shifts in excitonic energies using \erf{eqn:hamiltonian_matrix}.

The primary effects that invalidate the approximations leading to \erf{eqn:frenkel_hmat2} are electrostatic in nature.
The neglect of the $J_\text{trans}^\text{es}, J_\text{trans}^\text{gs}, J^\text{trans}_\text{es}, J^\text{trans}_\text{es}$ and the $J^\text{trans}_\text{trans}$ terms coupling the ground state $\ket{00}$ to the two-exciton state $\ket{11}$ (Heitler-London approximation) is valid since these are much less than $E_1 - E_0$ at all the inter-chromophore distance scales we examined.
However, the electrostatic corrections to the diagonal elements of \erf{eqn:hamiltonian_matrix}, $J_\text{gs}^\text{gs}, J_\text{es}^\text{gs}, J_\text{gs}^\text{es}, J_\text{es}^\text{es}$, are significant and cannot be neglected.
These are shifts to monomer energies due to the presence of the charges on the other chromophore.
These electrostatic shifts are dependent on the inter-chromophoric distance and the exact orientation of the chromophores.
Using the Mulliken partial atomic charge approach we are able to capture these electrostatic shifts and thereby get very good agreement with the TDDFT energies.
The static dipole for the ground state and excited state both lie nearly parallel to the transition dipole moment.
Consequently, the direction of the shifts is consistent with what is expected from the interaction of two electronic dipoles - the parallel dimers have a repulsive electrostatic effect while the anti-parallel dimers have an attractive effect.

While the asymmetric splitting is immediately captured by including these electrostatic distance-dependent shifts, scaling the ground-state/excited-state coulomb integrals by $c=0.66$ is necessary in order to achieve quantitative agreement with the TDDFT energies.
The value of this scaling factor is specific to a chromophore pair, but once determined for a particular orientation and distance separation, it holds for nearly all inter-chromophore separations and orientations.
We have confirmed this by explicit calculation of energies at additional orientations and distances not presented in Fig.~\ref{eigenvalue_energies}. 
We interpret this parameter as the screening of the Coulomb integral by the other electrons in the molecular dimer.

We remark here that similar electrostatic shifts of proximal chromophores were identified in Ref. \cite{Madjet:2006ce}.
Such electrostatic shifts to monomer energies can be a significant source of disorder in multi-chromophoric assemblies.
It is widely accepted that protein residues cause energetic shifts that are important for providing a favorable energetic landscape for energy transfer \cite{Muh.Mad.etal-2007}.
Our results indicate that in addition to the effect of the proteins, the electrostatic environment provided by neighboring chromophores should also be taken into account when calculating energetic shifts and disorder in multi-chromophoric arrays.
From a design perspective, this implies that the exact orientation and placement of chromophores are important not only for the precise engineering of the excitonic coupling between chromophores but also for engineering the energetic landscape.

\begin{figure}[htp!]
	\includegraphics{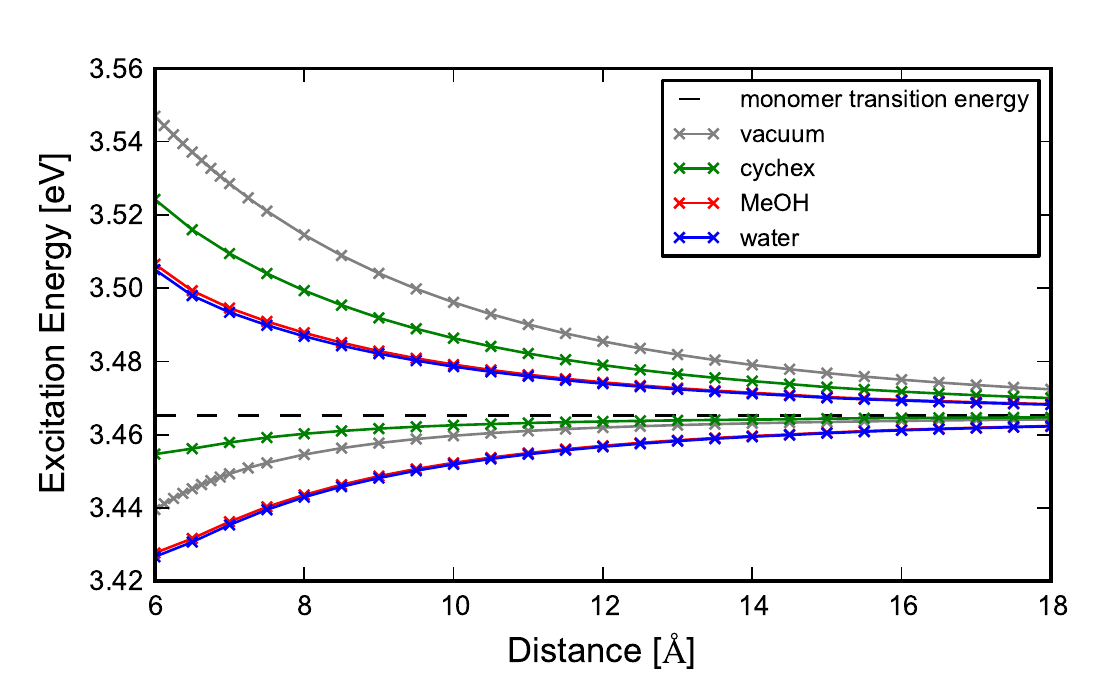}
	\caption{\small\label{pcm_graph}
	TDDFT excitation energies ($\tilde{E_1}-\tilde{E_0}$ and $\tilde{E_2}-\tilde{E_0}$) for the coupled excited states of the coumarin-343-MA dimer in various solvent environments (modeled using PCM).
	Only results for the parallel oriented dimer are shown.}
\end{figure}

We note that the electrostatic shifts identified here can be strongly affected by the polarity of the solvent \cite{Hsu:2001el}.
In particular, charge screening by a polar solvent can reduce the value of electrostatic integrals such as $J_\text{gs}^\text{es}$.
These integrals are likely to be more suppressed than the transition density integrals, e.g. $J_\text{trans}^\text{trans}$, and therefore the energy shifts can be suppressed even though the excitonic coupling (which is largely determined by $J_\text{trans}^\text{trans}$) may be only marginally effected by solvent screening \cite{Curutchet:2007ji}.
To demonstrate this we calculated the excitations energies of the coumarin-343-MA homo-dimer in various solvent environments modeled using a polarizable continuum model (PCM).
The results are shown in Fig. \ref{pcm_graph} for the parallel oriented dimer.
Clearly, as the polarity of the solvent increases the asymmetry of the excitation energy splitting around the monomer energy decreases.
This demonstrates that it is important to integrate information about the solvent environment when modeling excitonic properties of molecular aggregates; solvent polarity will dictate the amount of influence electrostatic effects have on excitonic energies.

\subsubsection{Exciton wave functions}
\begin{figure}[htp!]
	\vspace{-0.5cm}
	\includegraphics{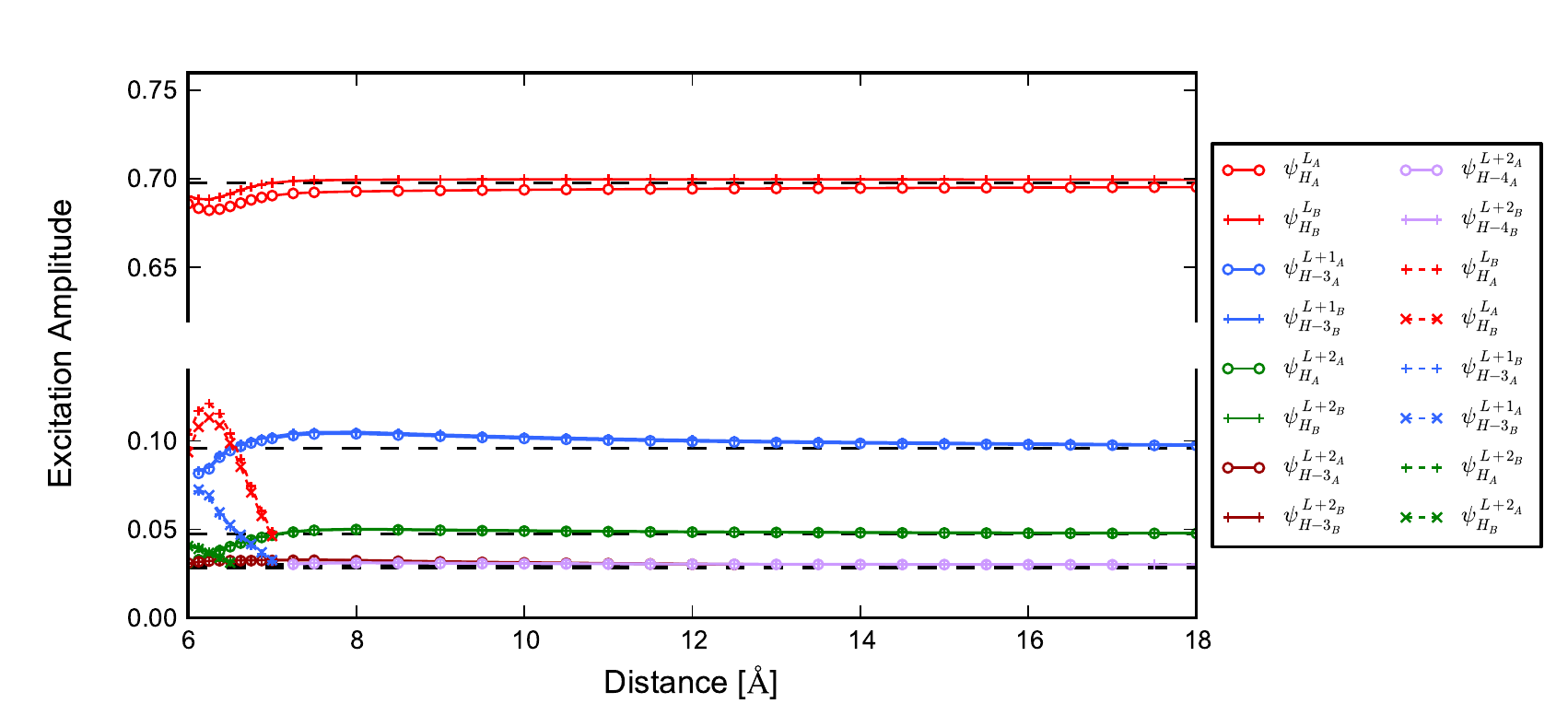}
	\caption{\small\label{single_particle_excitation_figure}
    The expansion coefficients of the parallel-oriented bright state shown as a function of distance.
 	Dashed lines represent the corresponding coefficients in the monomer TDDFT excited states.
    The top panel shows the dominant single-particle excitation, which corresponds to the excitation of an electron in the highest occupied molecular orbital (HOMO) to the lowest unoccupied molecular orbital (LUMO) of each monomer.
    The bottom panel shows other single-particle excitations which make up the TDDFT excited state.
    Molecular orbitals are labeled with respect to their energetic position below the HOMO ($H-n$) or above the LUMO ($L+n$).
	The excitations that are doubly degenerate (e.g. $\psi_{\text{H}_A}^{\text{L}_A}$ and $\psi_{\text{H}_B}^{\text{L}_B}$) have been averaged and plotted without their molecule index.}
\end{figure}

We now investigate the character of the TDDFT excited states to examine whether they are well described by products of monomer wavefunctions, as predicted by \erf{eqn:frenkel_eig}.
The tight-binding effective Hamiltonian description of the excited states can break down if either the dimer orbitals change with respect to the monomer orbitals, or the nature of the excited state changes significantly as a function of distance.
The TDDFT excited states are written as linear combinations of basis functions which represent single-particle excitations from the DFT ground state.
If the coefficients of this linear expansion are distance-dependent, the predictions of the effective Hamiltonian in \erf{eqn:frenkel_hmat} are invalid.
This is because the exciton wave functions in \erf{eqn:frenkel_eig} are constructed from symmetric and anti-symmetric combinations of the monomer states (i.e.~\erf{eqn:frenkel_eig}), and are therefore independent of the magnitude of the coupling energy $J$ and hence of the inter-chromophoric separation.
Fig.~\ref{single_particle_excitation_figure} shows these coefficients for the bright state of the parallel orientation of the monomers as a function of distance.
We see from this figure that the nature of the TDDFT excited state is relatively constant as a function of distance, until we get to small separation distances (below 8~\AA).
In particular, the dominant single-particle excitation, that from the highest occupied molecular orbital (HOMO) to the lowest unoccupied molecular orbital (LUMO) on each monomer, begins to change at intermolecular distances less than 8 \AA .
However, some of the other minor excitations which contribute to the TDDFT excited state begin to change gradually as a function of distance already at 12 \AA.
Some of the single-particle excitations which contribute below 7 \AA ~represent charge-transfer excitations from molecule A to molecule B.
The existence of charge transfer, especially between 6 and 7 \AA , is a feature present also for range-corrected TDDFT functionals (see Appendix).
However, it should be noted that TDDFT is known to be inaccurate in describing charge transfer, so we do not regard its quantitative predictions for charge transfer at these distance scales to be reliable.

\begin{figure}[htp!]
	\centering
	\vspace{-0.5cm}
	\includegraphics{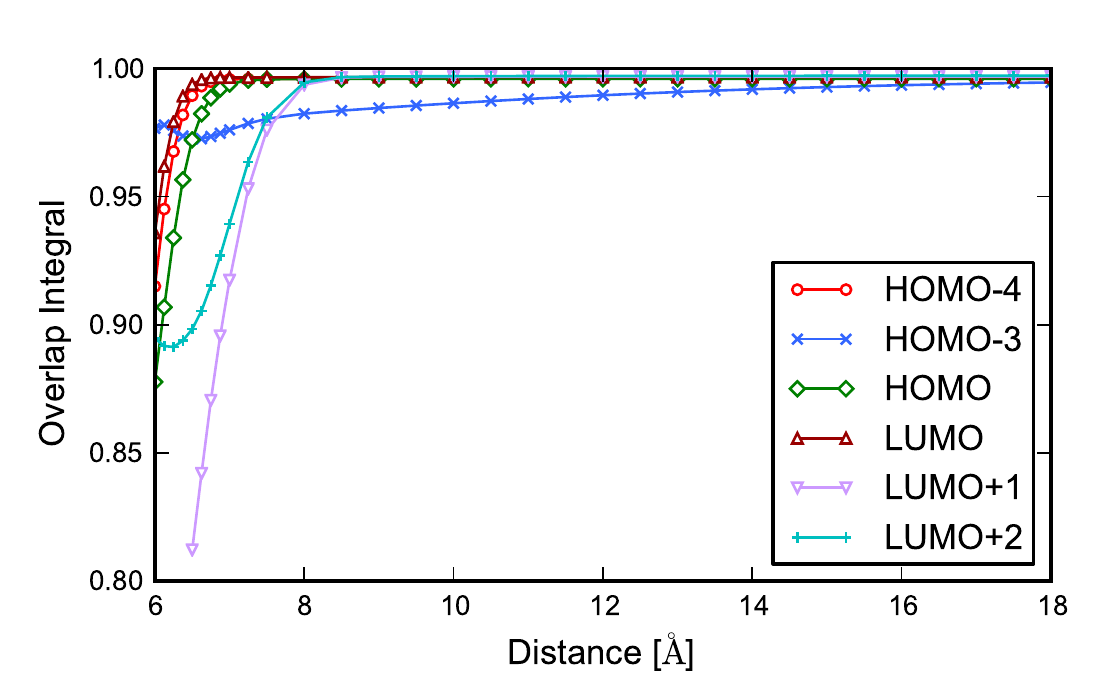}
	\caption{\small
		Overlaps of the bright state dimer molecular orbitals with their corresponding monomer molecular orbitals (See \erf{overlaps})
		\label{fig:mo_figure}}
\end{figure}

In Fig.~\ref{fig:mo_figure}, we show the overlap integral of the monomer molecular orbitals and the corresponding dimer orbitals,
\begin{equation}\label{overlaps}
	\int \phi_n^\text{dimer}(r)\phi_n^\text{monomer}(r)\mathrm{d}r.
\end{equation}
In the limit of infinite separation between the two dyes, each dimer molecular orbital is doubly degenerate, possessing unit overlap with a corresponding monomer molecular orbital.
For the HOMO and LUMO, which are the most important orbitals in the bright state (Fig.~\ref{single_particle_excitation_figure}), the correspondence between monomer and dimer MOs is almost perfect for distances greater than 7 \AA .
Between 6 and 7 \AA , these orbitals change by about 8\% .
Some orbitals change aready at larger distances, such as the LUMO+1 and the LUMO+2 orbitals, which begin to deform as the intermolecular distance is decreased below 8 \AA .
Finally, the HOMO-3, which with the LUMO+1 forms the next most dominant excitation in the bright state, changes continuously at distances less than 18~\AA; however, it changes by only 2\% and furthermore it does not form the majority of the excited state, so this effect is diminished in the excited state energies and couplings.

To summarize this investigation of excited state wave functions, we find that for coumarin-343-MA, it is reasonable to describe the dimer wavefunctions in a basis of monomer wavefunctions for separation distances greater than 8 \AA.
For smaller distances the B3LYP calculations indicate possible mixing in of charge transfer character into the excited state wavefunction, although the extent of this mixing is not conclusive from this level of calculation.

\section{Conclusions}
In this work we have made a critical assessment of the conventional effective Hamiltonian approach of modeling the excitonic properties of molecular aggregates, using the coumarin-343-MA dye as a case study.
Results from \textit{ab-initio} electronic structure calculations using TDDFT were compared with predictions from the conventional tight-binding effective Hamiltonian for a homo-dimer (the Heitler-London approximation).
Most interestingly, we found that the conventional effective Hamiltonian for a homo-dimer does not reproduce the asymmetric energy splittings calculated using TDDFT. 
In particular, the ideal dipole approximation of the excited state coupling was found to give a very inaccurate representation of the TDDFT energy splittings.
While the TDC method was found to perform much better, both the IDA and the TDC descriptions were found to be unable to reproduce the asymmetric nature of the splitting between bright and dark state energy levels that is predicted from TDDFT.
We showed that this is a result of ignoring non-negligible electrostatic energy shifts resulting from the proximity of the two chromophores.
We outlined a method for reincorporating these electrostatic shifts using a simple approach that only requires calculating Coulomb integrals based on Mulliken partial atomic charges.
This approach is an efficient method for forming more complete effective Hamiltonian descriptions that capture all the relevant physical effects; it only requires TDDFT calculations of dimers of chromophores (of each dimer combination of species present in the aggregate) and the remaining elements are accurately captured by Coulomb integrals.
We find that the combination of a TDC description of the transition density coupling together with proper incorporation of electrostatic shifts produces an excellent effective Hamiltonian description of excitonics in molecular aggregates.
We also demonstrated the importance of incorporating details of solvent polarity into the molecular aggregate model, since this also determines the degree of influence the electrostatic effects have on excitonic energies.

Additionally, we scrutinized the assumptions of the conventional Heitler-London tight-binding picture of excitonic coupling, by examining changes in the character of the excited state and the coupled molecular orbitals as a function of intermolecular distance.
These effects were determined to be small but nonzero for intermolecular distances greater than 7-8 \AA , while TDDFT predicts a significant departure from the Heitler-London picture at smaller distances.
It also predicts some charge-transfer character at these smaller distances.
In the future it would be useful to develop more reliable estimates of this charge-transfer character \cite{Scholes1996a} in order to analyze the interplay between excitonic and charge transfer states in chromophore arrays relevant to natural photosynthesis, such as, e.g., the bacterial reaction center~\cite{novoderezhkin2004coherent}.

We expect that our investigation and refinement of effective Hamiltonian descriptions of molecular aggregates will inform the modeling of large molecular aggregates formed by direct aggregation or aggregation by protein templated assembly.
Such aggregates show promise as the basis for next-generation light harvesting or sensing devices with tailored properties, and efficient modeling of their excitonic properties through effective Hamiltonians will be important for rational design and engineering of such devices.

\begin{acknowledgments}
We thank Matthew Francis and Dan Finley for useful discussions on virus-templated assembly for light harvesting.
We thank the Molecular Graphics and Computation Facility at UC-Berkeley for computational resources, which were made possible by the NSF under contract numbers CHE-0233882 and CHE-0840505.
Sandia National Laboratories is a multi-program laboratory managed and operated by Sandia Corporation, a wholly owned subsidiary of Lockheed Martin Corporation, for the United States Department of Energy's National Nuclear Security Administration under contract DE-AC04-94AL85000.
Financial support for this research was provided by the DARPA QuBE (Quantum Effects in Biological Environments) program under contract number N66001-10-1-4068 and by the NSF (grants CHE-0233882 and CHE-0840505).
\end{acknowledgments}

\appendix

\clearpage
\section{Orientation Dependence Calculations}
In order to sample the possible dimer orientations, we begin with the molecules held at a fixed intermolecular separation along the x-axis.
The transition dipole moment of each molecule is aligned along the z axis, with the plane of the molecule flat on the y-z plane as shown in the Parallel-$0^\circ$ geometry in Fig.~\ref{eigenvalue_energies}.
The orientation of each molecule can be characterized by the three angles: the roll angle (rotation of the molecule about the transition dipole moment axis), the polar angle $\theta$, and the azimuthal angle $\phi$.
These three angles correspond to the Euler angles $\alpha$, $\beta$, and $\gamma$, respectively.
We have found that varying the roll angle does not affect the magnitude of coupling very much, since the roll angle does not change the direction of the transition dipole moment, therefore we do not sample over these angles in the following calculations.

\begin{figure}[htp!]
\begin{center}
\subfigure[~Orientation angles for the chromophores.\label{fig:appendix1a}]{
	\includegraphics[width=3.3in]{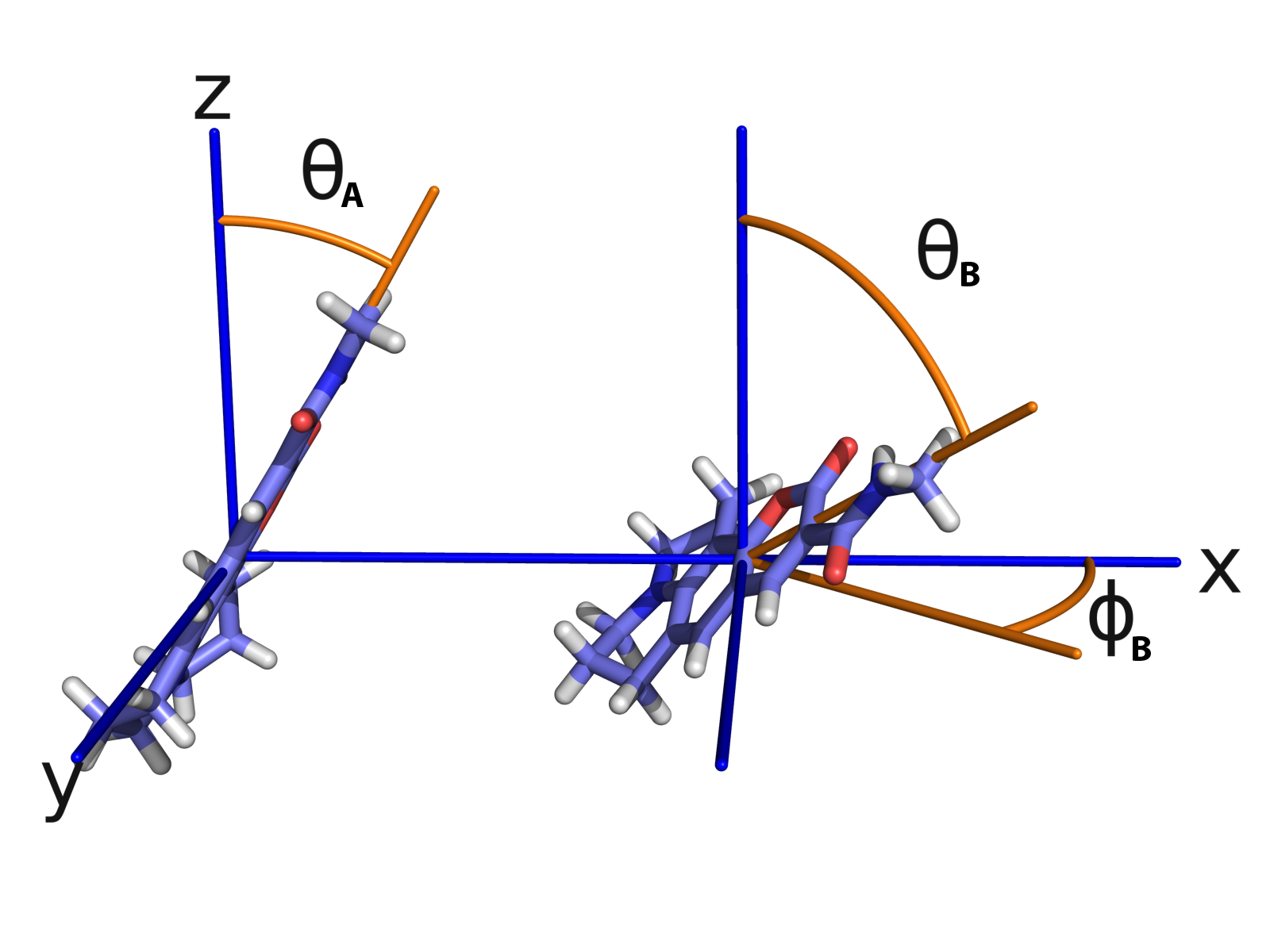}}
\subfigure[~Example of dimer configurations where the polar angle of the first dye is held at $30^\circ$, while sweeping over the various relative orientations of the second dye.\label{fig:appendix1b}]{
	\includegraphics[width=3.3in]{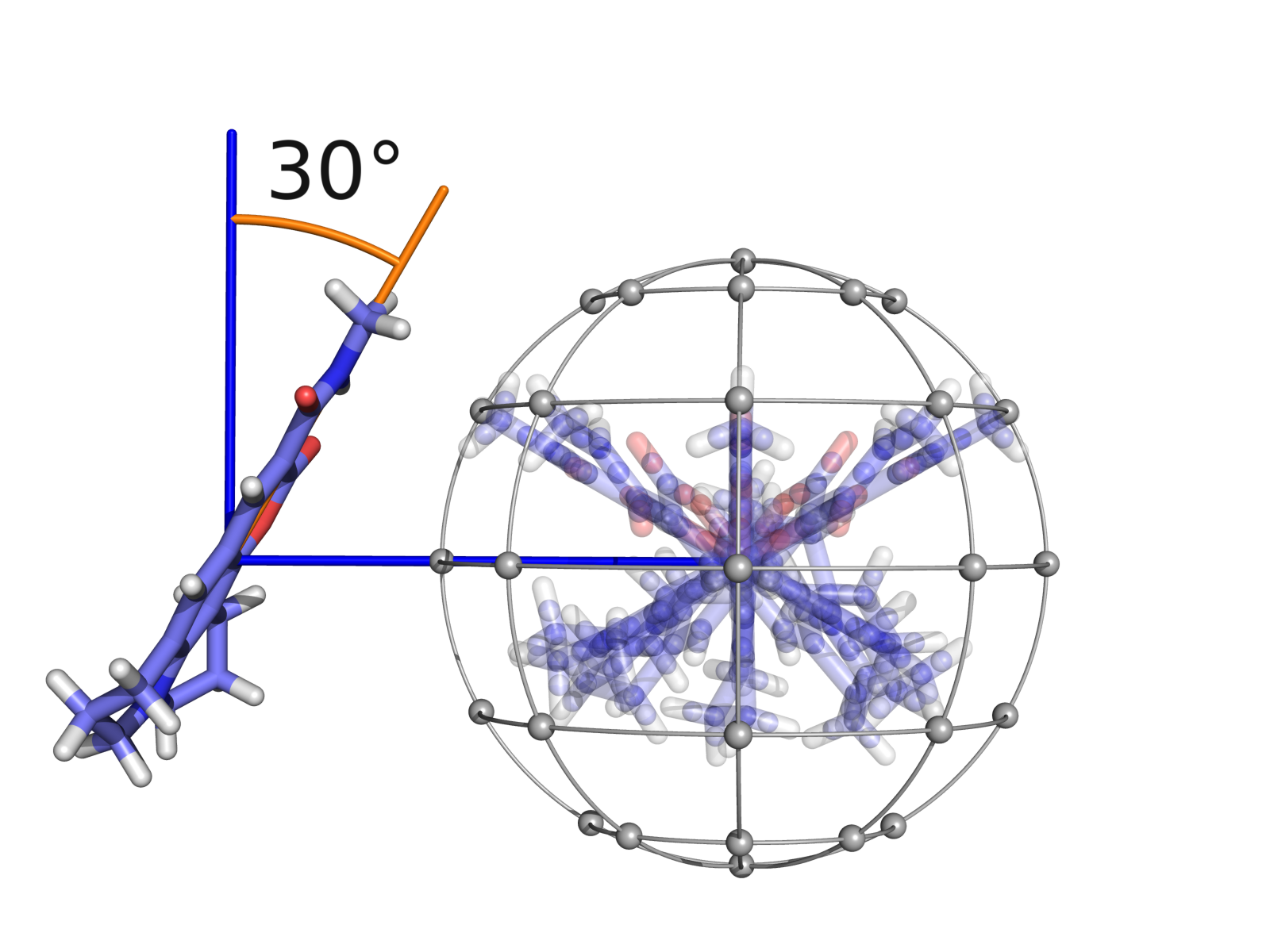}
	\vspace{4cm}}
\end{center}
	\caption{\small
        Definition and examples of the three angles used to sample the relative orientations between two molecules.
    }
\end{figure}

To create our dimer geometries, we first fix the azimuthal angle of molecule A by constraining its transition dipole moment to lie on the x-z plane.
We sample over the remaining 3 angular degrees of freedom: the polar angles $\theta_A$ and $\theta_B$, and the azimuthal angle $\phi_B$.
These angles are defined in Fig.~\ref{fig:appendix1a}.
Next, molecule A is rotated about the $y$-axis by $\theta_A$, and molecule B is rotated about the ($\hat{y}\cos{\phi_B}-\hat{x}\sin{\phi_B} $) axis by $\theta_B$.
In Fig.~\ref{fig:appendix1b}, we show examples of the resulting relative orientations possible for two chromophores.

\begin{figure}[htp!]
	\begin{tabular}{cccc}
		\hspace{-20pt}
		$\theta_A$	&	\hspace{-35pt}$J_{TDDFT}$	&	\hspace{-35pt}$J_{TDDFT} - J_{IDA}$	&	\hspace{-35pt}$J_{TDDFT} - J_{TDC}$\\
		\vspace{-15pt}
		\hspace{-20pt}
		\includegraphics[width=1in]{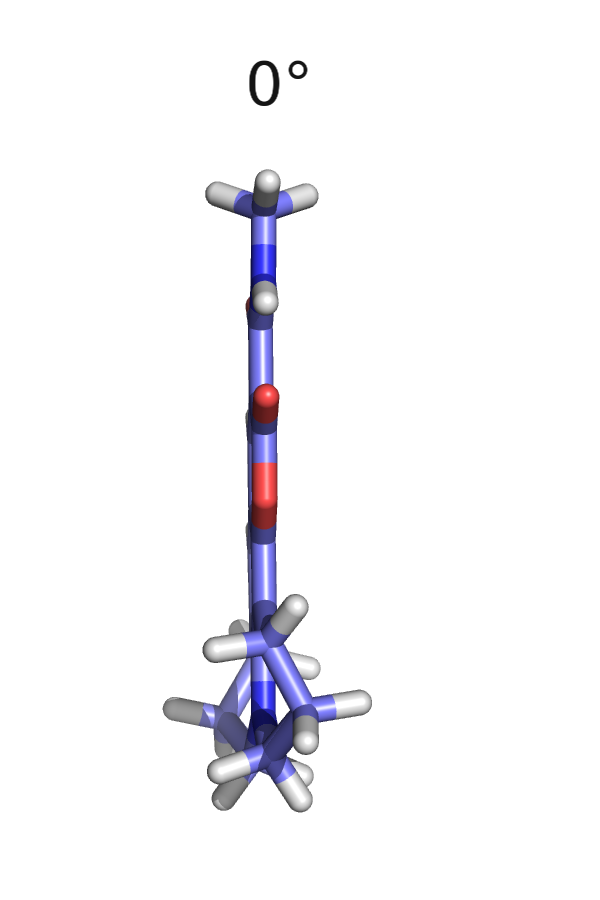}	&
		\hspace{-35pt}
		\includegraphics[width=2.2in]{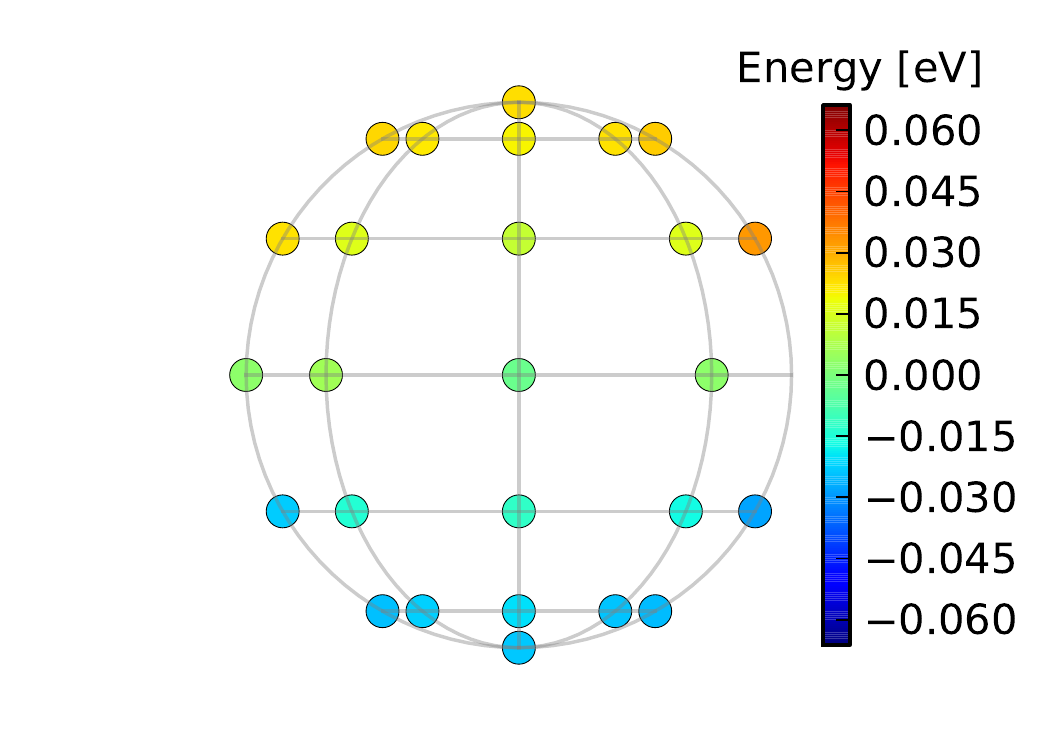}	&
		\hspace{-35pt}
		\includegraphics[width=2.2in]{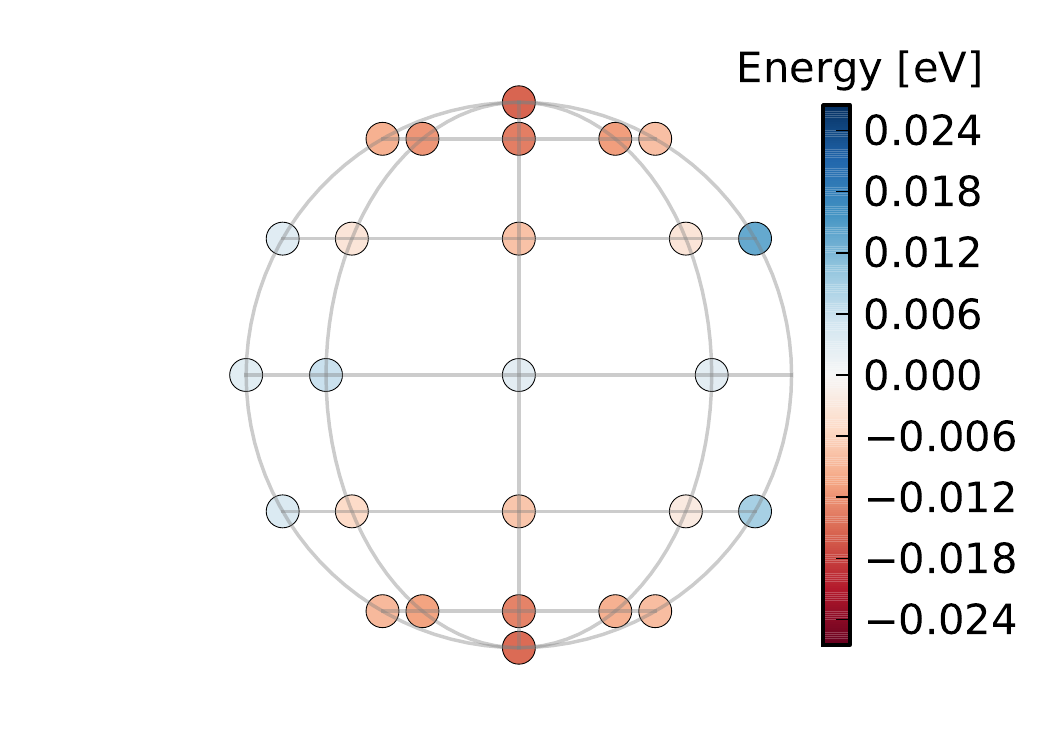}	&
		\hspace{-35pt}
		\includegraphics[width=2.2in]{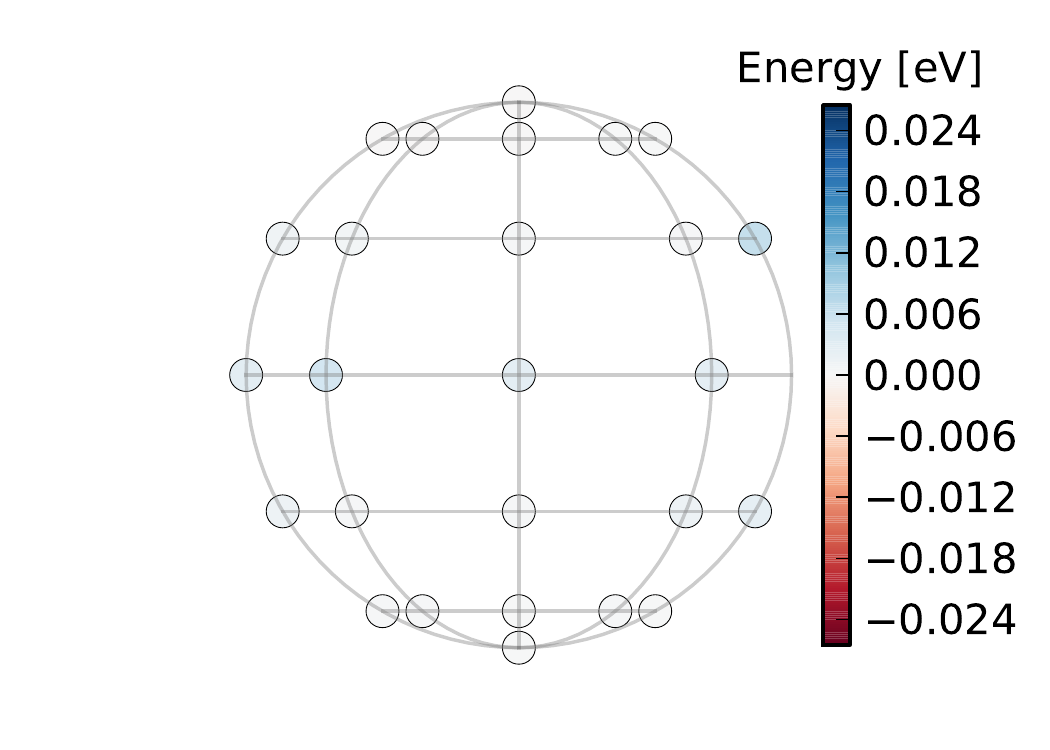}\\
		\vspace{-15pt}

		\hspace{-20pt}
		\includegraphics[width=1in]{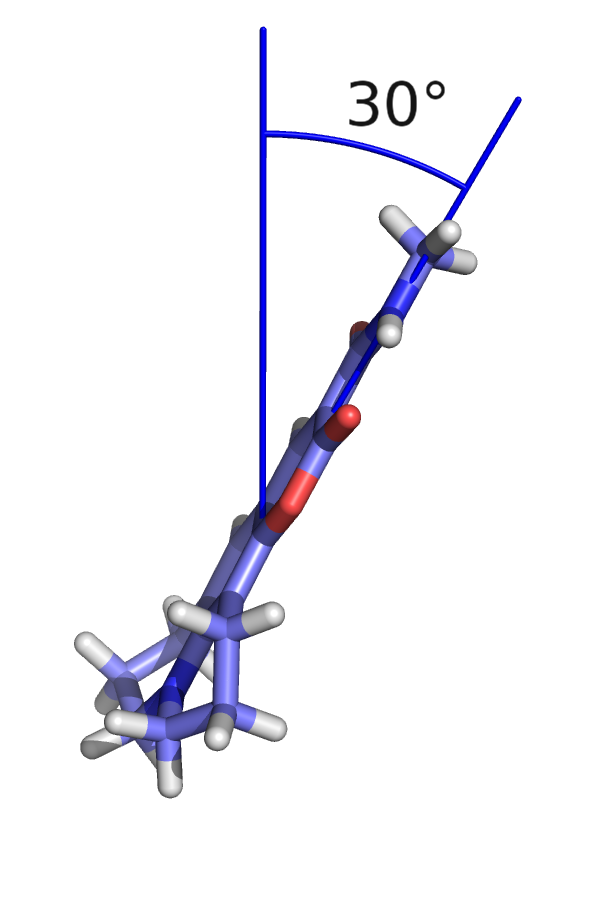}	&
		\hspace{-35pt}
		\includegraphics[width=2.2in]{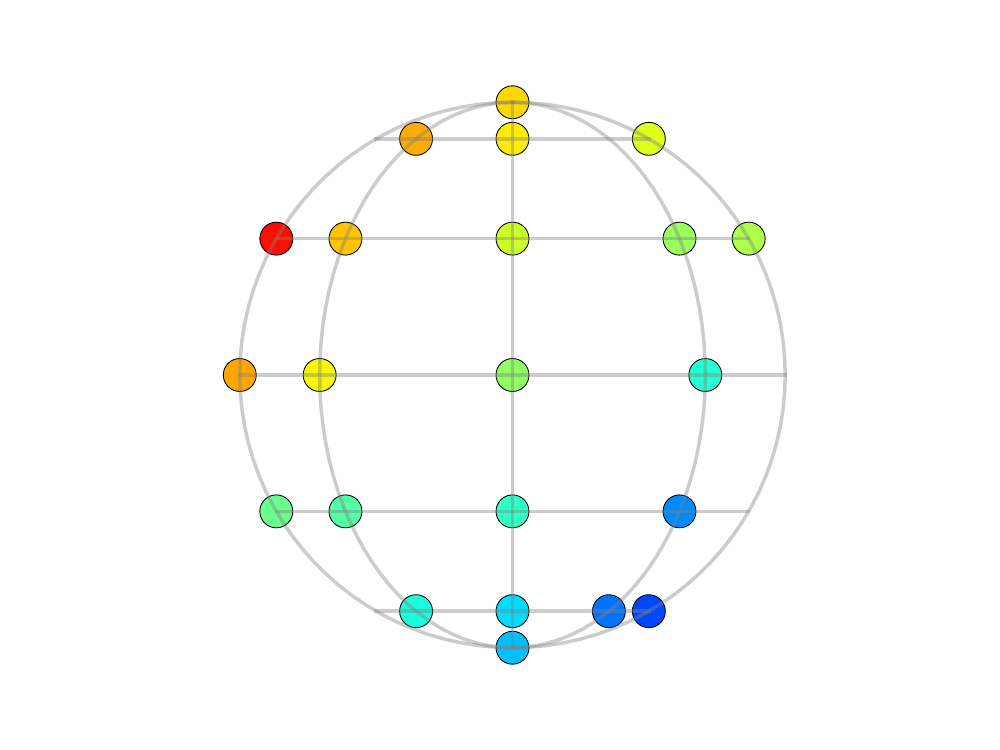}	&
		\hspace{-35pt}
		\includegraphics[width=2.2in]{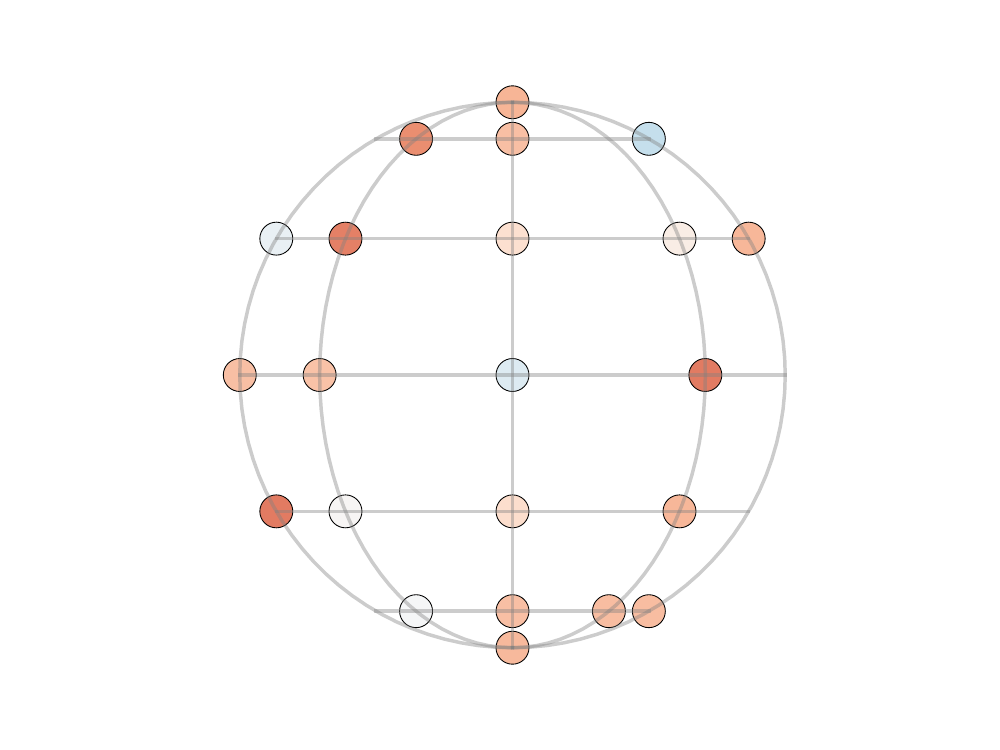}	&
		\hspace{-35pt}
		\includegraphics[width=2.2in]{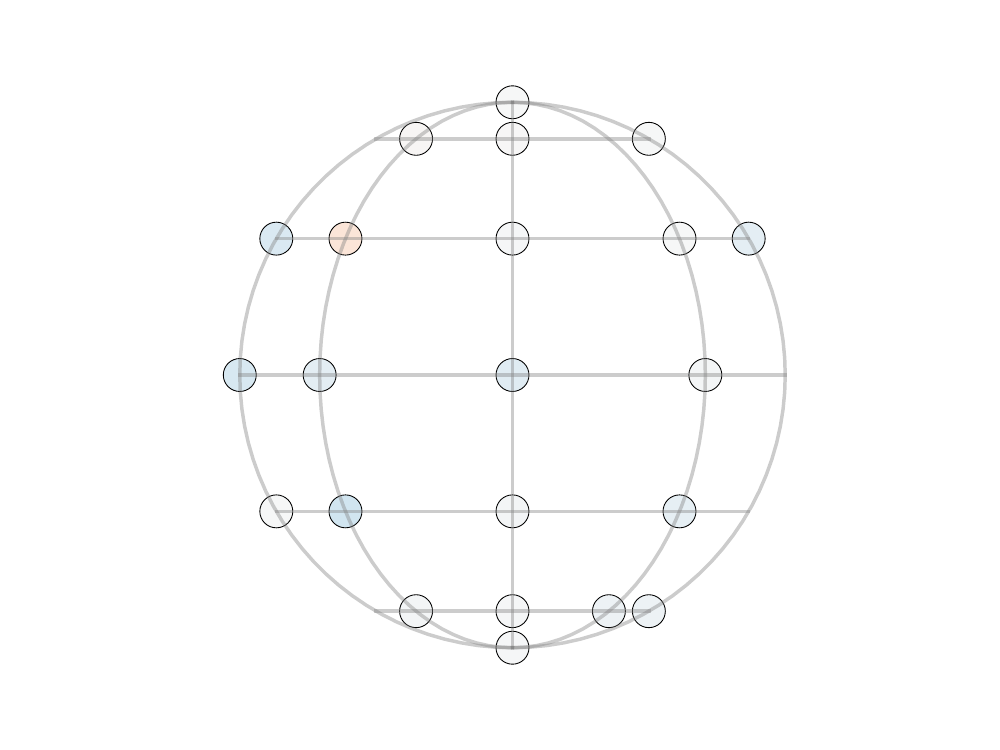}\\
		\vspace{-15pt}

		\hspace{-20pt}
		\includegraphics[width=1.5in]{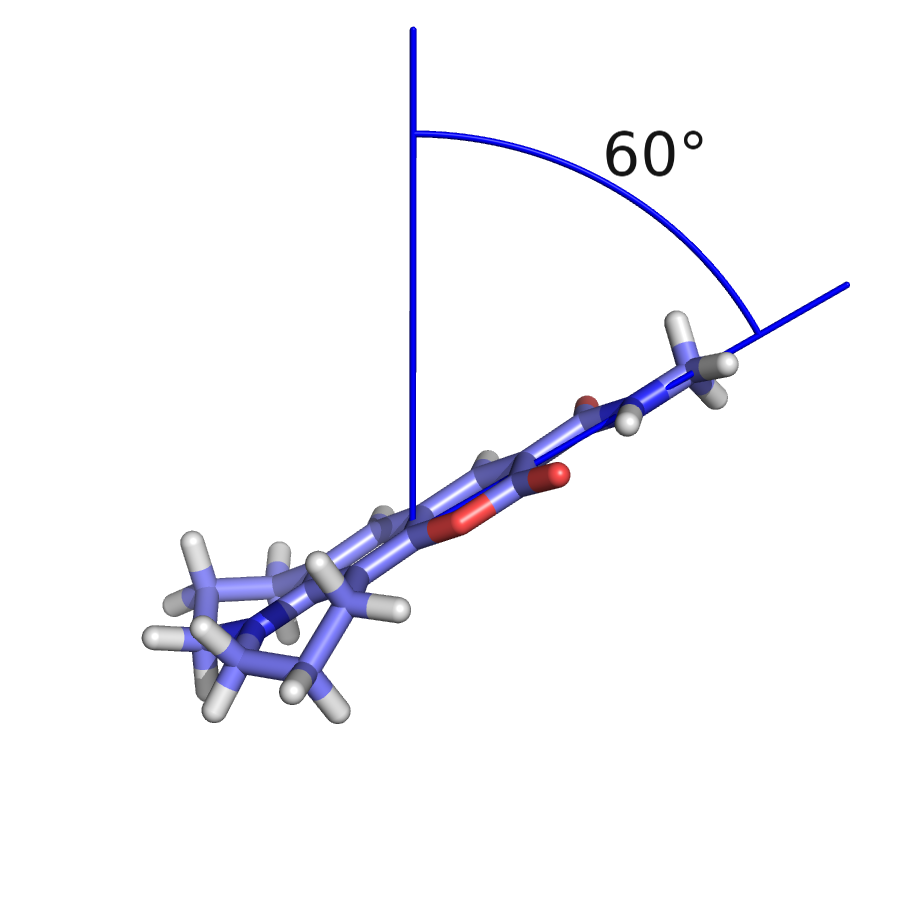}	&
		\hspace{-35pt}
		\includegraphics[width=2.2in]{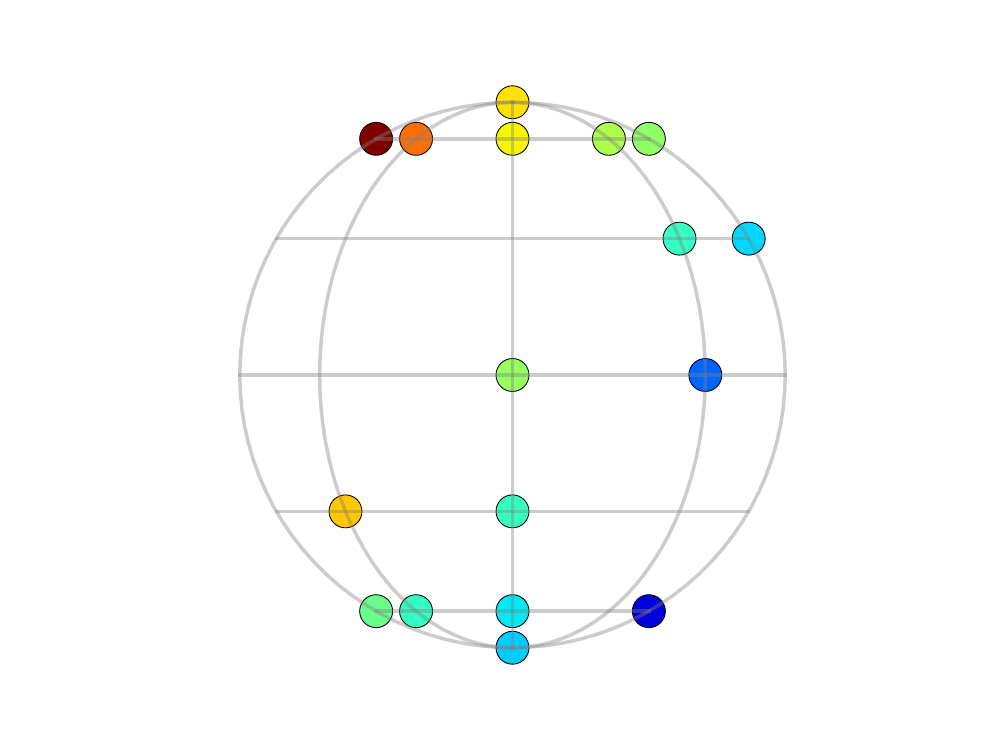}	&
		\hspace{-35pt}
		\includegraphics[width=2.2in]{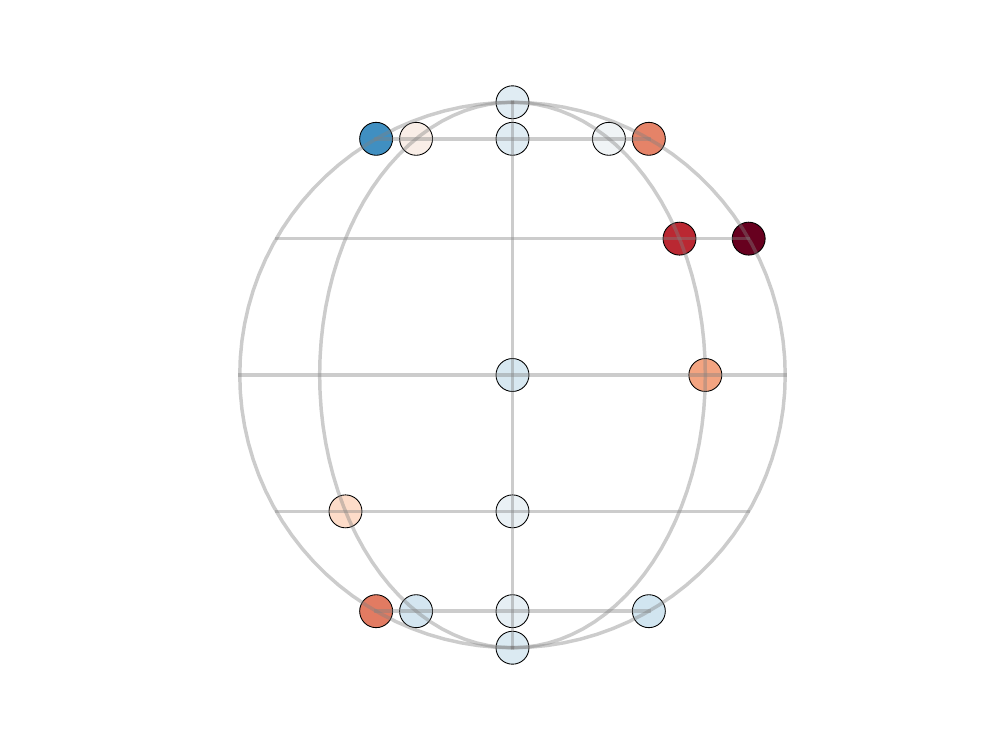}	&
		\hspace{-35pt}
		\includegraphics[width=2.2in]{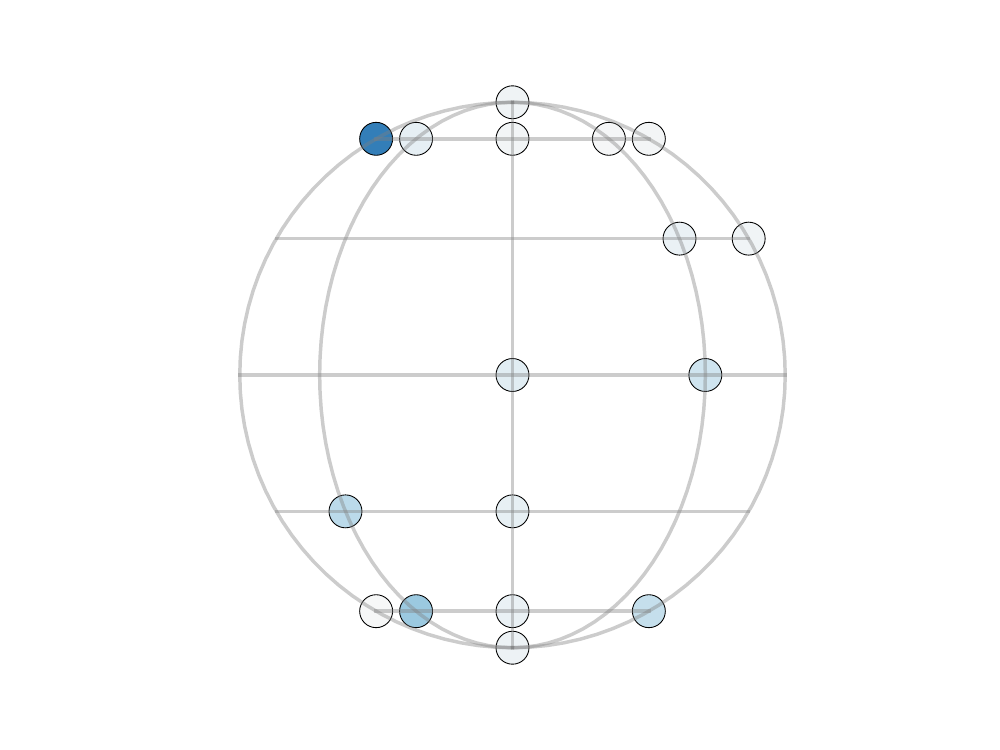}\\
		\vspace{-15pt}

		\hspace{-20pt}
		\includegraphics[width=1.58333in]{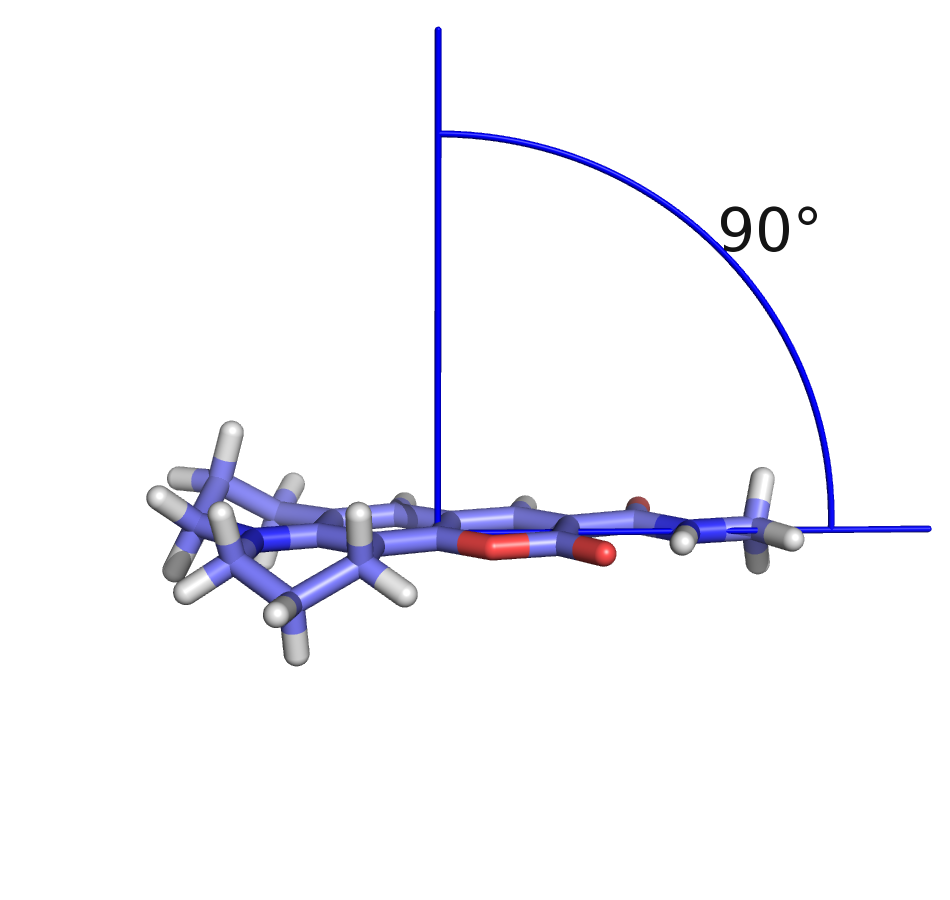}	&
		\hspace{-35pt}
		\includegraphics[width=2.2in]{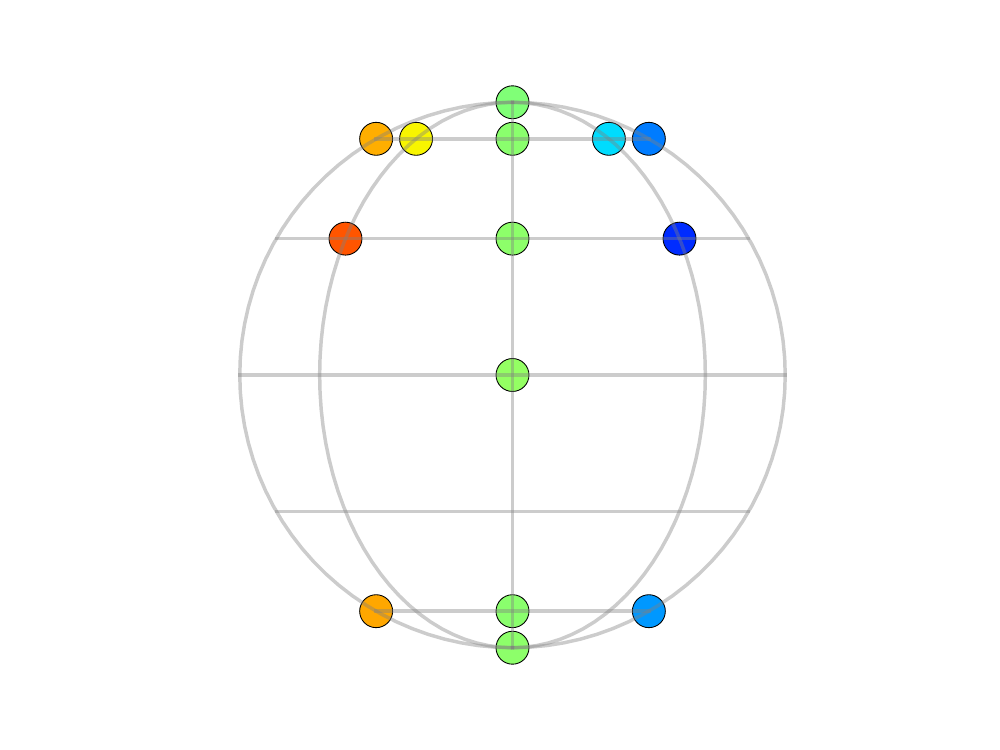}	&
		\hspace{-35pt}
		\includegraphics[width=2.2in]{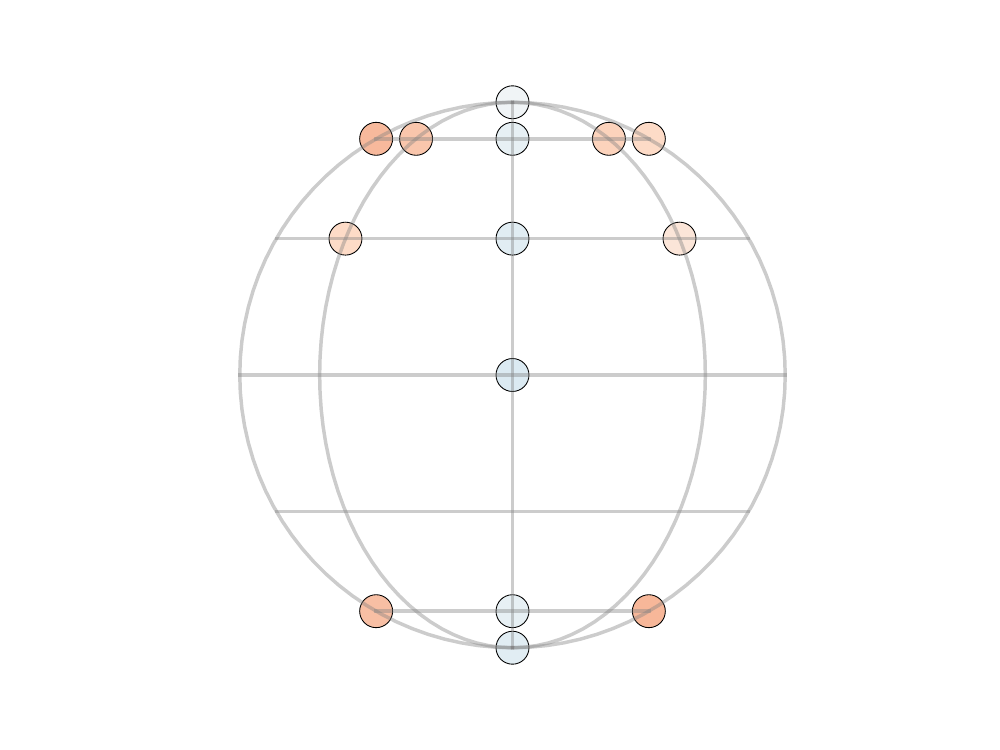}	&
		\hspace{-35pt}
		\includegraphics[width=2.2in]{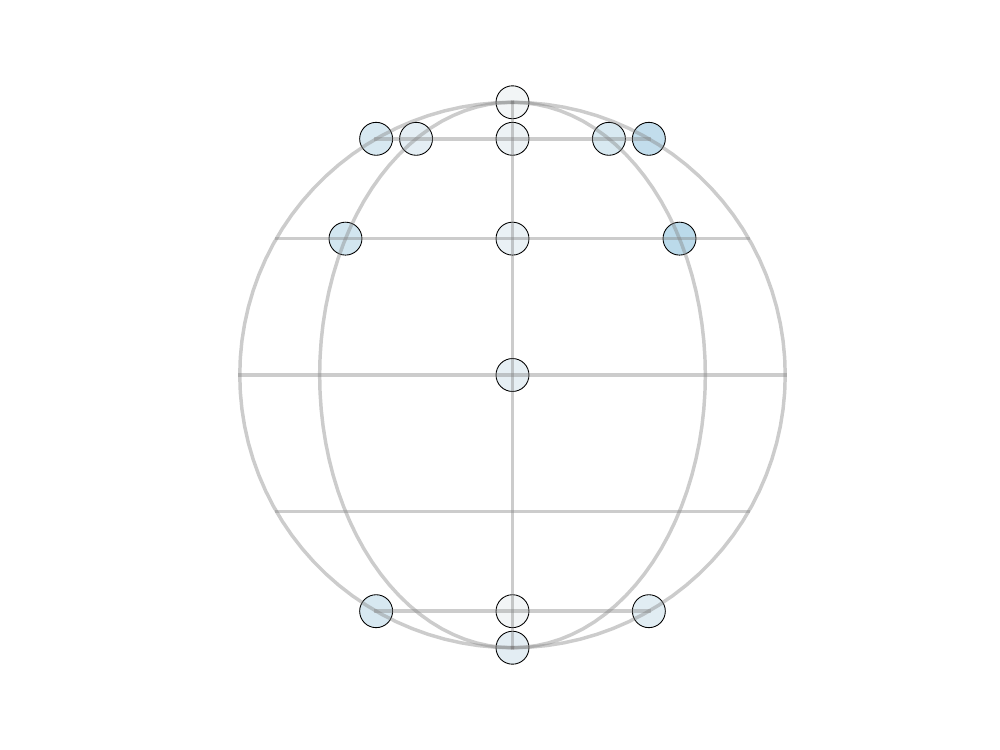}
	\end{tabular}
	\caption{\small
	Dimer relative orientational dependence of electronic coupling at 9 \text{\AA} separation.\\
	The first column depicts the polar orientation of the first molecule while the position on the polar plots on the right represents the orientation of the second molecule relative to the first (see Fig.~\ref{fig:appendix1b}).
	Columns 2-4 show the magnitude of $J_\text{TDDFT}$, the value of the coupling given by TDDFT (column 2), the error resulting from the IDA approximation to this (column 3) and the error resulting from the TDC estimate (column 4), as a function of the relative orientation.}
	\label{fig:appendix2}
\end{figure}

\begin{figure}[htp!]
	\begin{tabular}{cccc}
		\hspace{-20pt}
		$\theta_A$	&	\hspace{-35pt}$J_{TDDFT}$	&	\hspace{-35pt}$J_{TDDFT} - J_{IDA}$	&	\hspace{-35pt}$J_{TDDFT} - J_{TDC}$\\
		\vspace{-15pt}
		\hspace{-20pt}
		\includegraphics[width=1in]{figures/monomer0_fig.png}	&
		\hspace{-35pt}
		\includegraphics[width=2.2in]{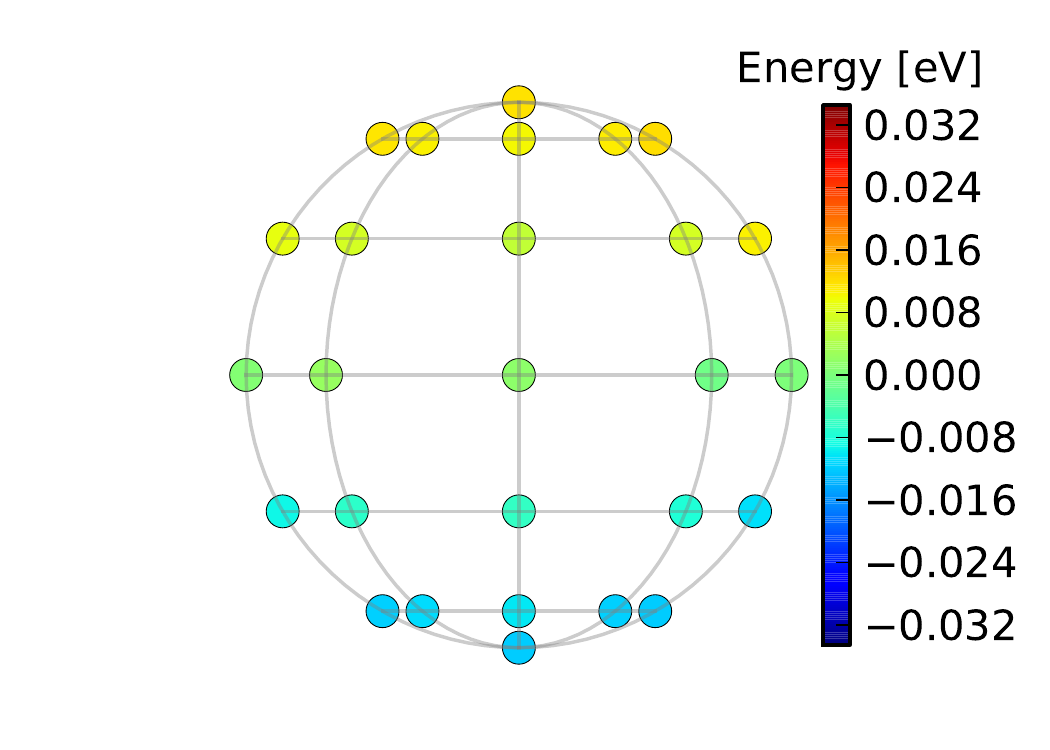}	&
		\hspace{-35pt}
		\includegraphics[width=2.2in]{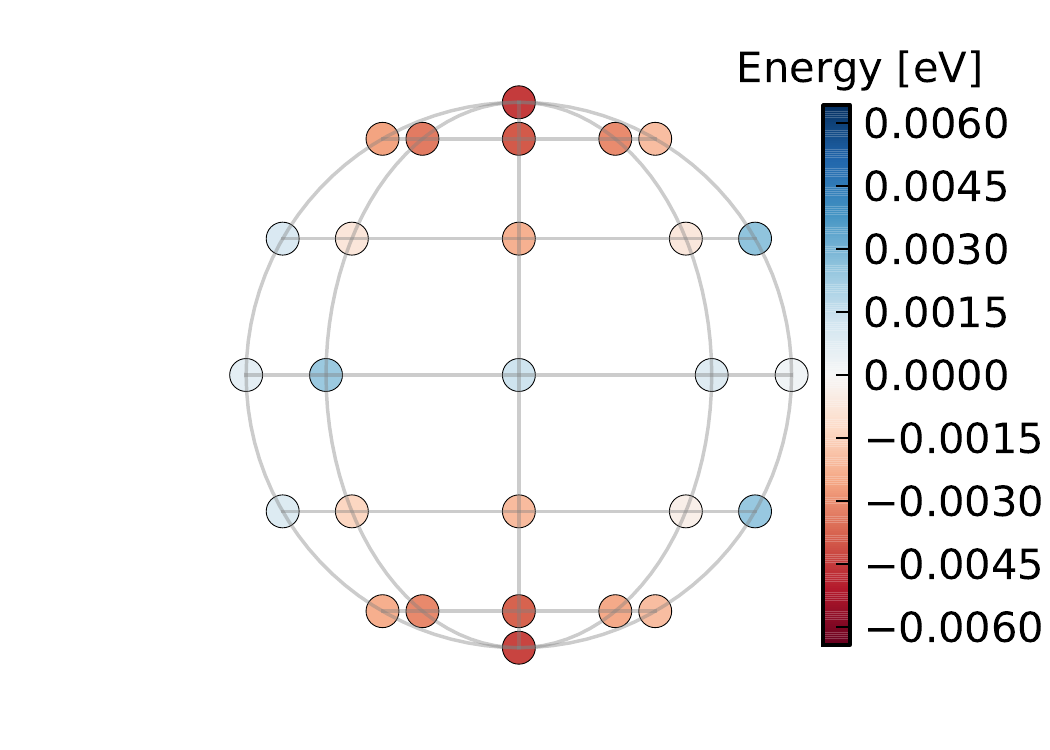}	&
		\hspace{-35pt}
		\includegraphics[width=2.2in]{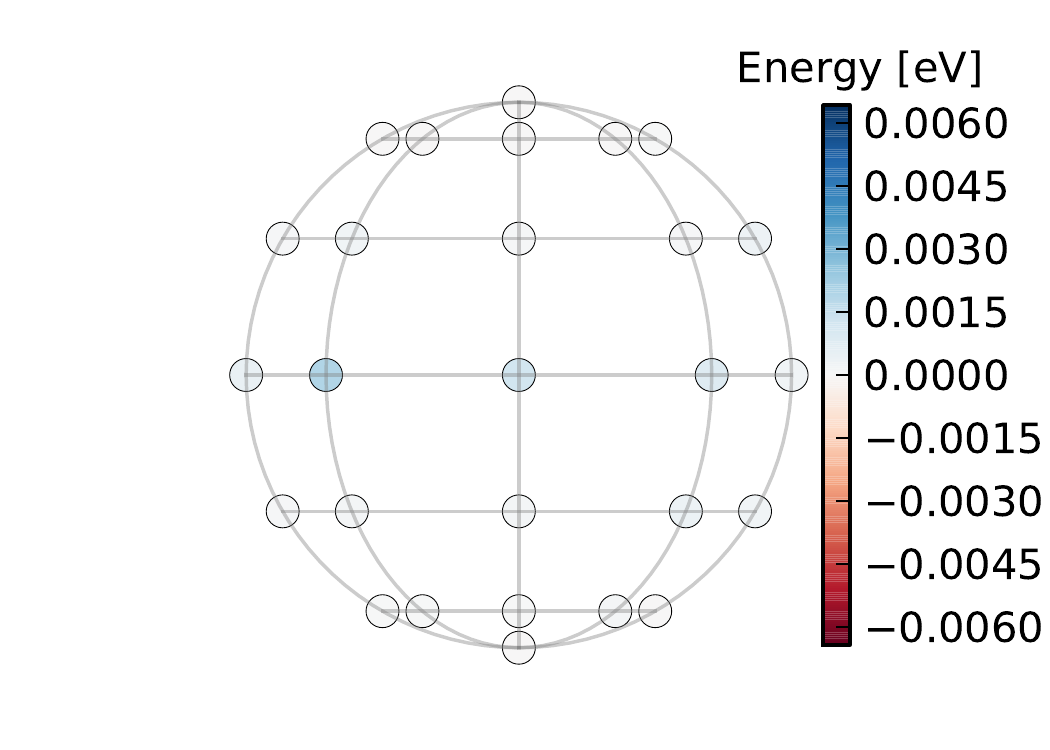}\\
		\vspace{-15pt}

		\hspace{-20pt}
		\includegraphics[width=1in]{figures/monomer30_fig.png}	&
		\hspace{-35pt}
		\includegraphics[width=2.2in]{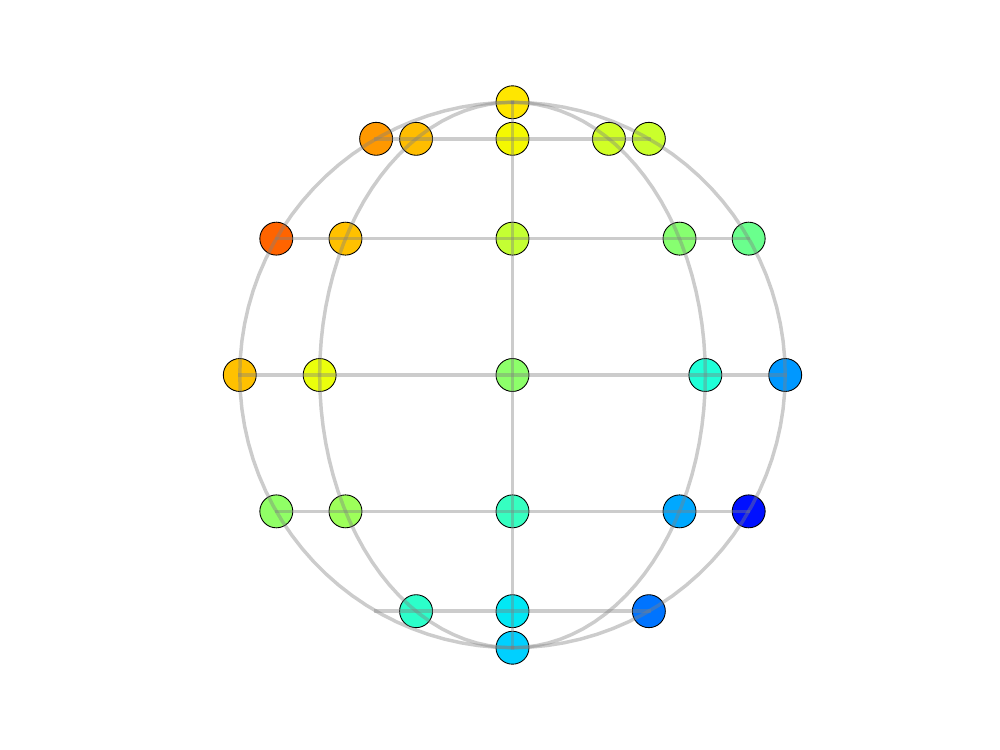}	&
		\hspace{-35pt}
		\includegraphics[width=2.2in]{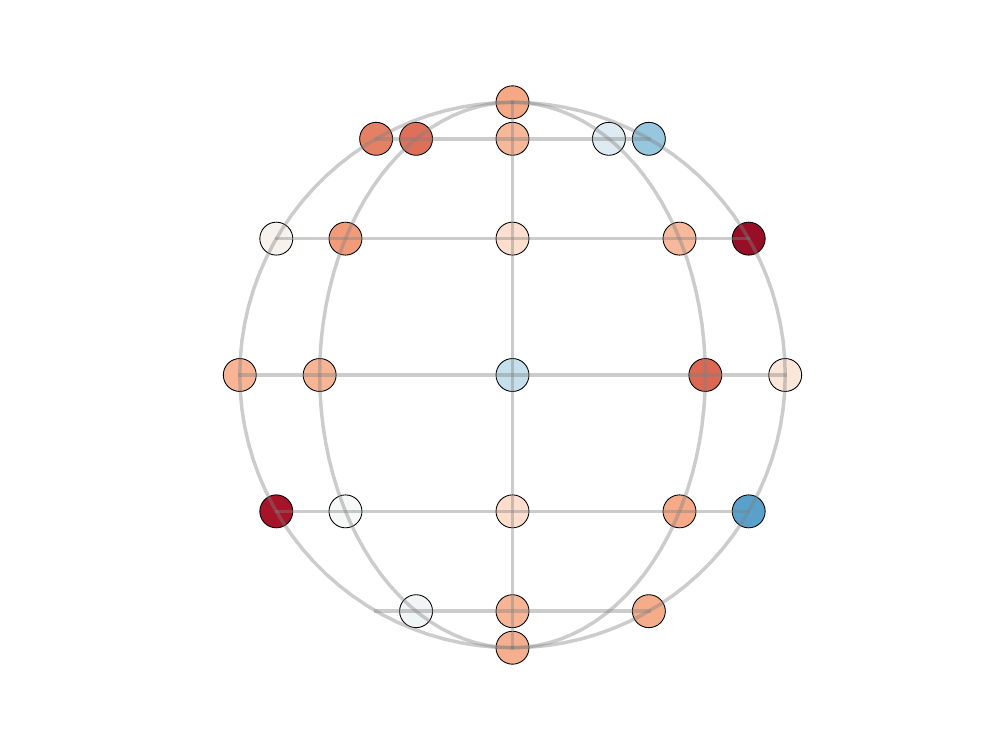}	&
		\hspace{-35pt}
		\includegraphics[width=2.2in]{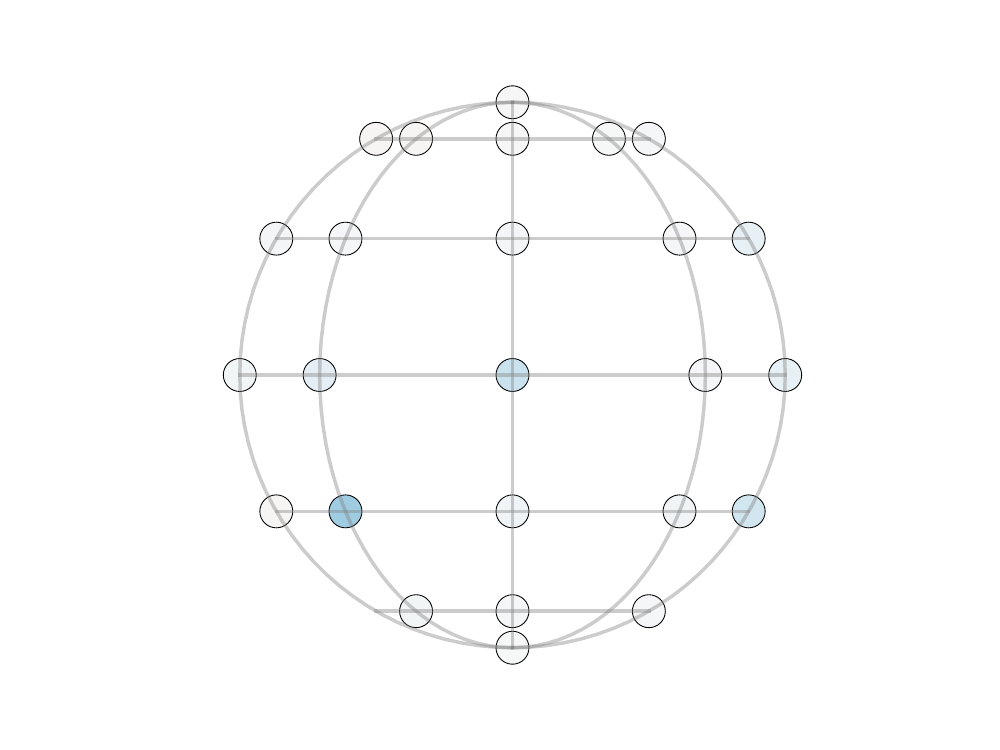}\\
		\vspace{-15pt}

		\hspace{-20pt}
		\includegraphics[width=1.5in]{figures/monomer60_fig.png}	&
		\hspace{-35pt}
		\includegraphics[width=2.2in]{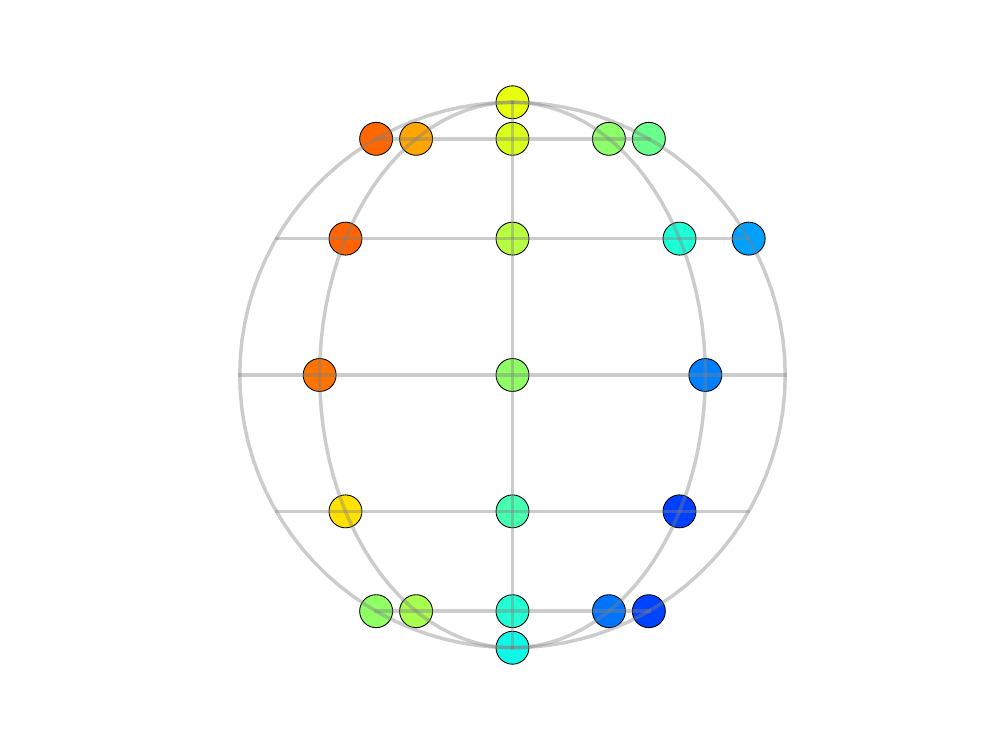}	&
		\hspace{-35pt}
		\includegraphics[width=2.2in]{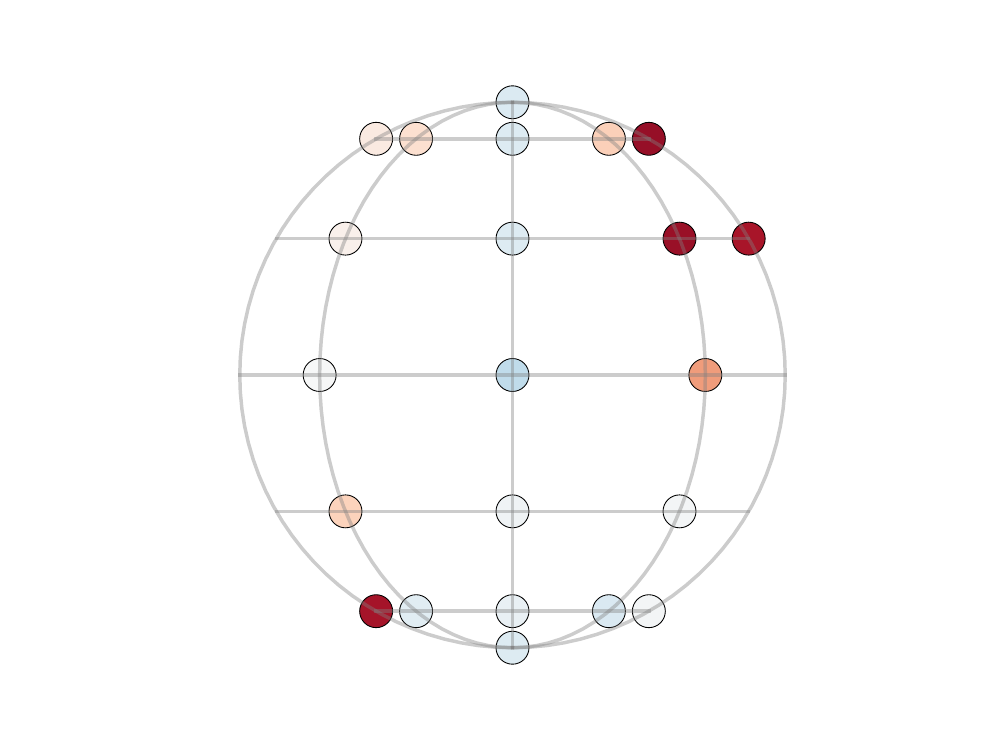}	&
		\hspace{-35pt}
		\includegraphics[width=2.2in]{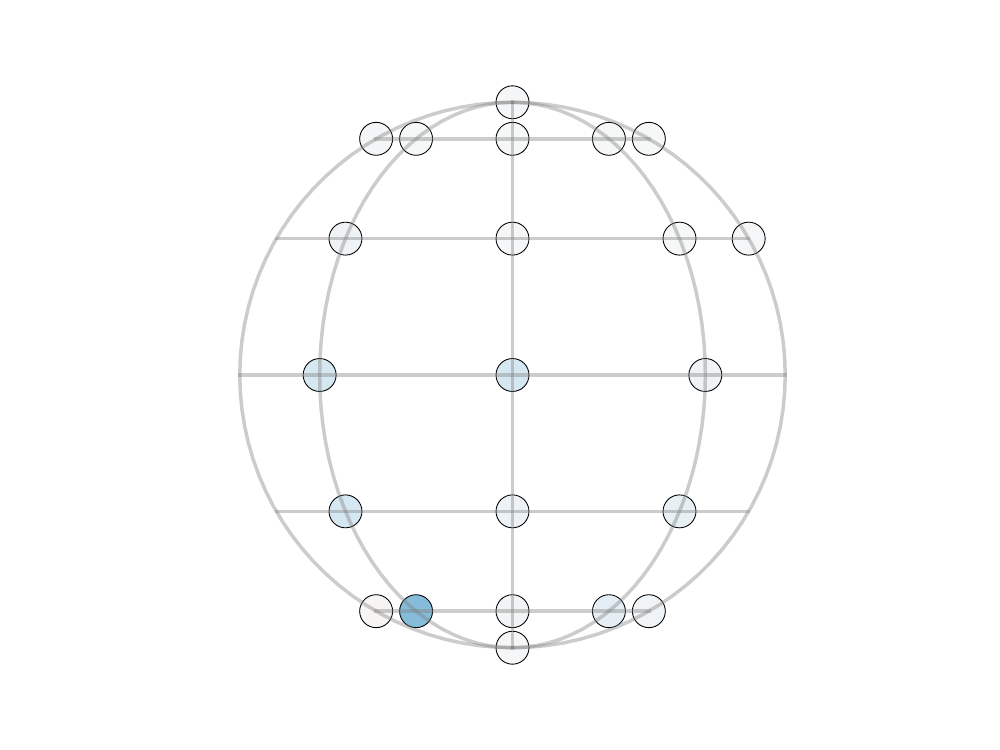}\\
		\vspace{-15pt}

		\hspace{-20pt}
		\includegraphics[width=1.58333in]{figures/monomer90_fig.png}	&
		\hspace{-35pt}
		\includegraphics[width=2.2in]{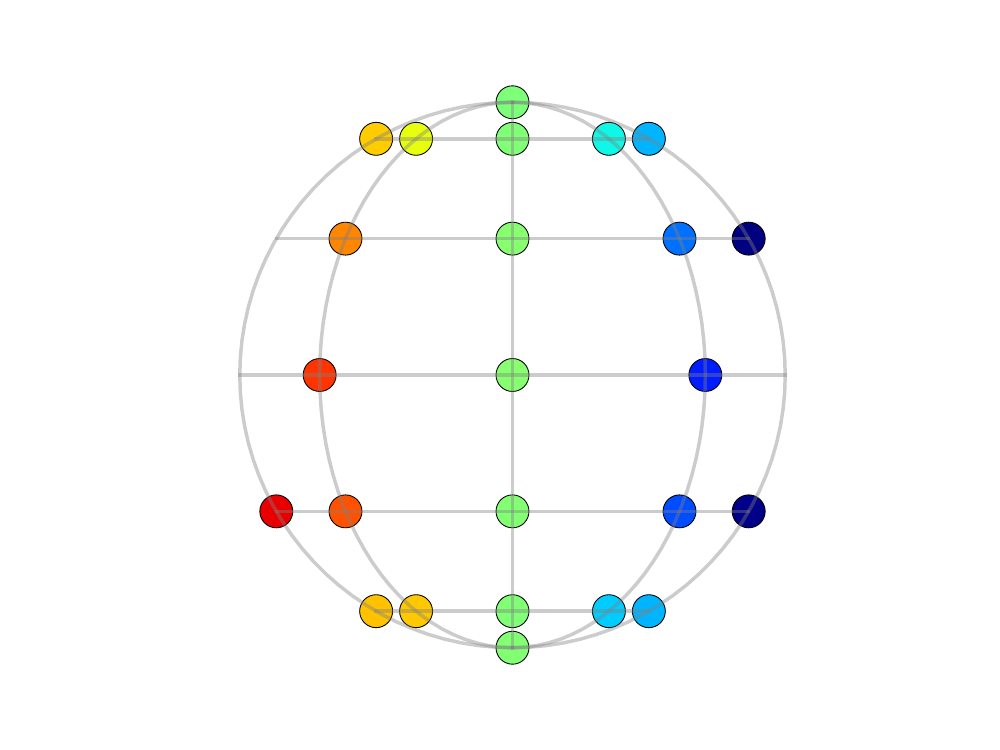}	&
		\hspace{-35pt}
		\includegraphics[width=2.2in]{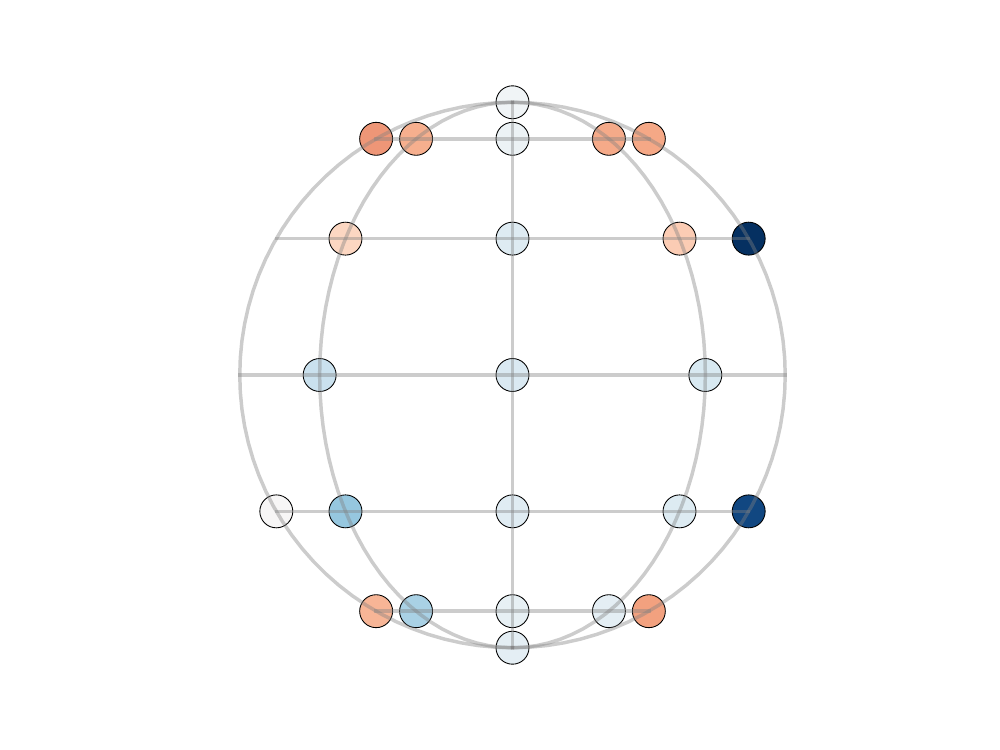}	&
		\hspace{-35pt}
		\includegraphics[width=2.2in]{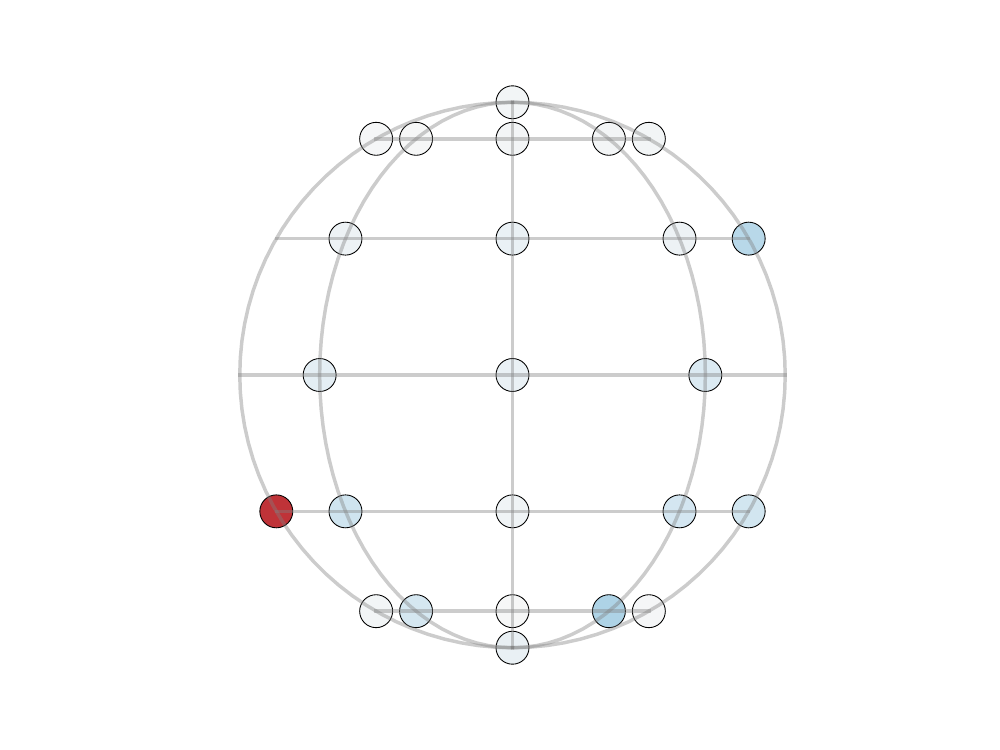}
	\end{tabular}
	\caption{\small
	Dimer relative orientational dependence of electronic coupling at 12 \text{\AA} separation.\\
	The first column depicts the polar orientation of the first molecule while the position on the polar plots on the right represents the orientation of the second molecule relative to the first (see Fig.~\ref{fig:appendix1b}).
	Columns 2-4 show the magnitude of $J_\text{TDDFT}$, the value of the coupling given by TDDFT (column 2), the error resulting from the IDA approximation to this (column 3) and the error resulting from the TDC estimate (column 4), as a function of the relative orientation.}
	\label{fig:appendix3}
\end{figure}

\begin{figure}[htp!]
	\begin{tabular}{cccc}
		\hspace{-20pt}
		$\theta_A$	&	\hspace{-35pt}$J_{TDDFT}$	&	\hspace{-35pt}$J_{TDDFT} - J_{IDA}$	&	\hspace{-35pt}$J_{TDDFT} - J_{TDC}$\\
		\vspace{-15pt}
		\hspace{-20pt}
		\includegraphics[width=1in]{figures/monomer0_fig.png}	&
		\hspace{-35pt}
		\includegraphics[width=2.2in]{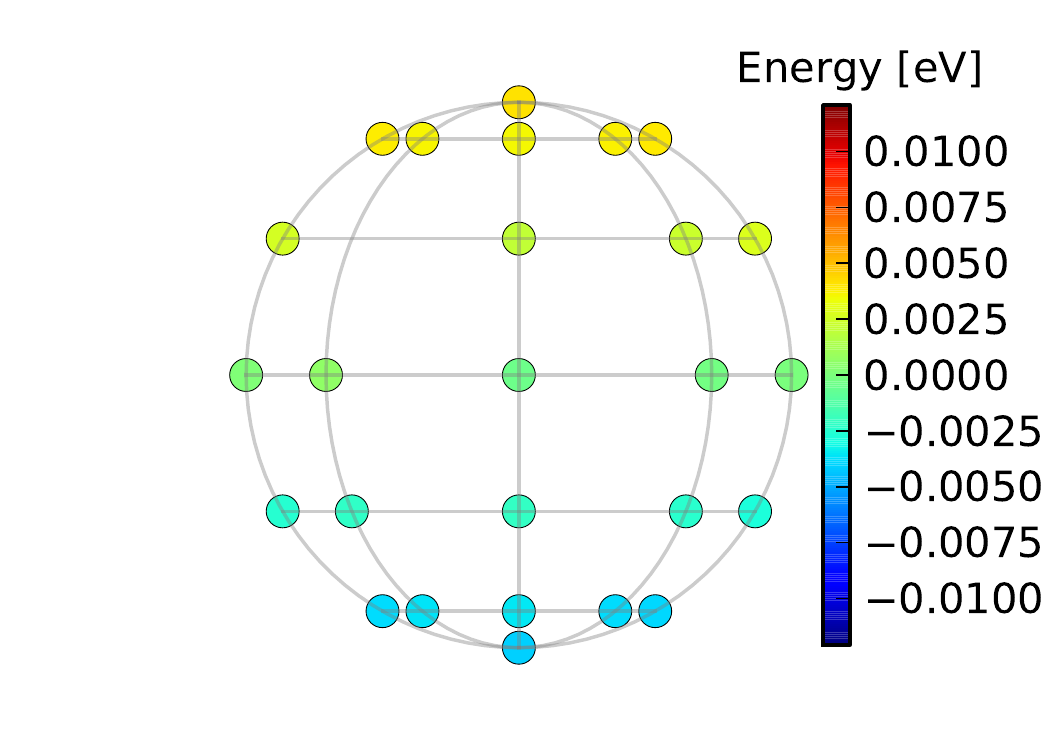}	&
		\hspace{-35pt}
		\includegraphics[width=2.2in]{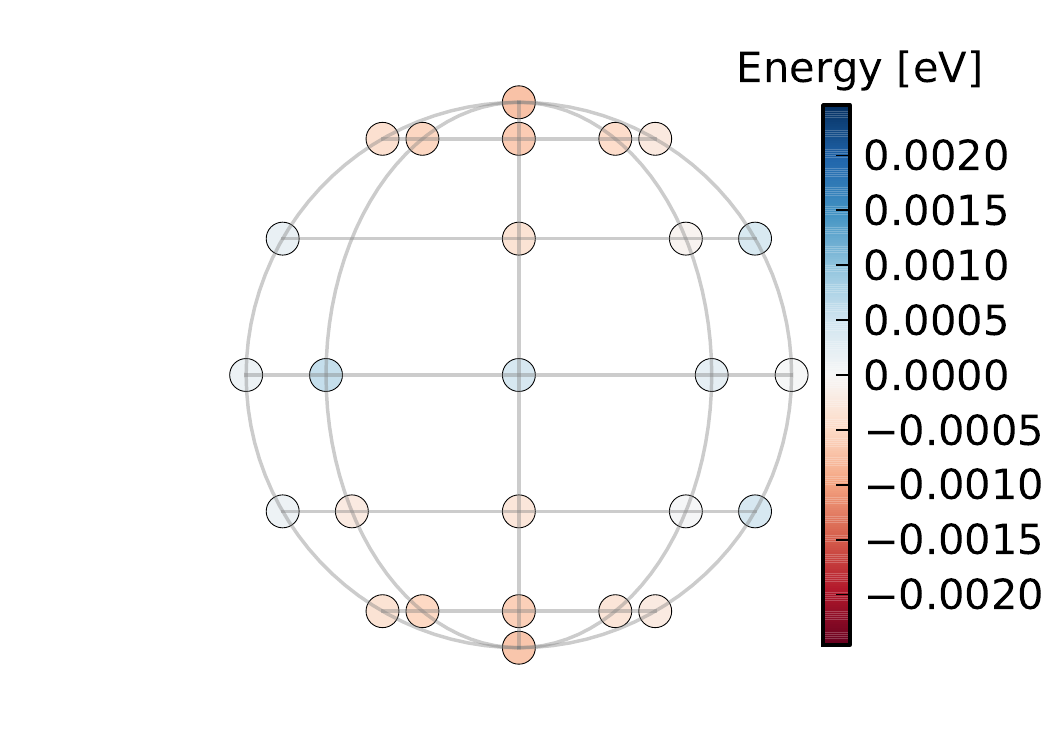}	&
		\hspace{-35pt}
		\includegraphics[width=2.2in]{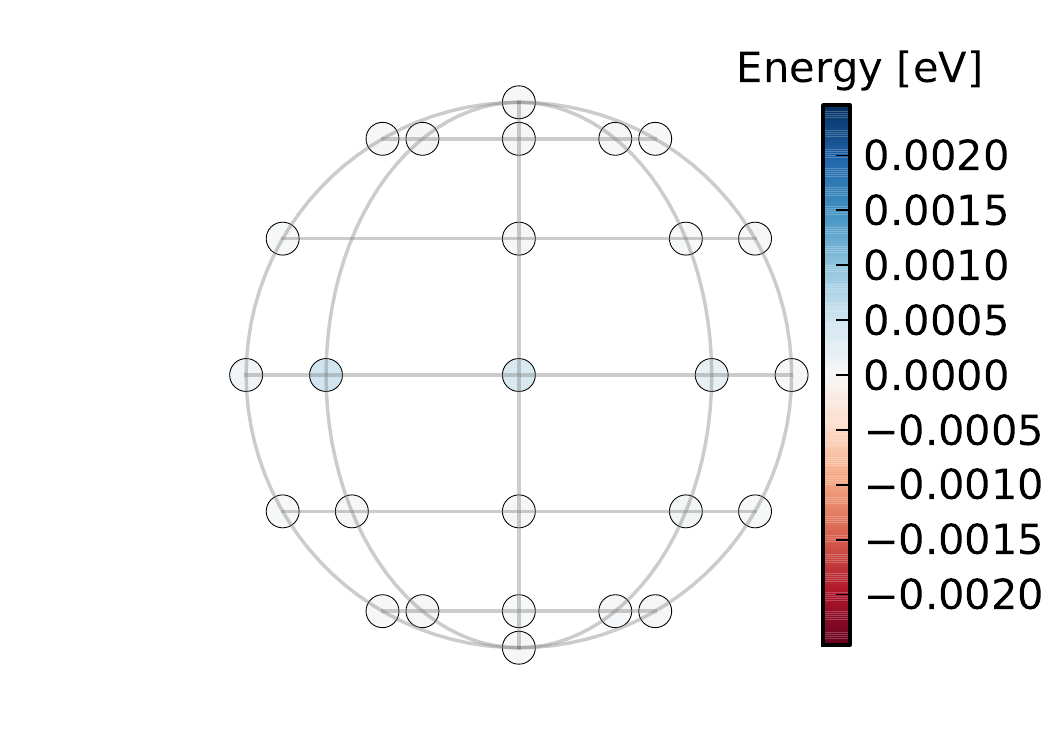}\\
		\vspace{-15pt}

		\hspace{-20pt}
		\includegraphics[width=1in]{figures/monomer30_fig.png}	&
		\hspace{-35pt}
		\includegraphics[width=2.2in]{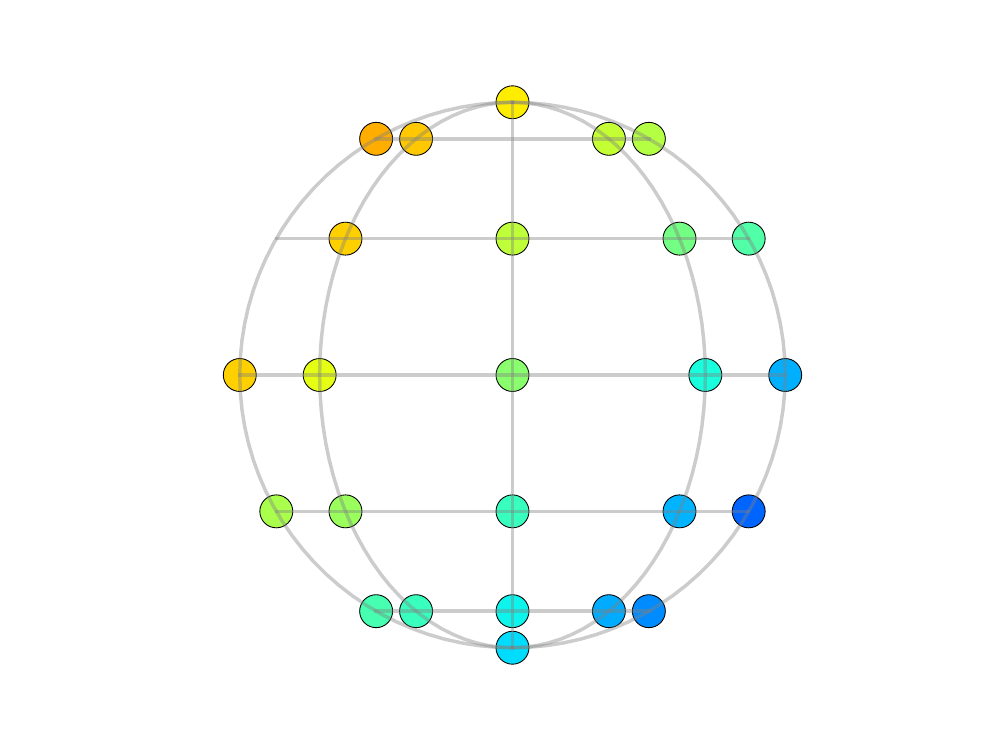}	&
		\hspace{-35pt}
		\includegraphics[width=2.2in]{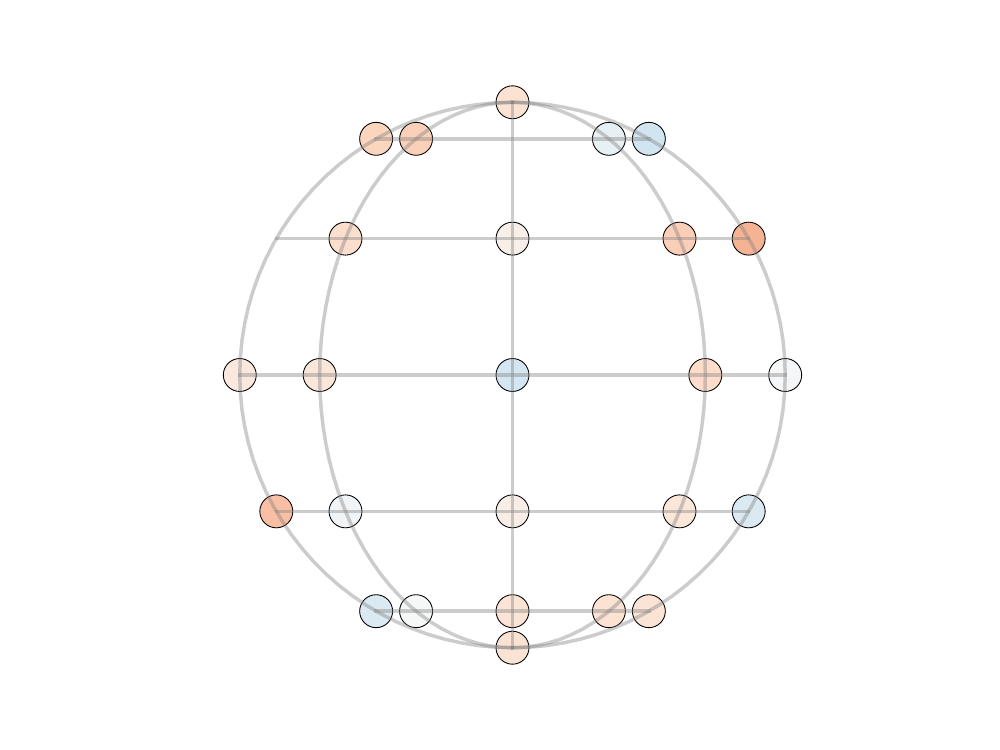}	&
		\hspace{-35pt}
		\includegraphics[width=2.2in]{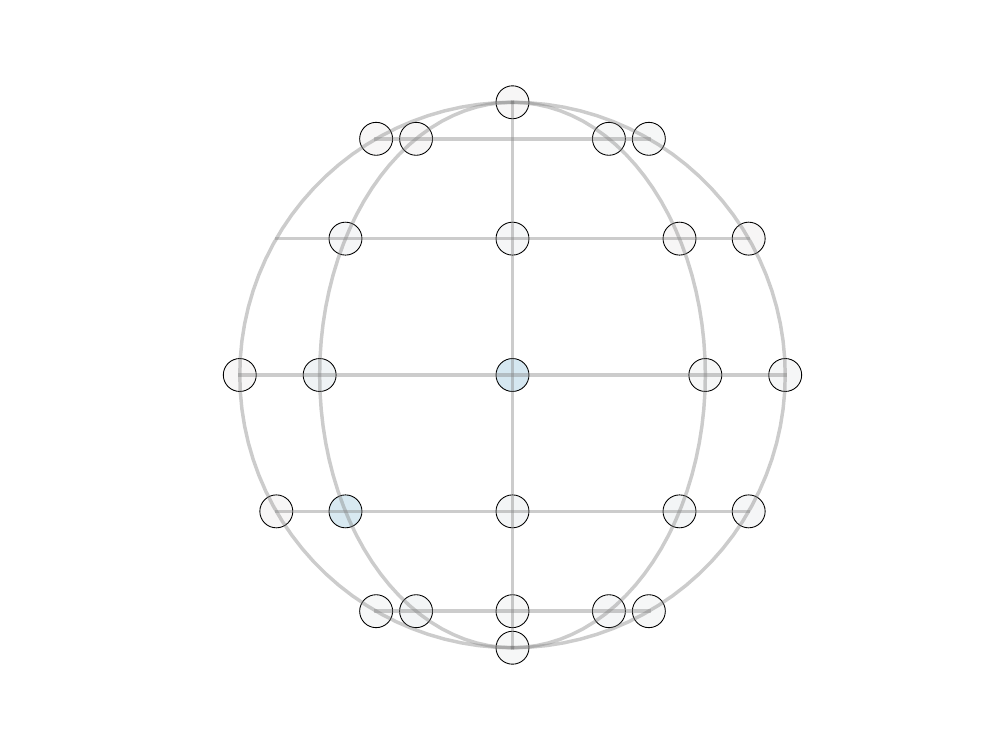}\\
		\vspace{-15pt}

		\hspace{-20pt}
		\includegraphics[width=1.5in]{figures/monomer60_fig.png}	&
		\hspace{-35pt}
		\includegraphics[width=2.2in]{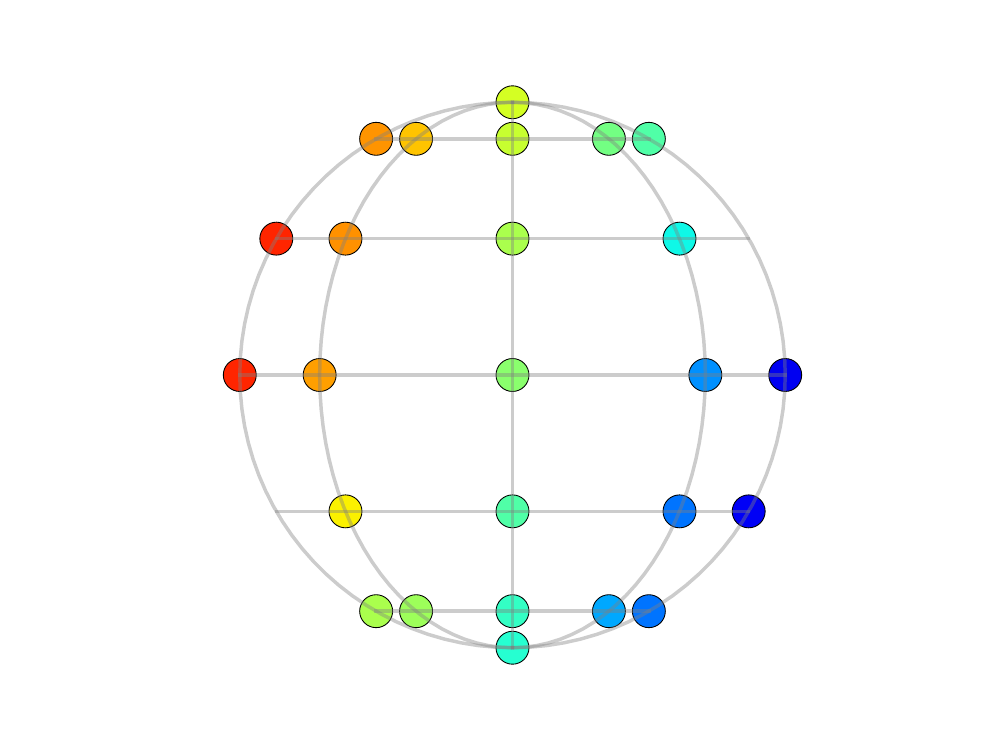}	&
		\hspace{-35pt}
		\includegraphics[width=2.2in]{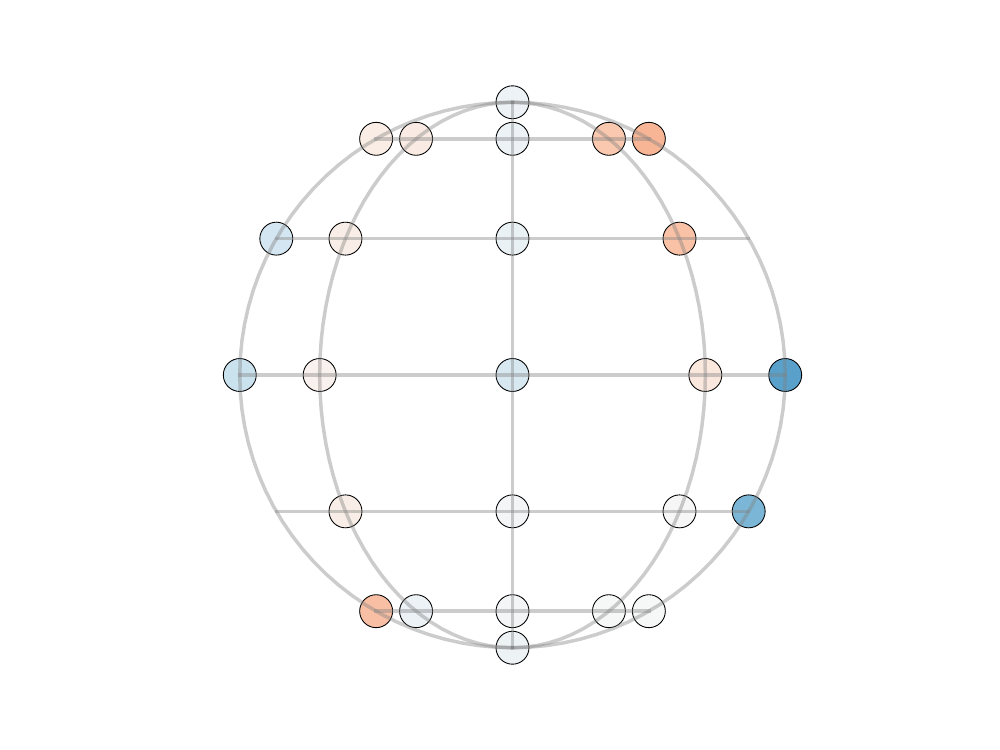}	&
		\hspace{-35pt}
		\includegraphics[width=2.2in]{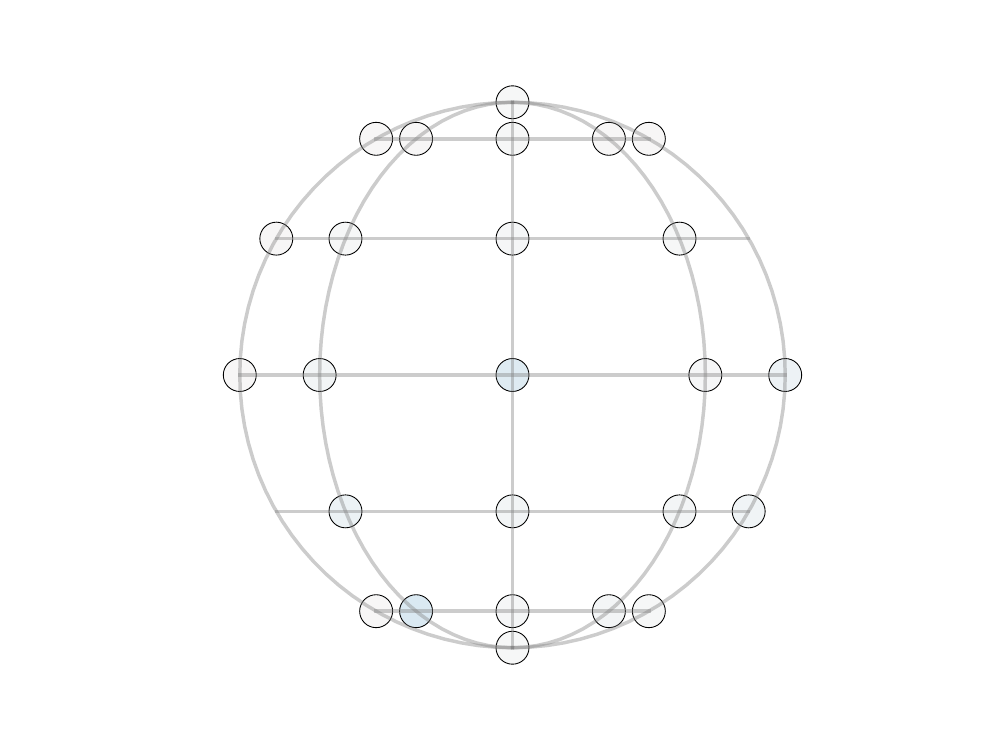}\\
		\vspace{-15pt}

		\hspace{-20pt}
		\includegraphics[width=1.58333in]{figures/monomer90_fig.png}	&
		\hspace{-35pt}
		\includegraphics[width=2.2in]{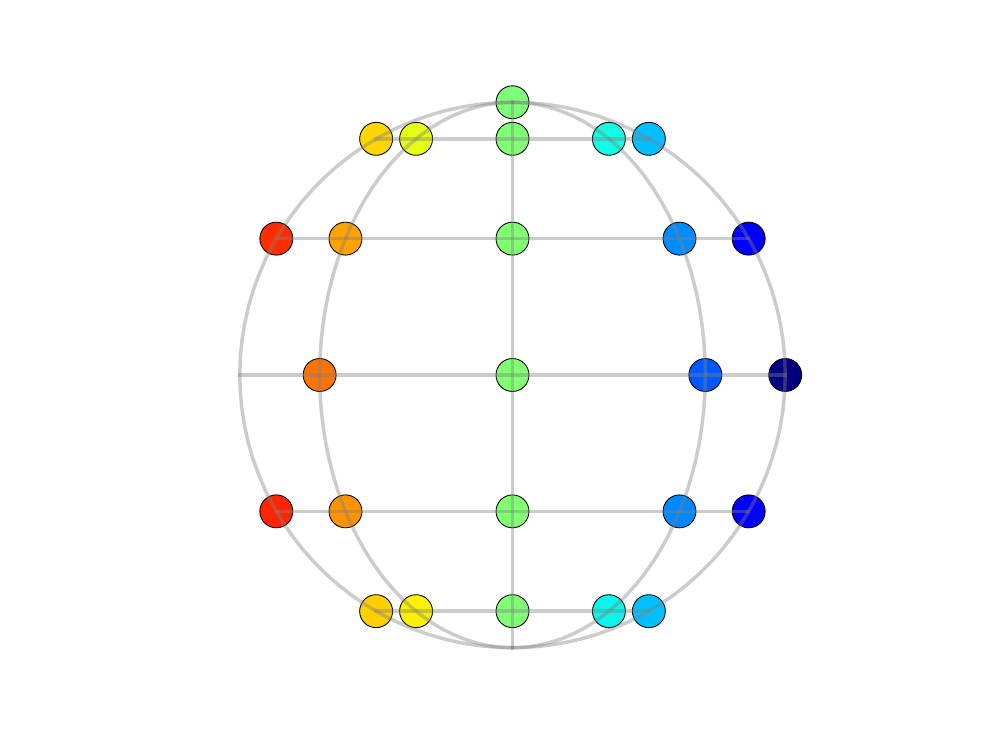}	&
		\hspace{-35pt}
		\includegraphics[width=2.2in]{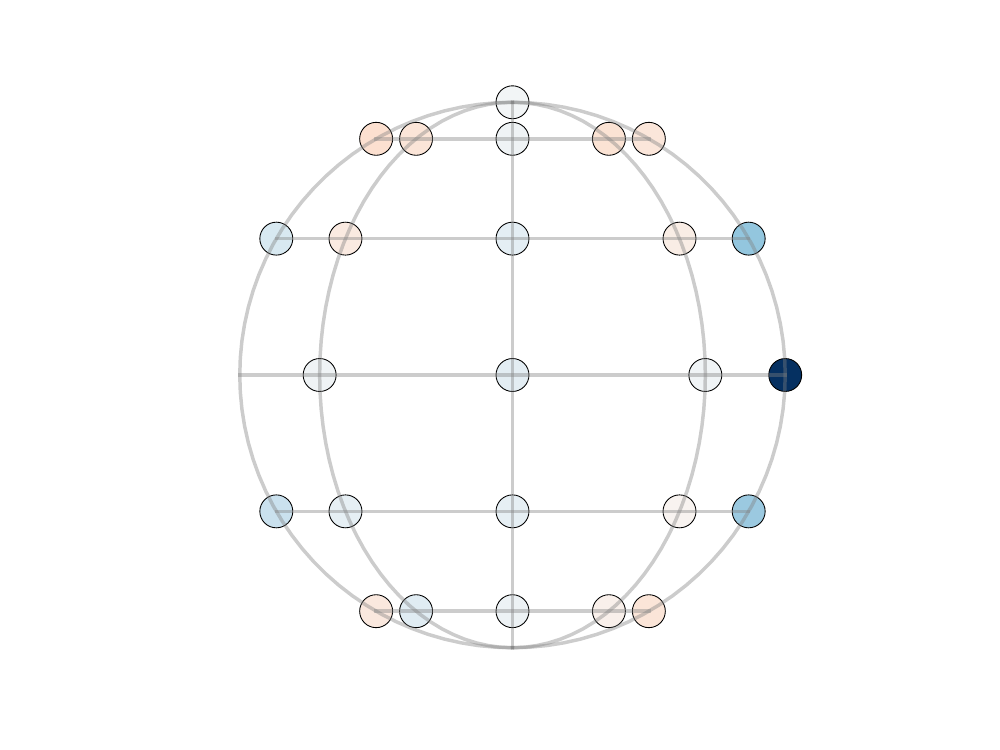}	&
		\hspace{-35pt}
		\includegraphics[width=2.2in]{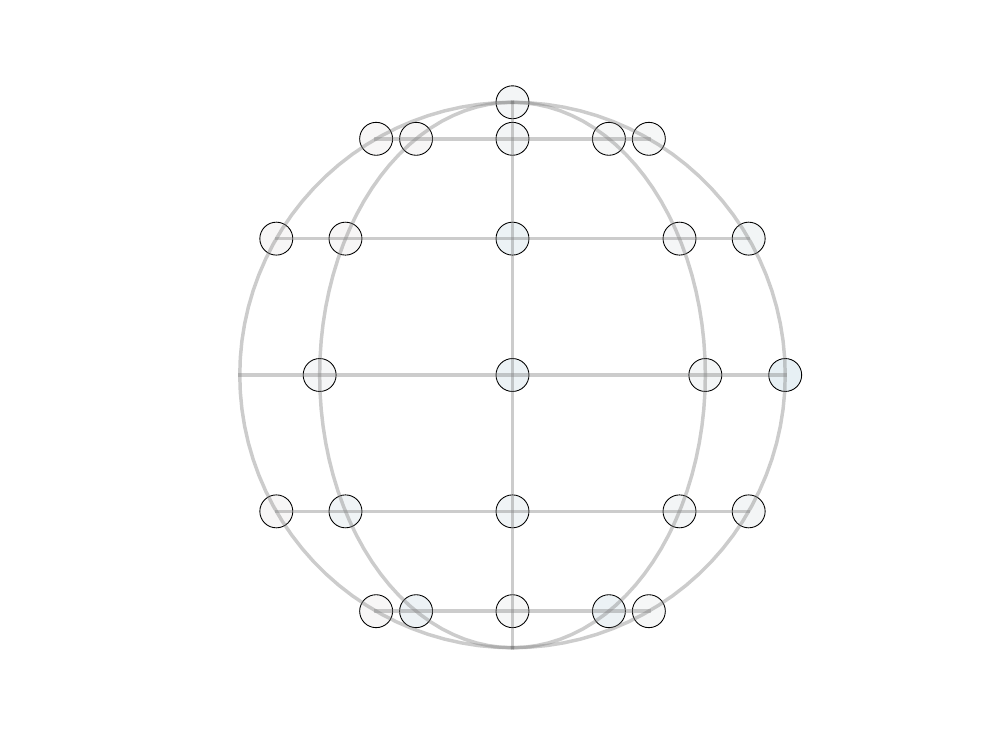}
	\end{tabular}
	\caption{\small 
	Dimer relative orientational dependence of electronic coupling at 18 \text{\AA} separation.\\
	The first column depicts the polar orientation of the first molecule while the position on the polar plots on the right represents the orientation of the second molecule relative to the first (see Fig.~\ref{fig:appendix1b}).
	Columns 2-4 show the magnitude of $J_\text{TDDFT}$, the value of the coupling given by TDDFT (column 2), the error resulting from the IDA approximation to this (column 3) and the error resulting from the TDC estimate (column 4), as a function of the relative orientation.}
	\label{fig:appendix4}
\end{figure}

In Figures~\ref{fig:appendix2} -- Figure~\ref{fig:appendix4}, we hold $\theta_A$ fixed and plot the coupling as a function of different relative orientations of the second dye.
Our sampling scheme leads to 42 possible relative orientations of the second dye.
Only half the sphere is shown because the reverse side was found to be quite symmetric due to the high symmetry of charge density across the plane of the page.

The data show that IDA can both over and underestimate the electronic coupling obtained from TDDFT.
In general, IDA overestimates for configurations that resemble H-aggregates (when $\theta_A$ is $0^\circ$) and underestimates for the configurations that resemble J-aggregates (when $\theta_A$ is $180^\circ$).

Since the IDA is by definition an approximation to the TDC coupling, the over and underestimation of IDA is due to the approximation of using a point dipole to approximate a charge density.
The TDC coupling itself seems to perform well over most configurations, giving good agreement with the TDDFT results.
There are a few data points for which TDC matches poorly to TDDFT; in these configurations, the spatial extent of the molecules overlaps enough to allow significant interaction between the two molecules.

\clearpage

\section{Charge Transfer states in B3LYP}
\begin{figure}[ht]
	\includegraphics[width=5in]{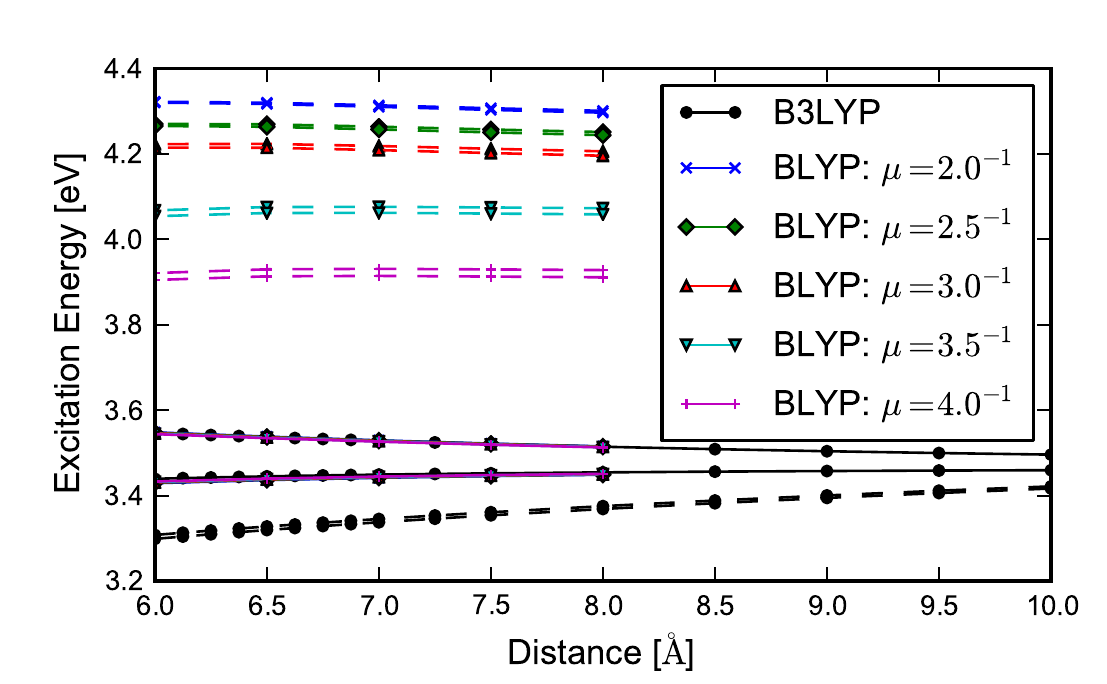}
	\caption{\label{cc_plot}
        \small Coumarin-343 energy levels predicted by B3LYP and range-corrected BLYP.
        The coupled exciton states (solid lines) for B3LYP and the range corrected calculations are shown overlapped, while the energies of the charge transfer states (dashed lines) varies based on the functional.
        The range corrected BLYP energies have been shifted such that the first excited state energies of the monomer calculations are all aligned to that of B3LYP.
        This is done to compare the energies of the charge transfer states relative to the coupled exciton states.}
\end{figure}
It is well known that TD-B3LYP is poor at predicting the energetics of charge transfer states, as are other functionals without 100\% Hartree-Fock exchange~\cite{Dreuw2004,Pan2009}.
Fig.~\ref{cc_plot} shows that TD-B3LYP predicts low-lying charge transfer states for the coumarin-343-MA dimer.
However, the range-corrected DFT calculations show that the charge transfer states are much higher in energy.
The splitting between the exciton states is consistent between B3LYP and the range corrected calculations.  
This suggests that the low lying charge transfer states predicted by B3LYP do not affect the character of the exciton states. 
Therefore, we may safely disregard these charge transfer states and use only the exciton states in our analysis.

\bibliographystyle{achemso}
\bibliography{Coumarin_DFT_Paper}

\providecommand*\mcitethebibliography{\thebibliography}
\csname @ifundefined\endcsname{endmcitethebibliography}
  {\let\endmcitethebibliography\endthebibliography}{}
\begin{mcitethebibliography}{70}
\providecommand*\natexlab[1]{#1}
\providecommand*\mciteSetBstSublistMode[1]{}
\providecommand*\mciteSetBstMaxWidthForm[2]{}
\providecommand*\mciteBstWouldAddEndPuncttrue
  {\def\EndOfBibitem{\unskip.}}
\providecommand*\mciteBstWouldAddEndPunctfalse
  {\let\EndOfBibitem\relax}
\providecommand*\mciteSetBstMidEndSepPunct[3]{}
\providecommand*\mciteSetBstSublistLabelBeginEnd[3]{}
\providecommand*\EndOfBibitem{}
\mciteSetBstSublistMode{f}
\mciteSetBstMaxWidthForm{subitem}{(\alph{mcitesubitemcount})}
\mciteSetBstSublistLabelBeginEnd
  {\mcitemaxwidthsubitemform\space}
  {\relax}
  {\relax}

\bibitem[Scholes et~al.(2005)Scholes, Ms, and Fleming]{Scholes2005}
Scholes,~G.~D.; Ms,~C.; Fleming,~G.~R. {Energy Transfer and Photosynthetic
  Light Harvesting}. In \emph{Advances in Chemical Physics}; Berry,~R.~S.,
  Jortner,~J., Eds.; Advances in Chemical Physics; John Wiley \& Sons, Inc.:
  Hoboken, NJ, USA, 2005; Vol.~i; Chapter 2, pp 57--129\relax
\mciteBstWouldAddEndPuncttrue
\mciteSetBstMidEndSepPunct{\mcitedefaultmidpunct}
{\mcitedefaultendpunct}{\mcitedefaultseppunct}\relax
\EndOfBibitem
\bibitem[Cheng and Fleming(2009)Cheng, and Fleming]{Cheng_Fleming_2009}
Cheng,~Y.-C.; Fleming,~G.~R. {Dynamics of Light Harvesting in Photosynthesis}.
  \emph{Annu. Rev. Phys. Chem.} \textbf{2009}, \emph{60}, 241--62\relax
\mciteBstWouldAddEndPuncttrue
\mciteSetBstMidEndSepPunct{\mcitedefaultmidpunct}
{\mcitedefaultendpunct}{\mcitedefaultseppunct}\relax
\EndOfBibitem
\bibitem[Scholes et~al.(2011)Scholes, Fleming, Olaya-Castro, and van
  Grondelle]{Scholes:2011qf}
Scholes,~G.~D.; Fleming,~G.~R.; Olaya-Castro,~A.; van Grondelle,~R. {Lessons
  From Nature About Solar Light Harvesting}. \emph{Nat Chem} \textbf{2011},
  \emph{3}, 763--774\relax
\mciteBstWouldAddEndPuncttrue
\mciteSetBstMidEndSepPunct{\mcitedefaultmidpunct}
{\mcitedefaultendpunct}{\mcitedefaultseppunct}\relax
\EndOfBibitem
\bibitem[Blankenship(2002)]{Bla-2002}
Blankenship,~R.~E. \emph{{Molecular Mechanisms of Photosynthesis}}; Blackwell
  Science, 2002\relax
\mciteBstWouldAddEndPuncttrue
\mciteSetBstMidEndSepPunct{\mcitedefaultmidpunct}
{\mcitedefaultendpunct}{\mcitedefaultseppunct}\relax
\EndOfBibitem
\bibitem[van Amerongen et~al.(2000)van Amerongen, Valkunas, and van
  Grondelle]{Ame.Val.etal-2000}
van Amerongen,~H.; Valkunas,~L.; van Grondelle,~R. \emph{{Photosynthetic
  Excitons}}; World Scientific, 2000\relax
\mciteBstWouldAddEndPuncttrue
\mciteSetBstMidEndSepPunct{\mcitedefaultmidpunct}
{\mcitedefaultendpunct}{\mcitedefaultseppunct}\relax
\EndOfBibitem
\bibitem[Scholes(2003)]{Scholes2003}
Scholes,~G.~D. {Long-Range Resonance Energy Transfer in Molecular Systems}.
  \emph{Annu. Rev. Phys. Chem.} \textbf{2003}, \emph{54}, 57--87\relax
\mciteBstWouldAddEndPuncttrue
\mciteSetBstMidEndSepPunct{\mcitedefaultmidpunct}
{\mcitedefaultendpunct}{\mcitedefaultseppunct}\relax
\EndOfBibitem
\bibitem[Scholes(1996)]{Scholes1996a}
Scholes,~G.~D. {Energy Transfer and Spectroscopic Characterization of
  Multichromophoric Assemblies}. \emph{J. Phys. Chem.} \textbf{1996},
  \emph{100}, 18731--18739\relax
\mciteBstWouldAddEndPuncttrue
\mciteSetBstMidEndSepPunct{\mcitedefaultmidpunct}
{\mcitedefaultendpunct}{\mcitedefaultseppunct}\relax
\EndOfBibitem
\bibitem[Cao and Silbey(2009)Cao, and Silbey]{Cao:2009kx}
Cao,~J.; Silbey,~R.~J. {Optimization of Exciton Trapping in Energy Transfer
  Processes}. \emph{J. Phys. Chem. A} \textbf{2009}, \emph{113},
  13825--38\relax
\mciteBstWouldAddEndPuncttrue
\mciteSetBstMidEndSepPunct{\mcitedefaultmidpunct}
{\mcitedefaultendpunct}{\mcitedefaultseppunct}\relax
\EndOfBibitem
\bibitem[Didraga et~al.(2002)Didraga, Klugkist, and Knoester]{Didraga:2002zr}
Didraga,~C.; Klugkist,~J.~A.; Knoester,~J. {Optical Properties of Helical
  Cylindrical Molecular Aggregates: The Homogeneous Limit}. \emph{J. Phys.
  Chem. B} \textbf{2002}, \emph{106}, 11474--11486\relax
\mciteBstWouldAddEndPuncttrue
\mciteSetBstMidEndSepPunct{\mcitedefaultmidpunct}
{\mcitedefaultendpunct}{\mcitedefaultseppunct}\relax
\EndOfBibitem
\bibitem[Didraga et~al.(2004)Didraga, Pug\v{z}lys, Hania, von Berlepsch,
  Duppen, Knoester, Pugz, and Berlepsch]{Didraga2004}
Didraga,~C.; Pug\v{z}lys,~A.; Hania,~P.~R.; von Berlepsch,~H.; Duppen,~K.;
  Knoester,~J.; Pugz,~A.; Berlepsch,~H.~V. {Structure, Spectroscopy, and
  Microscopic Model of Tubular Carbocyanine Dye Aggregates}. \emph{J. Phys.
  Chem. B} \textbf{2004}, \emph{108}, 14976--14985\relax
\mciteBstWouldAddEndPuncttrue
\mciteSetBstMidEndSepPunct{\mcitedefaultmidpunct}
{\mcitedefaultendpunct}{\mcitedefaultseppunct}\relax
\EndOfBibitem
\bibitem[Sarovar and Whaley(2013)Sarovar, and Whaley]{Sarovar:2013tc}
Sarovar,~M.; Whaley,~K.~B. {Design Principles and Fundamental Trade-Offs in
  Biomimetic Light Harvesting}. \emph{New J. Phys.} \textbf{2013}, \emph{15},
  013030\relax
\mciteBstWouldAddEndPuncttrue
\mciteSetBstMidEndSepPunct{\mcitedefaultmidpunct}
{\mcitedefaultendpunct}{\mcitedefaultseppunct}\relax
\EndOfBibitem
\bibitem[Scholes(2010)]{Sch-2010}
Scholes,~G.~D. {Quantum-Coherent Electronic Energy Transfer: Did Nature Think
  of It First?} \emph{J. Phys. Chem. Lett.} \textbf{2010}, \emph{1}, 2--8\relax
\mciteBstWouldAddEndPuncttrue
\mciteSetBstMidEndSepPunct{\mcitedefaultmidpunct}
{\mcitedefaultendpunct}{\mcitedefaultseppunct}\relax
\EndOfBibitem
\bibitem[Collini et~al.(2010)Collini, Wong, Wilk, Curmi, Brumer, and
  Scholes]{Collini2010}
Collini,~E.; Wong,~C.~Y.; Wilk,~K.~E.; Curmi,~P. M.~G.; Brumer,~P.;
  Scholes,~G.~D. {Coherently Wired Light-Harvesting in Photosynthetic Marine
  Algae at Ambient Temperature}. \emph{Nature} \textbf{2010}, \emph{463},
  644--647\relax
\mciteBstWouldAddEndPuncttrue
\mciteSetBstMidEndSepPunct{\mcitedefaultmidpunct}
{\mcitedefaultendpunct}{\mcitedefaultseppunct}\relax
\EndOfBibitem
\bibitem[Engel et~al.(2007)Engel, Calhoun, Read, Ahn, Mancal, Cheng,
  Blankenship, and Fleming]{Engel2007}
Engel,~G.~S.; Calhoun,~T.~R.; Read,~E.~L.; Ahn,~T.-K.; Mancal,~T.;
  Cheng,~Y.-C.; Blankenship,~R.~E.; Fleming,~G.~R. {Evidence for Wavelike
  Energy Transfer Through Quantum Coherence in Photosynthetic Systems}.
  \emph{Nature} \textbf{2007}, \emph{446}, 782--6\relax
\mciteBstWouldAddEndPuncttrue
\mciteSetBstMidEndSepPunct{\mcitedefaultmidpunct}
{\mcitedefaultendpunct}{\mcitedefaultseppunct}\relax
\EndOfBibitem
\bibitem[Sarovar et~al.(2010)Sarovar, Ishizaki, Fleming, and
  Whaley]{Sarovar:2010hc}
Sarovar,~M.; Ishizaki,~A.; Fleming,~G.~R.; Whaley,~K.~B. {Quantum Entanglement
  in Photosynthetic Light-Harvesting Complexes}. \emph{Nat. Phys.}
  \textbf{2010}, \emph{6}, 462--467\relax
\mciteBstWouldAddEndPuncttrue
\mciteSetBstMidEndSepPunct{\mcitedefaultmidpunct}
{\mcitedefaultendpunct}{\mcitedefaultseppunct}\relax
\EndOfBibitem
\bibitem[Ma et~al.(2008)Ma, Miller, Fleming, and Francis]{Ma2008}
Ma,~Y.-Z. Y.-Z.; Miller,~R.~A.; Fleming,~G.~R.; Francis,~M.~B. {Energy Transfer
  Dynamics in Light-Harvesting Assemblies Templated by the Tobacco Mosaic Virus
  Coat Protein}. \emph{J. Phys. Chem. B} \textbf{2008}, \emph{112},
  6887--92\relax
\mciteBstWouldAddEndPuncttrue
\mciteSetBstMidEndSepPunct{\mcitedefaultmidpunct}
{\mcitedefaultendpunct}{\mcitedefaultseppunct}\relax
\EndOfBibitem
\bibitem[Klug(1999)]{Klu-1999}
Klug,~A. {The Tobacco Mosaic Virus Particle: Structure and Assembly}.
  \emph{Phil. Trans. R. Soc. B} \textbf{1999}, \emph{354}, 531--5\relax
\mciteBstWouldAddEndPuncttrue
\mciteSetBstMidEndSepPunct{\mcitedefaultmidpunct}
{\mcitedefaultendpunct}{\mcitedefaultseppunct}\relax
\EndOfBibitem
\bibitem[Endo et~al.(2007)Endo, Fujitsuka, and Majima]{End.Fuj.etal-2007}
Endo,~M.; Fujitsuka,~M.; Majima,~T. {Porphyrin Light-Harvesting Arrays
  Constructed in the Recombinant Tobacco Mosaic Virus Scaffold}.
  \emph{Chemistry (Weinheim an der Bergstrasse, Germany)} \textbf{2007},
  \emph{13}, 8660--6\relax
\mciteBstWouldAddEndPuncttrue
\mciteSetBstMidEndSepPunct{\mcitedefaultmidpunct}
{\mcitedefaultendpunct}{\mcitedefaultseppunct}\relax
\EndOfBibitem
\bibitem[Nam et~al.(2010)Nam, Shin, Park, Magyar, Choi, Fantner, Nelson, and
  Belcher]{Nam:2010fw}
Nam,~Y.~S.; Shin,~T.; Park,~H.; Magyar,~A.~P.; Choi,~K.; Fantner,~G.;
  Nelson,~K.~A.; Belcher,~A.~M. {Virus-Templated Assembly of Porphyrins into
  Light-Harvesting Nanoantennae.} \emph{J. Am. Chem. Soc.} \textbf{2010},
  \emph{132}, 1462--3\relax
\mciteBstWouldAddEndPuncttrue
\mciteSetBstMidEndSepPunct{\mcitedefaultmidpunct}
{\mcitedefaultendpunct}{\mcitedefaultseppunct}\relax
\EndOfBibitem
\bibitem[Miller et~al.(2007)Miller, Presley, and Francis]{Miller2007}
Miller,~R.~A.; Presley,~A.~D.; Francis,~M.~B. {Self-Assembling Light-Harvesting
  Systems from Synthetically Modified Tobacco Mosaic Virus Coat Proteins}.
  \emph{J. Am. Chem. Soc.} \textbf{2007}, \emph{129}, 3104--9\relax
\mciteBstWouldAddEndPuncttrue
\mciteSetBstMidEndSepPunct{\mcitedefaultmidpunct}
{\mcitedefaultendpunct}{\mcitedefaultseppunct}\relax
\EndOfBibitem
\bibitem[Witus and Francis(2011)Witus, and Francis]{Witus:2011wz}
Witus,~L.~S.; Francis,~M.~B. {Using Synthetically Modified Proteins to Make New
  Materials.} \emph{Acc. Chem. Res.} \textbf{2011}, \emph{44}, 774--83\relax
\mciteBstWouldAddEndPuncttrue
\mciteSetBstMidEndSepPunct{\mcitedefaultmidpunct}
{\mcitedefaultendpunct}{\mcitedefaultseppunct}\relax
\EndOfBibitem
\bibitem[Scholes and Rumbles(2006)Scholes, and Rumbles]{Scholes2006a}
Scholes,~G.~D.; Rumbles,~G. {Excitons in Nanoscale Systems}. \emph{Nat. Mater.}
  \textbf{2006}, \emph{5}, 683--696\relax
\mciteBstWouldAddEndPuncttrue
\mciteSetBstMidEndSepPunct{\mcitedefaultmidpunct}
{\mcitedefaultendpunct}{\mcitedefaultseppunct}\relax
\EndOfBibitem
\bibitem[Knoester(2002)]{Knoester2002a}
Knoester,~J. {Optical Properties of Molecular Aggregates}. In \emph{Proceedings
  of the International School of Physics "Enrico Fermi" Course CXLIX};
  Agranovich,~V.~M., LaRocca,~G.~C., Eds.; Amsterdam: IOS Press, 2002; pp
  149--186\relax
\mciteBstWouldAddEndPuncttrue
\mciteSetBstMidEndSepPunct{\mcitedefaultmidpunct}
{\mcitedefaultendpunct}{\mcitedefaultseppunct}\relax
\EndOfBibitem
\bibitem[Krueger et~al.(1998)Krueger, Scholes, and Fleming]{Krueger:1998vn}
Krueger,~B.~P.; Scholes,~G.~D.; Fleming,~G.~R. {Calculation of Couplings and
  Energy-Transfer Pathways between the Pigments of LH2 by the ab Initio
  Transition Density Cube Method}. \emph{J. Phys. Chem. B} \textbf{1998},
  \emph{102}, 5378--5386\relax
\mciteBstWouldAddEndPuncttrue
\mciteSetBstMidEndSepPunct{\mcitedefaultmidpunct}
{\mcitedefaultendpunct}{\mcitedefaultseppunct}\relax
\EndOfBibitem
\bibitem[Kasha et~al.(1965)Kasha, Rawls, and {Ashraf El-Bayoumi}]{Kasha1965}
Kasha,~M.; Rawls,~H.~R.; {Ashraf El-Bayoumi},~M. {The Exciton Model in
  Molecular Spectroscopy}. \emph{Pure Appl. Chem.} \textbf{1965}, \emph{11},
  371--392\relax
\mciteBstWouldAddEndPuncttrue
\mciteSetBstMidEndSepPunct{\mcitedefaultmidpunct}
{\mcitedefaultendpunct}{\mcitedefaultseppunct}\relax
\EndOfBibitem
\bibitem[Howard et~al.(2004)Howard, Zutterman, Deroover, Lamoen, and {Van
  Alsenoy}]{Howard:2004kx}
Howard,~I.~A.; Zutterman,~F.; Deroover,~G.; Lamoen,~D.; {Van Alsenoy},~C.
  {Approaches to Calculation of Exciton Interaction Energies for a Molecular
  Dimer}. \emph{J. Phys. Chem. B} \textbf{2004}, \emph{108}, 19155--19162\relax
\mciteBstWouldAddEndPuncttrue
\mciteSetBstMidEndSepPunct{\mcitedefaultmidpunct}
{\mcitedefaultendpunct}{\mcitedefaultseppunct}\relax
\EndOfBibitem
\bibitem[Mu\~{n}oz Losa et~al.(2009)Mu\~{n}oz Losa, Curutchet, Krueger,
  Hartsell, and Mennucci]{Aurora-Mun-oz-Losa:sb}
Mu\~{n}oz Losa,~A.; Curutchet,~C.; Krueger,~B.~P.; Hartsell,~L.~R.;
  Mennucci,~B. {Fretting about FRET: Failure of the Ideal Dipole
  Approximation}. \emph{Biophys. J.} \textbf{2009}, \emph{96}, 4779--88\relax
\mciteBstWouldAddEndPuncttrue
\mciteSetBstMidEndSepPunct{\mcitedefaultmidpunct}
{\mcitedefaultendpunct}{\mcitedefaultseppunct}\relax
\EndOfBibitem
\bibitem[Scholes et~al.(2001)Scholes, Jordanides, and Fleming]{Scholes2001a}
Scholes,~G.~D.; Jordanides,~X.~J.; Fleming,~G.~R. {Adapting the F\"{o}rster
  Theory of Energy Transfer for Modeling Dynamics in Aggregated Molecular
  Assemblies}. \emph{J. Phys. Chem. B} \textbf{2001}, \emph{105},
  1640--1651\relax
\mciteBstWouldAddEndPuncttrue
\mciteSetBstMidEndSepPunct{\mcitedefaultmidpunct}
{\mcitedefaultendpunct}{\mcitedefaultseppunct}\relax
\EndOfBibitem
\bibitem[Madjet et~al.(2006)Madjet, Abdurahman, and Renger]{Madjet2006}
Madjet,~M.~E.; Abdurahman,~a.; Renger,~T. {Intermolecular Coulomb Couplings
  from Ab Initio Electrostatic Potentials: Application to Optical Transitions
  of Strongly Coupled Pigments in Photosynthetic Antennae and Reaction
  Centers}. \emph{J. Phys. Chem. B} \textbf{2006}, \emph{110}, 17268--81\relax
\mciteBstWouldAddEndPuncttrue
\mciteSetBstMidEndSepPunct{\mcitedefaultmidpunct}
{\mcitedefaultendpunct}{\mcitedefaultseppunct}\relax
\EndOfBibitem
\bibitem[Hsu et~al.(2001)Hsu, Fleming, Head-Gordon, and
  Head-Gordon]{Hsu:2001el}
Hsu,~C.-P.; Fleming,~G.~R.; Head-Gordon,~M.; Head-Gordon,~T. {Excitation Energy
  Transfer in Condensed Media}. \emph{J. Chem. Phys.} \textbf{2001},
  \emph{114}, 3065\relax
\mciteBstWouldAddEndPuncttrue
\mciteSetBstMidEndSepPunct{\mcitedefaultmidpunct}
{\mcitedefaultendpunct}{\mcitedefaultseppunct}\relax
\EndOfBibitem
\bibitem[Curutchet et~al.(2007)Curutchet, Scholes, Mennucci, and
  Cammi]{Curutchet:2007ji}
Curutchet,~C.; Scholes,~G.~D.; Mennucci,~B.; Cammi,~R. {How Solvent Controls
  Electronic Energy Transfer and Light Harvesting: Toward a Quantum-Mechanical
  Description of Reaction Field and Screening Effects}. \emph{J. Phys. Chem. B}
  \textbf{2007}, \emph{111}, 13253--13265\relax
\mciteBstWouldAddEndPuncttrue
\mciteSetBstMidEndSepPunct{\mcitedefaultmidpunct}
{\mcitedefaultendpunct}{\mcitedefaultseppunct}\relax
\EndOfBibitem
\bibitem[Sinnokrot and Sherrill(2006)Sinnokrot, and Sherrill]{Sinnokrot:2006ka}
Sinnokrot,~M.~O.; Sherrill,~C.~D. {High-Accuracy Quantum Mechanical Studies of
  Pi-Pi Interactions in Benzene Dimers}. \emph{J. Phys. Chem. A} \textbf{2006},
  \emph{110}, 10656--10668\relax
\mciteBstWouldAddEndPuncttrue
\mciteSetBstMidEndSepPunct{\mcitedefaultmidpunct}
{\mcitedefaultendpunct}{\mcitedefaultseppunct}\relax
\EndOfBibitem
\bibitem[Fink et~al.(2008)Fink, Pfister, Schneider, Zhao, and
  Engels]{Fink:2008he}
Fink,~R.~F.; Pfister,~J.; Schneider,~A.; Zhao,~H.; Engels,~B. {Ab Initio
  Configuration Interaction Description of Excitation Energy Transfer Between
  Closely Packed Molecules}. \emph{Chem. Phys.} \textbf{2008}, \emph{343},
  353--361\relax
\mciteBstWouldAddEndPuncttrue
\mciteSetBstMidEndSepPunct{\mcitedefaultmidpunct}
{\mcitedefaultendpunct}{\mcitedefaultseppunct}\relax
\EndOfBibitem
\bibitem[Fink et~al.(2008)Fink, Pfister, Zhao, and Engels]{Fink:2008jv}
Fink,~R.~F.; Pfister,~J.; Zhao,~H.~M.; Engels,~B. {Assessment of Quantum
  Chemical Methods and Basis Sets for Excitation Energy Transfer}. \emph{Chem.
  Phys.} \textbf{2008}, \emph{346}, 275--285\relax
\mciteBstWouldAddEndPuncttrue
\mciteSetBstMidEndSepPunct{\mcitedefaultmidpunct}
{\mcitedefaultendpunct}{\mcitedefaultseppunct}\relax
\EndOfBibitem
\bibitem[Zhao et~al.(2009)Zhao, Pfister, Settels, Renz, Kaupp, Dehm,
  W{\"u}rthner, Fink, and Engels]{Zhao:2009kd}
Zhao,~H.~M.; Pfister,~J.; Settels,~V.; Renz,~M.; Kaupp,~M.; Dehm,~V.~C.;
  W{\"u}rthner,~F.; Fink,~R.~F.; Engels,~B. {Understanding Ground- and
  Excited-State Properties of Perylene Tetracarboxylic Acid Bisimide Crystals
  by Means of Quantum Chemical Computations}. \emph{J. Am. Chem. Soc.}
  \textbf{2009}, \emph{131}, 15660--15668\relax
\mciteBstWouldAddEndPuncttrue
\mciteSetBstMidEndSepPunct{\mcitedefaultmidpunct}
{\mcitedefaultendpunct}{\mcitedefaultseppunct}\relax
\EndOfBibitem
\bibitem[Liu et~al.(2011)Liu, Settels, Harbach, Dreuw, Fink, and
  Engels]{Liu:2011fk}
Liu,~W.; Settels,~V.; Harbach,~P. H.~P.; Dreuw,~A.; Fink,~R.~F.; Engels,~B.
  {Assessment of TD-DFT- and TD-HF-based Approaches for the Prediction of
  Exciton Coupling Parameters, Potential Energy Curves, and Electronic
  Characters of Electronically Excited Aggregates}. \emph{J. Comput. Chem.}
  \textbf{2011}, \emph{32}, 1971--1981\relax
\mciteBstWouldAddEndPuncttrue
\mciteSetBstMidEndSepPunct{\mcitedefaultmidpunct}
{\mcitedefaultendpunct}{\mcitedefaultseppunct}\relax
\EndOfBibitem
\bibitem[Stephens et~al.(1994)Stephens, Devlin, Chabalowski, and Frisch]{B3LYP}
Stephens,~P.~J.; Devlin,~F.~J.; Chabalowski,~C.~F.; Frisch,~M.~J. Ab Initio
  Calculation of Vibrational Absorption and Circular Dichroism Spectra Using
  Density Functional Force Fields. \emph{J. Phys. Chem.} \textbf{1994},
  \emph{98}, 11623\relax
\mciteBstWouldAddEndPuncttrue
\mciteSetBstMidEndSepPunct{\mcitedefaultmidpunct}
{\mcitedefaultendpunct}{\mcitedefaultseppunct}\relax
\EndOfBibitem
\bibitem[Ditchfield et~al.(1971)Ditchfield, Hehre, and Pople]{631g_h}
Ditchfield,~R.; Hehre,~W.~J.; Pople,~J.~A. Self-Consistent Molecular-Orbital
  Methods. IX. An Extended Gaussian-Type Basis for Molecular-Orbital Studies of
  Organic Molecules. \emph{J. Chem. Phys.} \textbf{1971}, \emph{54}, 724\relax
\mciteBstWouldAddEndPuncttrue
\mciteSetBstMidEndSepPunct{\mcitedefaultmidpunct}
{\mcitedefaultendpunct}{\mcitedefaultseppunct}\relax
\EndOfBibitem
\bibitem[Hehre et~al.(1972)Hehre, Ditchfield, and Pople]{631g_cnof}
Hehre,~W.~J.; Ditchfield,~R.; Pople,~J.~A. Self---Consistent Molecular Orbital
  Methods. XII. Further Extensions of Gaussian---Type Basis Sets for Use in
  Molecular Orbital Studies of Organic Molecules. \emph{J. Chem. Phys.}
  \textbf{1972}, \emph{56}, 2257\relax
\mciteBstWouldAddEndPuncttrue
\mciteSetBstMidEndSepPunct{\mcitedefaultmidpunct}
{\mcitedefaultendpunct}{\mcitedefaultseppunct}\relax
\EndOfBibitem
\bibitem[Hariharan and Pople(1973)Hariharan, and Pople]{631g_polarization}
Hariharan,~P.; Pople,~J. The Influence of Polarization Functions on Molecular
  Orbital Hydrogenation Energies. \emph{Theo. Chim. Acta} \textbf{1973},
  \emph{28}, 213\relax
\mciteBstWouldAddEndPuncttrue
\mciteSetBstMidEndSepPunct{\mcitedefaultmidpunct}
{\mcitedefaultendpunct}{\mcitedefaultseppunct}\relax
\EndOfBibitem
\bibitem[Runge and Gross(1984)Runge, and Gross]{TDDFT}
Runge,~E.; Gross,~E. K.~U. Density-Functional Theory for Time-Dependent
  Systems. \emph{Phys. Rev. Lett.} \textbf{1984}, \emph{52}, 997\relax
\mciteBstWouldAddEndPuncttrue
\mciteSetBstMidEndSepPunct{\mcitedefaultmidpunct}
{\mcitedefaultendpunct}{\mcitedefaultseppunct}\relax
\EndOfBibitem
\bibitem[Gross and Kohn(1990)Gross, and Kohn]{TDDFT_Rev}
Gross,~E.; Kohn,~W. Time-Dependent Density-Functional Theory. In \emph{Density
  Functional Theory of Many-Fermion Systems}; L\"owdin,~P.-O., Ed.; Adv. Quant.
  Chem.; Academic Press, 1990; Vol.~21; pp 255 -- 291\relax
\mciteBstWouldAddEndPuncttrue
\mciteSetBstMidEndSepPunct{\mcitedefaultmidpunct}
{\mcitedefaultendpunct}{\mcitedefaultseppunct}\relax
\EndOfBibitem
\bibitem[Zhao and Truhlar(2006)Zhao, and Truhlar]{M06HF}
Zhao,~Y.; Truhlar,~D.~G. Density Functional for Spectroscopy:‚Äâ No
  Long-Range Self-Interaction Error, Good Performance for Rydberg and
  Charge-Transfer States, and Better Performance on Average than B3LYP for
  Ground States. \emph{J. Phys. Chem. A} \textbf{2006}, \emph{110}, 13126\relax
\mciteBstWouldAddEndPuncttrue
\mciteSetBstMidEndSepPunct{\mcitedefaultmidpunct}
{\mcitedefaultendpunct}{\mcitedefaultseppunct}\relax
\EndOfBibitem
\bibitem[Peverati and Truhlar(2011)Peverati, and Truhlar]{Peverati2011}
Peverati,~R.; Truhlar,~D.~G. {Improving the Accuracy of Hybrid Meta-GGA Density
  Functionals by Range Separation}. \emph{J. Phys. Chem. Lett.} \textbf{2011},
  \emph{2}, 2810--2817\relax
\mciteBstWouldAddEndPuncttrue
\mciteSetBstMidEndSepPunct{\mcitedefaultmidpunct}
{\mcitedefaultendpunct}{\mcitedefaultseppunct}\relax
\EndOfBibitem
\bibitem[Peverati and Truhlar(2012)Peverati, and Truhlar]{Peverati2012}
Peverati,~R.; Truhlar,~D.~G. {M11-L: A Local Density Functional that Provides
  Improved Accuracy for Electronic Structure Calculations in Chemistry and
  Physics}. \emph{J. Phys. Chem. Lett.} \textbf{2012}, \emph{3}, 117--124\relax
\mciteBstWouldAddEndPuncttrue
\mciteSetBstMidEndSepPunct{\mcitedefaultmidpunct}
{\mcitedefaultendpunct}{\mcitedefaultseppunct}\relax
\EndOfBibitem
\bibitem[Perdew et~al.(1996)Perdew, Burke, and Ernzerhof]{PBE}
Perdew,~J.~P.; Burke,~K.; Ernzerhof,~M. Generalized Gradient Approximation Made
  Simple. \emph{Phys. Rev. Lett.} \textbf{1996}, \emph{77}, 3865--3868\relax
\mciteBstWouldAddEndPuncttrue
\mciteSetBstMidEndSepPunct{\mcitedefaultmidpunct}
{\mcitedefaultendpunct}{\mcitedefaultseppunct}\relax
\EndOfBibitem
\bibitem[Perdew et~al.(1997)Perdew, Burke, and Ernzerhof]{PBE_Err}
Perdew,~J.~P.; Burke,~K.; Ernzerhof,~M. Generalized Gradient Approximation Made
  Simple [Phys. Rev. Lett. 77, 3865 (1996)]. \emph{Phys. Rev. Lett.}
  \textbf{1997}, \emph{78}, 1396--1396\relax
\mciteBstWouldAddEndPuncttrue
\mciteSetBstMidEndSepPunct{\mcitedefaultmidpunct}
{\mcitedefaultendpunct}{\mcitedefaultseppunct}\relax
\EndOfBibitem
\bibitem[Clark et~al.(1983)Clark, Chandrasekhar, Spitznagel, and
  Schleyer]{321G_diffuse}
Clark,~T.; Chandrasekhar,~J.; Spitznagel,~G.~W.; Schleyer,~P. V.~R. Efficient
  Diffuse Function-Augmented Basis Sets For Anion Calculations. III. The 3-21+G
  Basis Set for First-Row Elements, Li-F. \emph{J. Comput. Chem.}
  \textbf{1983}, \emph{4}, 294\relax
\mciteBstWouldAddEndPuncttrue
\mciteSetBstMidEndSepPunct{\mcitedefaultmidpunct}
{\mcitedefaultendpunct}{\mcitedefaultseppunct}\relax
\EndOfBibitem
\bibitem[Krishnan et~al.(1980)Krishnan, Binkley, Seeger, and Pople]{6311g}
Krishnan,~R.; Binkley,~J.~S.; Seeger,~R.; Pople,~J.~A. Self-Consistent
  Molecular Orbital Methods. XX. A Basis Set for Correlated Wave Functions.
  \emph{J. Chem. Phys.} \textbf{1980}, \emph{72}, 650\relax
\mciteBstWouldAddEndPuncttrue
\mciteSetBstMidEndSepPunct{\mcitedefaultmidpunct}
{\mcitedefaultendpunct}{\mcitedefaultseppunct}\relax
\EndOfBibitem
\bibitem[Frisch et~al.(1984)Frisch, Pople, and Binkley]{6311g_polarization}
Frisch,~M.~J.; Pople,~J.~A.; Binkley,~J.~S. Self-Consistent Molecular Orbital
  Methods 25. Supplementary Functions for Gaussian Basis Sets. \emph{J. Chem.
  Phys.} \textbf{1984}, \emph{80}, 3265\relax
\mciteBstWouldAddEndPuncttrue
\mciteSetBstMidEndSepPunct{\mcitedefaultmidpunct}
{\mcitedefaultendpunct}{\mcitedefaultseppunct}\relax
\EndOfBibitem
\bibitem[Piecuch et~al.(2002)Piecuch, Kucharski, Kowalski, and
  Musia≈Ç]{EOMCCSD1}
Piecuch,~P.; Kucharski,~S.~A.; Kowalski,~K.; Musia≈Ç,~M. Efficient Computer
  Implementation of the Renormalized Coupled-Cluster Methods: The R-CCSD[T],
  R-CCSD(T), CR-CCSD[T], and CR-CCSD(T) Approaches. \emph{Comput. Phys.
  Commun.} \textbf{2002}, \emph{149}, 71\relax
\mciteBstWouldAddEndPuncttrue
\mciteSetBstMidEndSepPunct{\mcitedefaultmidpunct}
{\mcitedefaultendpunct}{\mcitedefaultseppunct}\relax
\EndOfBibitem
\bibitem[Kowalski and Piecuch(2004)Kowalski, and Piecuch]{EOMCCSD2}
Kowalski,~K.; Piecuch,~P. New Coupled-Cluster Methods With Singles, Doubles,
  and Noniterative Triples for High Accuracy Calculations of Excited Electronic
  States. \emph{J. Chem. Phys.} \textbf{2004}, \emph{120}, 1715\relax
\mciteBstWouldAddEndPuncttrue
\mciteSetBstMidEndSepPunct{\mcitedefaultmidpunct}
{\mcitedefaultendpunct}{\mcitedefaultseppunct}\relax
\EndOfBibitem
\bibitem[Wloch et~al.(2005)Wloch, Gour, Kowalski, and Piecuch]{EOMCCSD3}
Wloch,~M.; Gour,~J.~R.; Kowalski,~K.; Piecuch,~P. Extension of Renormalized
  Coupled-Cluster Methods Including Triple Excitations to Excited Electronic
  States of Open-Shell Molecules. \emph{J. Chem. Phys.} \textbf{2005},
  \emph{122}, 214107\relax
\mciteBstWouldAddEndPuncttrue
\mciteSetBstMidEndSepPunct{\mcitedefaultmidpunct}
{\mcitedefaultendpunct}{\mcitedefaultseppunct}\relax
\EndOfBibitem
\bibitem[Ivanic and Ruedenberg(2001)Ivanic, and Ruedenberg]{MCSCF_Direct}
Ivanic,~J.; Ruedenberg,~K. Identification of Deadwood in Configuration Spaces
  Through General Direct Configuration Interaction. \emph{Theo. Chem. Acc.}
  \textbf{2001}, \emph{106}, 339\relax
\mciteBstWouldAddEndPuncttrue
\mciteSetBstMidEndSepPunct{\mcitedefaultmidpunct}
{\mcitedefaultendpunct}{\mcitedefaultseppunct}\relax
\EndOfBibitem
\bibitem[Cossi and Barone(2001)Cossi, and Barone]{TDDFT_CPCM}
Cossi,~M.; Barone,~V. Time-Dependent Density Functional Theory for Molecules in
  Liquid Solutions. \emph{J. Chem. Phys.} \textbf{2001}, \emph{115}, 4708\relax
\mciteBstWouldAddEndPuncttrue
\mciteSetBstMidEndSepPunct{\mcitedefaultmidpunct}
{\mcitedefaultendpunct}{\mcitedefaultseppunct}\relax
\EndOfBibitem
\bibitem[Tawada et~al.(2004)Tawada, Tsuneda, Yanagisawa, Yanai, and
  Hirao]{uBLYP}
Tawada,~Y.; Tsuneda,~T.; Yanagisawa,~S.; Yanai,~T.; Hirao,~K. A
  Long-Range-Corrected Time-Dependent Density Functional Theory. \emph{J. Chem.
  Phys.} \textbf{2004}, \emph{120}, 8425\relax
\mciteBstWouldAddEndPuncttrue
\mciteSetBstMidEndSepPunct{\mcitedefaultmidpunct}
{\mcitedefaultendpunct}{\mcitedefaultseppunct}\relax
\EndOfBibitem
\bibitem[Schmidt et~al.(1993)Schmidt, Baldridge, Boatz, Elbert, Gordon, Jensen,
  Koseki, Matsunaga, Nguyen, Su, Windus, Dupuis, and Montgomery]{GAMESS}
Schmidt,~M.~W.; Baldridge,~K.~K.; Boatz,~J.~A.; Elbert,~S.~T.; Gordon,~M.~S.;
  Jensen,~J.~H.; Koseki,~S.; Matsunaga,~N.; Nguyen,~K.~A.; Su,~S.;
  Windus,~T.~L.; Dupuis,~M.; Montgomery,~J.~A. General Atomic and Molecular
  Electronic Structure System. \emph{J. Comput. Chem.} \textbf{1993},
  \emph{14}, 1347\relax
\mciteBstWouldAddEndPuncttrue
\mciteSetBstMidEndSepPunct{\mcitedefaultmidpunct}
{\mcitedefaultendpunct}{\mcitedefaultseppunct}\relax
\EndOfBibitem
\bibitem[M.S.Gordon and M.W.Schmidt(2005)M.S.Gordon, and M.W.Schmidt]{GAMESS2}
M.S.Gordon,; M.W.Schmidt, In \emph{Theory and Applications of Computational
  Chemistry, the First Forty Years}; C.E.Dykstra,, G.Frenking,, K.S.Kim,,
  G.E.Scuseria,, Eds.; Elsevier, 2005; p 1167\relax
\mciteBstWouldAddEndPuncttrue
\mciteSetBstMidEndSepPunct{\mcitedefaultmidpunct}
{\mcitedefaultendpunct}{\mcitedefaultseppunct}\relax
\EndOfBibitem
\bibitem[Shao et~al.(2006)Shao, Molnar, Jung, Kussmann, Ochsenfeld, Brown,
  Gilbert, Slipchenko, Levchenko, O{'}Neill, DiStasio~Jr, Lochan, Wang, Beran,
  Besley, Herbert, Yeh~Lin, Van~Voorhis, Hung~Chien, Sodt, Steele, Rassolov,
  Maslen, Korambath, Adamson, Austin, Baker, Byrd, Dachsel, Doerksen, Dreuw,
  Dunietz, Dutoi, Furlani, Gwaltney, Heyden, Hirata, Hsu, Kedziora, Khalliulin,
  Klunzinger, Lee, Lee, Liang, Lotan, Nair, Peters, Proynov, Pieniazek,
  Min~Rhee, Ritchie, Rosta, David~Sherrill, Simmonett, Subotnik, Lee
  Woodcock~III, Zhang, Bell, Chakraborty, Chipman, Keil, Warshel, Hehre,
  Schaefer~III, Kong, Krylov, Gill, and Head-Gordon]{QCHEM}
Shao,~Y. et~al.  Advances in Methods and Algorithms in a Modern Quantum
  Chemistry Program Package. \emph{Phys. Chem. Chem. Phys.} \textbf{2006},
  \emph{8}, 3172\relax
\mciteBstWouldAddEndPuncttrue
\mciteSetBstMidEndSepPunct{\mcitedefaultmidpunct}
{\mcitedefaultendpunct}{\mcitedefaultseppunct}\relax
\EndOfBibitem
\bibitem[L\"{o}wdin(1955)]{Lowdin1955}
L\"{o}wdin,~P.-O. {Quantum Theory of Many-Particle Systems. I. Physical
  Interpretations by Means of Density Matrices, Natural Spin-Orbitals, and
  Convergence Problems in the Method of Configurational Interaction}.
  \emph{Physical Review} \textbf{1955}, \emph{97}, 1474--1489\relax
\mciteBstWouldAddEndPuncttrue
\mciteSetBstMidEndSepPunct{\mcitedefaultmidpunct}
{\mcitedefaultendpunct}{\mcitedefaultseppunct}\relax
\EndOfBibitem
\bibitem[Davydov(1964)]{Davydov1964}
Davydov,~A.~S. {The Theory of Molecular Excitons}. \emph{Sov. Phys. Usp.}
  \textbf{1964}, \emph{7}, 145--178\relax
\mciteBstWouldAddEndPuncttrue
\mciteSetBstMidEndSepPunct{\mcitedefaultmidpunct}
{\mcitedefaultendpunct}{\mcitedefaultseppunct}\relax
\EndOfBibitem
\bibitem[Agranovich(2008)]{Agranovich2008}
Agranovich,~V. \emph{{Excitations in Organic Solids}}; Oxford University Press,
  2008\relax
\mciteBstWouldAddEndPuncttrue
\mciteSetBstMidEndSepPunct{\mcitedefaultmidpunct}
{\mcitedefaultendpunct}{\mcitedefaultseppunct}\relax
\EndOfBibitem
\bibitem[Agranovich and Basko(2000)Agranovich, and Basko]{Agr.Bas-2000}
Agranovich,~V.~M.; Basko,~D.~M. {Frenkel excitons beyond the Heitler-London
  approximation}. \emph{J. Chem. Phys.} \textbf{2000}, \emph{112},
  8156--8162\relax
\mciteBstWouldAddEndPuncttrue
\mciteSetBstMidEndSepPunct{\mcitedefaultmidpunct}
{\mcitedefaultendpunct}{\mcitedefaultseppunct}\relax
\EndOfBibitem
\bibitem[Mulliken(1955)]{Mulliken1955}
Mulliken,~R.~S. {Electronic Population Analysis on LCAO[Single Bond]MO
  Molecular Wave Functions. I}. \emph{J. Chem. Phys.} \textbf{1955}, \emph{23},
  1833\relax
\mciteBstWouldAddEndPuncttrue
\mciteSetBstMidEndSepPunct{\mcitedefaultmidpunct}
{\mcitedefaultendpunct}{\mcitedefaultseppunct}\relax
\EndOfBibitem
\bibitem[Madjet et~al.(2006)Madjet, Abdurahman, and Renger]{Madjet:2006ce}
Madjet,~M.~E.; Abdurahman,~A.; Renger,~T. {Intermolecular Coulomb Couplings
  from Ab Initio Electrostatic Potentials: Application to Optical Transitions
  of Strongly Coupled Pigments in Photosynthetic Antennae and Reaction
  Centers}. \emph{J. Phys. Chem. B} \textbf{2006}, \emph{110},
  17268--17281\relax
\mciteBstWouldAddEndPuncttrue
\mciteSetBstMidEndSepPunct{\mcitedefaultmidpunct}
{\mcitedefaultendpunct}{\mcitedefaultseppunct}\relax
\EndOfBibitem
\bibitem[Muh et~al.(2007)Muh, Madjet, Adolphs, Abdurahman, Rabenstein,
  Ishikita, Knapp, and Renger]{Muh.Mad.etal-2007}
Muh,~F.; Madjet,~M.~E.; Adolphs,~J.; Abdurahman,~A.; Rabenstein,~B.;
  Ishikita,~H.; Knapp,~E.-W.; Renger,~T. {$\alpha$-Helices Direct Excitation
  Energy Flow in the Fenna-Matthews-Olson Protein}. \emph{Proc. Nat. Acad. Sc.}
  \textbf{2007}, \emph{104}, 16862\relax
\mciteBstWouldAddEndPuncttrue
\mciteSetBstMidEndSepPunct{\mcitedefaultmidpunct}
{\mcitedefaultendpunct}{\mcitedefaultseppunct}\relax
\EndOfBibitem
\bibitem[Novoderezhkin et~al.(2004)Novoderezhkin, Yakovlev, van Grondelle, and
  Shuvalov]{novoderezhkin2004coherent}
Novoderezhkin,~V.~I.; Yakovlev,~A.~G.; van Grondelle,~R.; Shuvalov,~V.~A.
  Coherent Nuclear and Electronic Dynamics in Primary Charge Separation in
  Photosynthetic Reaction Centers: a Redfield Theory Approach. \emph{J. Phys.
  Chem. B} \textbf{2004}, \emph{108}, 7445--7457\relax
\mciteBstWouldAddEndPuncttrue
\mciteSetBstMidEndSepPunct{\mcitedefaultmidpunct}
{\mcitedefaultendpunct}{\mcitedefaultseppunct}\relax
\EndOfBibitem
\bibitem[Dreuw and Head-Gordon(2004)Dreuw, and Head-Gordon]{Dreuw2004}
Dreuw,~A.; Head-Gordon,~M. {Failure of Time-Dependent Density Functional Theory
  for Long-Range Charge-Transfer Excited States: the
  Zincbacteriochlorin-Bacteriochlorin and Bacteriochlorophyll-Spheroidene
  Complexes}. \emph{J. Am. Chem. Soc.} \textbf{2004}, \emph{126},
  4007--16\relax
\mciteBstWouldAddEndPuncttrue
\mciteSetBstMidEndSepPunct{\mcitedefaultmidpunct}
{\mcitedefaultendpunct}{\mcitedefaultseppunct}\relax
\EndOfBibitem
\bibitem[Pan et~al.(2009)Pan, Gao, Liang, and Zhao]{Pan2009}
Pan,~F.; Gao,~F.; Liang,~W.; Zhao,~Y. {Nature of Low-Lying Excited States in
  H-Aggregated Perylene Bisimide Dyes: Results of TD-LRC-DFT and the Mixed
  Exciton Model}. \emph{J. Phys. Chem. B} \textbf{2009}, \emph{113},
  14581--7\relax
\mciteBstWouldAddEndPuncttrue
\mciteSetBstMidEndSepPunct{\mcitedefaultmidpunct}
{\mcitedefaultendpunct}{\mcitedefaultseppunct}\relax
\EndOfBibitem
\end{mcitethebibliography}

\end{document}